\documentclass[10pt,aps,prc,twocolumn,amsmath,superscriptaddress,preprintnumbers]
{revtex4-1}
\usepackage{color,graphicx}
\usepackage{mathptmx}                
\usepackage{dcolumn}                 
\usepackage{bm}                      
\usepackage{ulem,soul}
\usepackage{hyperref}                

\begin{document}

\preprint{INT-PUB-17-029}

\title{The equation of state for dense nucleonic matter from a metamodeling. I. Foundational aspects}

\author{J\'er\^ome Margueron}
\affiliation{Institute for Nuclear Theory, University of Washington, Seattle, Washington 98195, USA}
\affiliation{Institut de Physique Nucl\'eaire de Lyon, CNRS/IN2P3, Universit\'e de Lyon, Universit\'e Claude Bernard Lyon 1, F-69622 Villeurbanne Cedex, France}

\author{Rudiney Hoffmann Casali}
\affiliation{Institut de Physique Nucl\'eaire de Lyon, CNRS/IN2P3, Universit\'e de Lyon, Universit\'e Claude Bernard Lyon 1, F-69622 Villeurbanne Cedex, France}
\affiliation{Departamento de F\'isica, Instituto Tecnol\'ogico de Aeron\'autica, CTA, 12228900, S\~ao Jos\'e dos Campos, SP, Brazil}

\author{Francesca Gulminelli}
\affiliation{CNRS, ENSICAEN, UMR6534, LPC ,F-14050 Caen cedex, France}

\date{\today}

\begin{abstract}
A metamodeling for the nucleonic equation of state (EOS), inspired from a Taylor expansion around the saturation density of symmetric nuclear matter, is proposed and parameterized in terms of the empirical parameters.
The present knowledge of nuclear empirical parameters is  first reviewed in order to estimate their average values and associated uncertainties, and thus defining the parameter space of the metamodeling.
They are divided into isoscalar and isovector type, and ordered according to their power in the density expansion.
The goodness of the metamodeling is analyzed against the predictions of the original models.
In addition, since no correlation among the empirical parameters is assumed a priori, all arbitrary density dependences can be explored, which might not be accessible in existing functionals. 
Spurious correlations due to the assumed functional form are also removed.
This meta-EOS allows direct relations between the uncertainties on the empirical parameters and the density dependence of the nuclear equation of state and its derivatives, and the mapping between the two can be done with standard Bayesian techniques. 
A sensitivity analysis shows that the more influential empirical parameters are the isovector parameters $L_{sym}$ and $K_{sym}$, and that laboratory constraints at super-saturation densities are essential to reduce the present uncertainties.
The present metamodeling for the EOS for nuclear matter is proposed for further applications in neutron stars and supernova matter.
\end{abstract}

\maketitle

\section{Introduction}
\label{Sec: Introduction}

Since the discovery of neutron stars (NS) in 1967~\cite{Bell1967,Hewish1969a,Hewish1969b}, 
the accurate prediction of the nuclear equation of state (EOS) has become of great importance 
and a lot of efforts, both from the theoretical and the experimental side, have been devoted to this aim. 
The seminal work of Tolman, Oppenheimer and Volkov in 1939 had proved that considering only the
kinetic contribution of nucleons to nuclear matter equation of state provides a limit in the maximum mass
of neutron stars of about 0.7M$_\odot$~\cite{Tolman1939,Oppenheimer1939}.
This contradicts the present observations for the canonical NS mass which is of the order
of 1.44M$_\odot$~\cite{Lattimer2005b}, as well as the recent observations proving the existence of about 
2M$_\odot$ NS~\cite{Demorest2010a,Antoniadis2013}.
These observational data clearly demonstrate the importance of the nuclear interaction for the understanding 
of the global properties of neutron stars.

Several ab-initio approaches have been developed for the accurate prediction of NS equation of state, see for
instance Refs.~\cite{Baldo2012,Gandolfi2014b} for recent reviews.
More recently, new nuclear potentials (chiral EFT) have been developed offering the possibility to perform calculations
in perturbation theory~\cite{Tews2013,Drischler2016} and they have been implemented as well in Quantum Monte-Carlo (QMC)
methods~\cite{Roggero2014,Wlazlowski2014,Lynn2017}. 
These potentials have also been applied to the NS EOS, see for instance Ref.~\cite{Hebeler2013,Tews2017}.
While there is a convergence in the prediction of these EOS at low density, such method might fail above the saturation 
density of nuclear matter because an expansion in supposedly small parameters is no longer really valid there.
In addition, there are larger and larger deviations between the different predictions above saturation density, 
mainly because of the different treatments of the many-body correlations and the different nuclear interactions, see for 
instance Ref.~\cite{Dutra2012} for a detailed comparison of some of these approaches.

With the development of x-ray observations of the thermal emission from the surface of neutron stars, it was envisioned that the nuclear EOS 
may be directly determined from observational data such as NS radii~\cite{Read2009,Ozel2009,Steiner2010a,Ozel2010,Guillot2011,Guillot2013,Steiner2013,Lattimer2014,Guillot2014,Guillot2016,Raithel2016,Raithel2017}.
In all these papers, the nuclear EOS is  expressed in terms of a reduced number of parameters, such as for instance
matching densities of piecewise polytropes first introduced in Ref.~\cite{Read2009}.
The use of polytropes, while extremely simple and not to far from the model predictions, does not allow however a simple 
connection to the present nuclear physics knowledge, such as nuclear saturation and empirical parameters
nor can bring information concerning matter composition, such as the proton fraction.
It is therefore interesting to extend these ideas towards a more complementary approach between astrophysical and nuclear experiments constraints.

Other approaches for the nuclear EOS are derived from some simple nuclear interaction, such as for instance Skyrme-type
contact interactions complemented by a density dependent term~\cite{Hebeler2013,Alam2014}.
While extremely useful and simple, the density dependent term usually brings correlations among the nuclear empirical parameters
which may be unphysical~\cite{Khan2012,Khan2013}.
Non-relativistic Skyrme-type EOS~\cite{Roggero2015,Rrapaj2016} as well as relativistic ones~\cite{Santos2015,Moustakidis2017} can be selected according to their ability to reproducing \textsl{ab-initio} calculations.

A third modeling of the nuclear EOS is based on a Taylor expansion of the nuclear EOS around saturation density~\cite{Steiner2010a}
or a Fermi momentum expansion~\cite{Bulgac2015}.
This kind of approach offers a unique possibility to incorporate in the nuclear EOS the best knowledge issued from nuclear
physics, reducing the number of free parameters.
{The Taylor expansion allows the separation of the low order derivatives, which are better determined by nuclear experiments, from the high orders ones, which } are best determined by NS observations. 
Indeed, the higher order parameters are more sensitive to the EOS at the highest densities, which are difficult to access from nuclear laboratory experiments.

In this paper a metamodel, or a "model of a model"~\cite{Book:Kleijnen}, for the nucleonic EOS is proposed and analyzed.
Metamodels are practical solution to solve complex and numerical issues and/or to facilitate optimization under uncertainty. 
They are therefore often used to provide fast approximations to the results of more complex problems, and to perform comparative analysis of different models belonging to the class covered by the metamodeling.
Conceptually, metamodels build a hyper-surface from a limited amount of input and output data and approximate the output over a much wider parameter space, see Refs.~\cite{Simpson2001,Jin2003,Jin2005} for an  overview of metamodeling techniques.
Metamodels have to be evaluated with respect to the goodness and there is no proof of existence or of uniqueness in general.
A metamodel is always associated to a given model or class of models. In the present application, we will consider homogeneous nucleonic EOS.
In principle, different metamodels can be introduced to represent different model classes, e.g. nucleonic EOS against high density phase transition EOS.
The goodness of the data adjustment with respect to one of these classes can, for instance, be analyzed by introducing Bayesian factors~\cite{Book:Kleijnen}.
We introduce the concept of a metamodel for the nucleonic EOS since it present several interesting advantages:
i) it provides a unique mapping of  very different existing EOS with many different input parameters,
ii) it provides a flexible approach that can interpolate continuously between existing EOS,
iii) as a consequence, it may orientate the preferred input parameters towards values which are not among the existing EOS,
iv) it allows the definition of a generic model where the nuclear physics knowledge acquired from laboratory experiments can be simply encoded as input parameters,
v) it includes in its parameter space the results of complex \textsl{ab-initio} models, and can thus be used to extract the constraints on the EOS imposed by them,
vi) and finally, combined with the Bayesian framework, it facilitates the estimation of the experimental and theoretical error bars into confidence levels for the astrophysics observables.
In this paper, we introduce and analyze the properties of  this nucleonic metamodeling,  while the connection with NS observables is performed in a second paper~\cite{Margueron2017b}.
Further extensions of this approach to the description of non-homogeneous matter and/or of dense matter phase transitions can easily be developed in the future from the present framework.

The present paper is organized as follows: in Sec.~\ref{Sec:empirical}, a review of the experimental  information on the nuclear empirical parameters is performed, and their uncertainties are estimated.
To this aim, predictions from relativistic and non-relativistic, phenomenological and ab-initio interactions, are compiled and compared, and uncertainties are obtained from a statistical analysis.
The metamodeling is formulated in Sec.~\ref{Sec:eeos}, presenting different options for the Taylor expansion.  The quality of the different strategies is estimated by comparing the convergence of predictions with respect to a reference EOS. 
Sec.~\ref{sec:system} explores the flexibility of the meta-EOS. 
We show that this metamodeling can very accurately reproduce a large number of existing EOS, and at the same time it can explore density dependences which are not accessible to usual phenomenological functionals because of the imposed functional form.
In that section, it is also shown that the huge uncertainty in higher order empirical parameters can only be reduced if extra empirical information is added on 
 a second higher density-reference point, in addition to the saturation density. 
One of the advantages of the present meta-EOS is the fact that no a-priori correlations are imposed on the empirical parameters.
{The physical correlations can be added \textsl{a-posteriori} as illustrated in the second paper~\cite{Margueron2017b}.} 
We perform a sensitivity analysis of the meta-EOS to the different empirical parameters by varying them one-by-one according to their uncertainties. 
This is done in  Sec.~\ref{sec:eos}, where we show that the most influential parameters are the isovector ones, namely $L_{sym}$, $K_{sym}$ and $Q_{sym}$. 
This stresses once again the need of experimental constraints at high density on asymmetric matter, typically from high energy heavy ion collisions with rare isotopic beams.
Finally, conclusions and outlooks are presented in Sec.~\ref{Sec:Conclusions}.

\section{Empirical characterization of the nuclear equation of state}
\label{Sec:empirical}

In the following we analyze the properties of nuclear matter composed of neutrons and protons with
different isoscalar (is) density $n_0=n_n+n_p$ and isovector (iv) density 
$n_1=n_n-n_p$, where $n_{n/p}$ is the neutron/proton density defined as,
\begin{equation}
n_{n/p} 
= \frac{1}{3\pi^2} k_{F_{n/p}}^3 ,
\end{equation}
where $k_{F_{n/p}}$ is the neutron/proton Fermi energy.
Isospin asymmetric nuclear matter (ANM) can also be defined in terms of the asymmetry parameter
$\delta=n_1/n_0$.
The two boundaries $\delta=0$ and 1 correspond to symmetric nuclear matter (SNM) and to 
pure neutron matter (PNM).
The saturation density of SNM is defined as the density at which the symmetric matter pressure is zero and
it is denoted as $n_{sat}$.

The general properties of relativistic and non-relativistic nuclear interactions are often characterized in
terms of the nuclear empirical parameters, defined as the coefficients of the following series expansion in the
parameter $x=(n_0-n_{sat})/(3n_{sat})$~\cite{Piekarewicz2009},
\begin{eqnarray}
 e_{is} &=& E_{sat}+\frac{1}{2}K_{sat}x^{2}+\frac{1}{3!}Q_{sat}x^{3}+\frac{1}{4!}Z_{sat}x^{4}+...,
 \label{eq:eis}\\
 e_{iv} &=& E_{sym}+L_{sym}x+\frac{1}{2}K_{sym}x^{2}+\frac{1}{3!}Q_{sym}x^{3} 
+\frac{1}{4!}Z_{sym}x^{4}+...,
\label{eq:eiv} \nonumber \\
 \end{eqnarray} 
where the isoscalar energy $e_{is}$ and the isovector energy $e_{iv}$ enter into the definition of
the energy per nucleon in nuclear matter, defined as
\begin{eqnarray}
 e(n_0,n_1)&=&e_{is}(n_0)+\delta^{2} e_{iv}(n_0).
\label{eq:bindingenergy} 
\end{eqnarray} 
The isovector energy $e_{iv}$ is often called the symmetry energy $S(n_0)=e_{iv}(n_0)$.
Note that this definition implies a parabolic approximation for the isospin dependence, while  
the proper definition is given by the second derivative with respect to $\delta$ around symmetric matter, 
see  Eq.~(\ref{eq:symmetryenergy}) below.

The empirical parameters entering the series expansion (\ref{eq:eis}) and (\ref{eq:eiv}) are
separated into two channels~\cite{Ducoin2010,Dutra2012}: the isoscalar channel which defines the saturation energy $E_{sat}$,
the saturation density $n_{sat}$, the incompressibility modulus $K_{sat}$, the isoscalar skewness $Q_{sat}$, and
the isoscalar kurtosis $Z_{sat}$; and the isovector channel which defines the symmetry energy 
$E_{sym}$, the slope $L_{sym}$, the isovector incompressibility $K_{sym}$, the isovector skewness $Q_{sym}$, and
the isovector kurtosis $Z_{sym}$.
There is no unique nomenclature for the empirical parameters, but in principle, Eqs.~(\ref{eq:eis})-(\ref{eq:eiv}) 
makes our's unambiguous.
A very clear synthesis of the various terminologies used in the literature is discussed in the appendix of 
Ref.~\cite{Piekarewicz2009}.

The energy per nucleon~(\ref{eq:bindingenergy}) can be expressed in the following compact form~\cite{Ducoin2010,Ducoin2011},
\begin{eqnarray}
e(n_0,n_1)&=& \sum_{\alpha\ge 0} \frac{1}{\alpha!} \left( c_\alpha^{is} + c_\alpha^{iv}\delta^2\right) x^\alpha ,
\label{eq:etotisiv}
\end{eqnarray} 
where the coefficients $c_\alpha^{is/iv}$ are the empirical parameters introduced in Eqs.~(\ref{eq:eis})-(\ref{eq:eiv})~\cite{Piekarewicz2009}.
Note that the coefficient $c_1^{is}=0$ due to the choice of the saturation density $n_{sat}$ as the reference density in the definition
of $x$.
Consequently, choosing an arbitrary density as reference in the definition of $x$ would lead to a non-vanishing $c_1^{is}$, and $n_{sat}$ would be determined by the isoscalar empirical parameters.
The total number of free parameters is thus conserved: considering $c_1^{is}$ or $n_{sat}$ as isoscalar empirical parameter for $\alpha=1$,
it is 2 per exponent $\alpha$.

The empirical properties are determined from nuclear physics experiments such as measurements of nuclear 
masses, of charge-density profiles, from analysis of collective modes (ISGMR, IVGDR, etc...). 
More details in the experimental determinations of the empirical parameters are presented in Sec.~\ref{Sec:empirical:exp}.

The series expansion (\ref{eq:etotisiv}) in the parameter $x$ is in principle infinite, and it is
not guaranteed that this expansion converges.
The convergence property is however analyzed in Sec.~\ref{Sec:eeos}, and anticipating the results,
it is shown that in a density range going up to 4$n_{sat}$ an order by order convergence for the binding
energy, the pressure and the sound velocity is found. 
This result is tested for a large number of nuclear interactions in Sec.~\ref{sec:system}.
In this section, we therefore concentrate on the experimental determination of the first terms in the 
expansion (\ref{eq:etotisiv}).


The expansion in the asymmetry parameter $\delta$ in Eq.~(\ref{eq:etotisiv}) does not include terms beyond second order in $\delta$. 
Note however that small corrections may appear, such as for instance those induced by the $T=0$ pairing or quarteting~\cite{Myers1969,Reinhard2006},
which have been considered in recent works, see for instance Refs.~\cite{Cai2012,Moustakidis2012,Seif2014,Bulgac2015}.
\textsl{Ab-initio} approaches  show that the energy per nucleon in homogeneous asymmetric nuclear matter is mostly quadratic in $\delta$~\cite{Zuo2002,Vidana2009}, and   
residual non-quadraticities are mostly related to the kinetic energy part of the total energy (including the effective mass splitting)~\cite{Ducoin2011}.
This is also confirmed by an analysis of various finite-range nuclear forces~\cite{Tsukioka2017}.
For this reason, in Sec.~\ref{Sec:eeos} we will replace the global expansion (\ref{eq:bindingenergy}) by an expression where the contribution of the kinetic energy is expressed separately,  and limit the parabolic approximation to the interaction part.

In the following, we first review the "experimental" determination of the first parameters in Eq.~(\ref{eq:etotisiv}), hereafter called "low order".
In sec.~\ref{Sec:empirical:exp} we list a large, but certainly not extensive, amount of referenced analyses where authors have optimized their models
on specific experimental data to extract some of the empirical parameters.
We call these analysis "experimental" by opposition of the generic determination which is presented in sec.~\ref{sec:empiricalparameters}.
In the generic approach, the parameters are directly deduced from a set of models known by their ability to reasonably well predict a large number of nuclear properties, such as masses and radii at least.
The generic approach is supposed to provide an upper bound on the empirical parameters uncertainties.
For the low order empirical parameters, a good overlap is found between the "experimental" analysis and the generic one.
The advantage of the generic analysis is that it could also provide an estimation of the uncertainties associated to the high order empirical parameters
which are yet quite unknown.

\subsection{Experimental determination of the nuclear empirical parameters}
\label{Sec:empirical:exp}

There is a very important experimental and theoretical program aiming at a better estimation of the nuclear empirical parameters.
For this reason, some of the empirical quantities are rather well determined.
This concerns essentially the first terms of the series expansion~(\ref{eq:etotisiv}), such as
the saturation energy, the saturation density, the incompressibility modulus and the symmetry energy.
We have grouped them in the so-called group A and presented them in Tab.~\ref{table:empiric1}.
The other empirical parameters are less well known, and we will show that this second group
of nuclear empirical parameters can be divided into two sub-groups: the one for which we can give a range of 
variation compatible with our experimental knowledge, the so-called group B shown in Tab.~\ref{table:empiric2}, 
and a group of parameters which are yet quite undetermined and not presently accessible by nuclear 
experiments, the so-called group C.
In the following, we review the experimental determination of the empirical parameters for the group A and B.
Let us however notice that the following review is not exhaustive but more illustrative. 
The aim of the subsection is to justify the current estimation of these empirical parameters.

The values reported in Tab.~\ref{table:empiric1} are extracted from experimental analysis and can therefore
be considered as closely related to nuclear data.
They are not directly determined from experimental data since these quantities are not accessible to experimental
probes without the use of a theoretical model.
For instance, the saturation density 
is extrapolated from fits of finite nuclei density profiles.
An additional difficulty comes from the fact that   the isoscalar density is not directly measurable from electron scattering in finite nuclei, and
the relation between the charge density and the total density is thus performed via a theoretical model.
{The neutron density can be determined in a relatively model-independent way by measurement of the parity-violating electron scattering asymmetry from $^{208}$Pb. This is the aim of the PREX experiment at Jefferson Lab~\cite{Abrahamyan2012}.}

\begin{table}[tb]
\centering
\setlength{\tabcolsep}{2pt}
\renewcommand{\arraystretch}{1.2}
\begin{ruledtabular}
\begin{tabular}{cccccc}
Model    & Ref. & $E_{sat}$ & $n_{sat}$ & $K_{sat}$ & $E_{sym}$ \\
          & &  MeV         & fm$^{-3}$ & MeV & MeV \\
\hline
El. scatt.  & Wang-99~\cite{Wang1999} & & 0.1607 & 235 & \\
 & & & & $\pm$15 & \\
\hline
LDM & Myers-66~\cite{Myers1966} & -15.677 & 0.136$^\dagger$ & 295 & 28.06 \\
LDM & Royer-08~\cite{Royer2008} & -15.5704 & 0.133$^\dagger$ & & 23.45 \\
LSD & Pomorski-03~\cite{Pomorski2003} & -15.492 & 0.142$^\dagger$ & & 28.82 \\
DM & Myers-77~\cite{Book:Myers1977} & -15.96 & 0.145$^\dagger$ & 240 & 36.8 \\
FRDM & Buchinger-01~\cite{Buchinger2001} & & 0.157 & & \\
                                             & & & $\pm$0.004 & & \\
INM & Satpathy-99~\cite{Satpathy1999} & -16.108 & 0.1620 & 288 & \\
        &                                 &              &             & $\pm$20 & \\
\hline
DF-Skyrme & Tondeur-86~\cite{Tondeur1986} & & 0.158 & & \\
DF-Skyrme & Klupfel-09~\cite{Klupfel2009} & -15.91 & 0.1610 & 222 & 30.7 \\
                   & & $\pm$0.06 & $\pm$0.0013 & $\pm$8 & $\pm$1.4 \\
DF-BSK2 & Goriely-02~\cite{Goriely2002a} & -15.79 & 0.1575 & 234 & 28.0 \\
DF-BSK24, & Goriely-15~\cite{Goriely2015} & -16.045 & 0.1575 & 245 & 30.0 \\
 28,29                  & & $\pm$0.005 & $\pm$0.0004 & &  \\
DF-Skyrme & McDonnell-15~\cite{McDonnell2015} & -15.75 & 0.160 & 220 & 29 \\
                   & & $\pm$0.25 & $\pm$0.005 & $\pm$20 & $\pm$1 \\
DF-NLRMF & NL3$^\ast$~\cite{RMFNL:NL3s} & -16.3 & 0.15 & 258 & 38.7 \\
DF-NLRMF & PK~\cite{RMF:PK} & -16.27 & 0.148 & 283 & 37.7 \\
DF-DDRMF & DDME1,2~\cite{RMFDD:DDME1,RMFDD:DDME2} & -16.17 & 0.152 & 247 & 32.7 \\
                   & & $\pm$0.03 & $\pm$0.00 & $\pm$3 & $\pm$0.4 \\
DF-DDRMF & PK~\cite{RMF:PK} & -16.27 & 0.150 & 262 & 36.8 \\
\hline
present & &  -15.8 & 0.155 & 230 & 32 \\
estimation & & $\pm$0.3 & $\pm$0.005 & $\pm$20 & $\pm$ 2 \\
\end{tabular}
\end{ruledtabular}
$^\dagger$ value determined from $r_0$.
\caption{Group A: saturation energy $E_{sat}$, density $n_{sat}$, incompressibility $K_{sat}$ and
symmetry energy $E_{sym}$ estimated from various analysis of experimental data.
See text for more details.}
\label{table:empiric1}
\end{table}

The values for the saturation energy reported in Tab.~\ref{table:empiric1}  are remarkably stable 
 in the different analysis.
From Tab.~\ref{table:empiric1}, the current value of $E_{sat}$ is estimated to be -15.8$\pm$0.3~MeV.
Let us mention a recent estimation of $E_{sat}$ and its uncertainty based on Liquid Drop Models (LDM) and the frequency-domain bootstrap method~\cite{Bertsch2017}.
The obtained value is $-15.56\pm0.17$~MeV, which is slightly lower, but still compatible with our current estimation.

The saturation density is more difficult to determine from the analysis presented in Tab.~\ref{table:empiric1}.
The value estimated from LDM is lower than the one obtained from Density Functional (DF) models,
which are supposed to provide the more accurate determination of the saturation density.
This is confirmed by the fact that the values extracted from the Droplet Model (DM) and the Finite-Range Droplet Model (FRDM), which are 
more realistic than the original LDM~\cite{Book:Myers1977}, are closer to the ones extracted by DF.
We have selected the DF models for which the value for the saturation density was not assumed
\textsl{a priori} in the fitting protocol to global properties of finite nuclei such as binding energies and 
charge radii.
The value obtained for the saturation density could therefore be considered as a prediction of these models.
In summary, we consider the following current estimation of 
$n_{sat}$=0.155$\pm$0.005~fm$^{-3}$.
Note that the error in the determination of these quantities was larger some decades ago, see for instance
\cite{Haensel1981}.

The incompressibility modulus $K_{sat}$ given in Tab.~\ref{table:empiric1} varies from
210~MeV up to 300~MeV, revealing here also the difficulty to estimate this quantity from
experimental data as well as its model dependence.
A more systematical review of the various theoretical predictions for $K_{sat}$ is presented in Ref.~\cite{Stone2014}.
The determination  of the incompressibility modulus from the LDM is usually not very {accurate}~\cite{Blaizot1980,Stone2014}.
A better determination can be obtained
from a method proposed by Blaizot~\cite{Blaizot1980,Blaizot1995}, 
based on the correlation between the isoscalar Giant Monopole Resonance (ISGMR) energy and the empirical parameter $K_{sat}$.
This estimation remains however quite model dependent, and for instance, a lower value 
$K_{sat}\approx 210$~MeV is obtained for the BCP functional~\cite{Baldo2013} and Gogny 
interactions~\cite{Blaizot1980}, while a higher value $K_{sat}\approx 250-270$~MeV is predicted from 
RMF approaches~\cite{Lalazissis1997,MA2001173,PhysRevC.68.024310}.
A part of this model dependence can be understood from the violations of self-consistency in some early calculations~\cite{Schlomo2002}.
This model dependence might also reveal a more complex correlation in terms of several empirical parameters, 
instead of the single one proposed by Blaizot.
It was indeed shown that the ISGMR is also sensitive to symmetry properties, and information on $K_{sat}$ cannot be easily deconvoluted from information on 
$K_{sym}$~\cite{Colo2004}.
For a deeper review, see Ref.~\cite{Colo2008}.
It was also recently shown that 
higher order isoscalar parameters also play a role, and the correlation analysis should be performed in terms of several empirical parameters instead of only one, such as for instance $K_{sat}$ and $Q_{sat}$~\cite{Khan2012,Khan2013}.
The value of $Q_{sat}$ is yet quite undetermined, 
and most of the model dependence in the determination of $K_{sat}$ can be  attributed to the uncertainties in $Q_{sat}$~\cite{Khan2013}. 
In other words, a better estimation of $Q_{sat}$ would refine the estimation of $K_{sat}$ based on the correlation with the ISGMR.
From a LDM approach separating the bulk contribution ($K_{sat}$) from the surface one (largely influenced by $Q_{sat}$), the importance of the surface properties for the determination of $K_{sat}$ was pointed out as well~\cite{Stone2014}.
An estimation of $K_{sat}=230\pm$40~MeV was given in Ref.~\cite{Khan2012,Khan2013} where the error-bar contain the maximum and minimum possible value for $K_{sat}$. 
It is therefore larger than a 1$\sigma$ uncertainty, where 1$\sigma$ is the error-bar accounting for 68\% of the models around the centroid.
In summary, the current estimation of $K_{sat}$  can be given as 
{230$\pm$20~MeV, where the error-bar corresponds to 1$\sigma$ uncertainty.}

\begin{figure}[tb]
\begin{center}
\includegraphics[angle=0,width=1.0\linewidth]{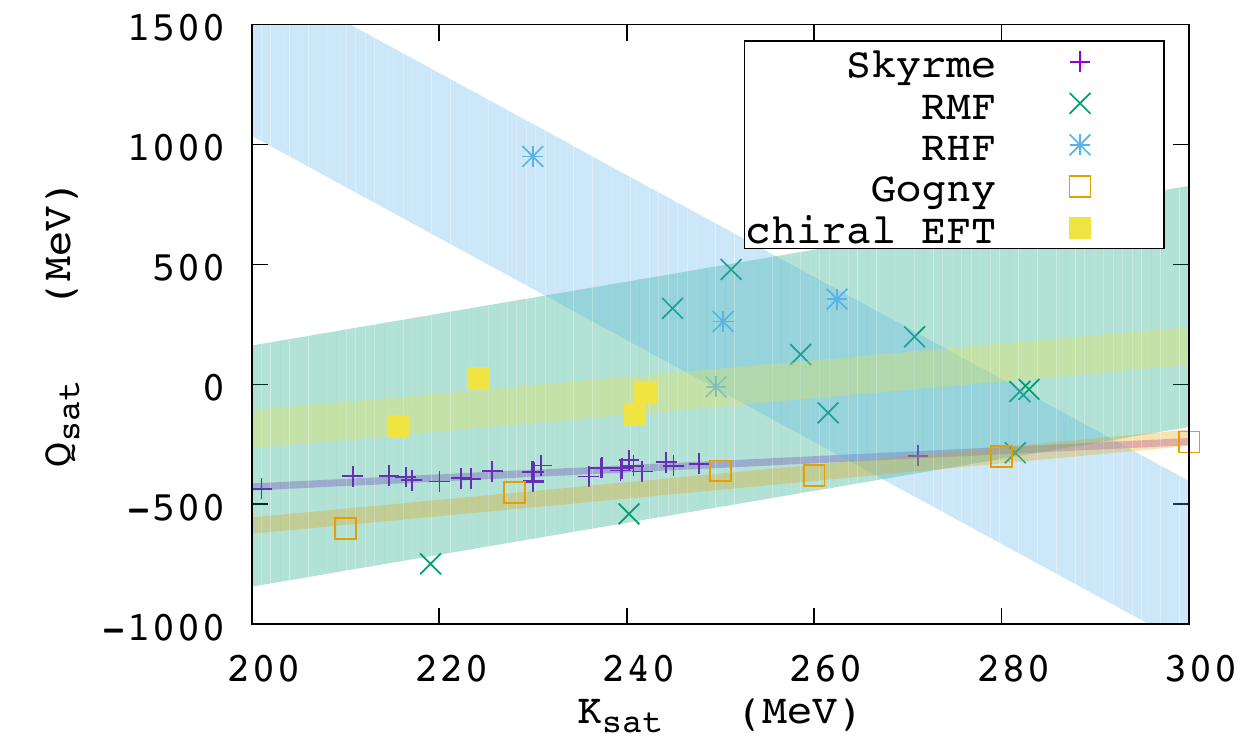}
\end{center}
\caption{(Color online) Correlation between the empirical parameters $K_{sat}$ and $Q_{sat}$
for different kind of nuclear interactions: Skyrme, Gogny, RMF, and RHF.
Points from EFT approach are also plotted. 
The points are obtained from Tabs.~\ref{tab:mean-table-models-XEFT}-\ref{table:empiricalrhf}, except for the Gogny model which is extracted from Ref.~\cite{Khan2013},
and the colored bands come from fits of the data including their dispersion considering 67\% of the best models.}
\label{fig:plotKQ} 
\end{figure}

It is interesting to observe the correlations between the empirical parameters $K_{sat}$ and $Q_{sat}$ represented
in Fig.~\ref{fig:plotKQ}.
This correlation is shown for Skyrme models (purple line), RMF models (light-green area), RHF 
models (light-blue area), Gogny models (orange line), and chiral EFT predictions (Yellow line).
The correlation bands for each models are shown for clarity.
They are obtained assuming a linear correlation between the values of $K_{sat}$ and $Q_{sat}$, and the width of the bands are determined from the $1-\sigma$ deviation.
The strongest correlation is found for the Skyrme and Gogny models, already suggested in Ref.~\cite{Khan2013}, and the
origin of this correlation can be found in the so-called $t_3$ density dependent terms which dominates in $K_{sat}$ and 
$Q_{sat}$.
It is however interesting to remark that also the relativistic models (RMF and RHF) exhibit a correlation between 
these empirical parameters, even if its origin is less easy to analyze.
In addition, the very different correlations between the various kinds of models shown in Fig.~\ref{fig:plotKQ}
indicate a strong model dependence of the correlation, that might not reflect a physical property.
Since $K_{sat}$ and $Q_{sat}$ governs the density dependence of the equation of state in SM and around saturation 
density, the correlation shown in Fig.~\ref{fig:plotKQ} indicates that models do not explore all possible density 
dependences.

This is one of the main motivations of the present work: in the following Sec.~\ref{sec:eos}, we propose a metamodeling which can explore the full parameter space (including $K_{sat}$ and $Q_{sat}$), with no a-priori restriction. {Physical correlations could be added by imposing some constraints to the metamodeling, as illustrated in the second paper~\cite{Margueron2017b}.}

While the binding energies $E_{sat}$ are predicted in a quite narrow interval for the various models 
presented in Tab.~\ref{table:empiric1}, the symmetry energy varies substantially between LDM, DM and 
DF models.
This might be because the value for the symmetry energy is very strongly related to the value of the 
slope of the symmetry energy $L_{sym}$ in many models~\cite{Carbone2010,Roca-Maza2013a,Danielewicz2014}.
The quantity which matters in the fit to experimental energies seems to be more closely related
to the symmetry energy at the average density of nuclei, at around $(2/3)n_{sat}$~\cite{Trippa2008}.
In addition, it has also been observed that RMF models prefer large values for the symmetry energy, 
such as 34-36~MeV, and it has been proposed that the symmetry energy and the incompressibility 
modulus $K_{sat}$ are correlated in DF models~\cite{Colo2004}.
Furthermore, a recent analysis of the bulk and surface contributions of the symmetry energy have shown 
that the sign of surface contribution depends strongly on the choice for the asymmetry parameter: the global
asymmetry parameter or the bulk asymmetry parameter, which contains a correction from the neutron 
skin~\cite{Aymard2014}. 
Considering this large model dependence, the current estimation of $E_{sym}$ is approximately 
32$\pm$2~MeV and this is in agreement with other estimations, see 
Refs.~\cite{Chabanat1997,Bender2003a,BALi2008,Danielewicz2009}.

\begin{table}[tb]
\centering
\setlength{\tabcolsep}{2pt}
\renewcommand{\arraystretch}{1.2}
\begin{ruledtabular}
\begin{tabular}{cccccc}
Model    & Ref. & $Q_{sat}$ & $L_{sym}$ & $K_{sym}$ & $K_{\tau}$ \\
        &  &  MeV         & MeV & MeV & MeV \\
\hline
DF-Skyrme & Berdichevsky-88~\cite{Berdichevsky1988} & 30 & 0 & & \\
DF-Skyrme & Farine-97~\cite{Farine1997} & -700 & & & \\
  & & $\pm$ 500 & & & \\
DF-Skyrme & Alam-14~\cite{Alam2014} & -344 & 65 & -23 & -322 \\
  & & $\pm$ 46 & $\pm$14 & $\pm$73 & $\pm$34\\
DF-Skyrme & McDonnell-15~\cite{McDonnell2015} &  & 40 & &  \\
                   & & & $\pm$20 &  &  \\
DF-NLRMF & NL3$^\ast$~\cite{RMFNL:NL3s} & 124 & 123 & 106 & -690 \\
DF-NLRMF & PK~\cite{RMF:PK} & -25 & 116 & 55 & -630 \\
DF-DDRMF & DDME1,2~\cite{RMFDD:DDME1,RMFDD:DDME2} & 400 & 53 & -94 & -500 \\
                   & & $\pm$80 & $\pm$3 & $\pm$7 & $\pm$7 \\
DF-DDRMF & PK~\cite{RMF:PK} & -119 & 79.5 & -50 & -491 \\
Correlation & Centelles-09~\cite{Centelles2009} & & 70 & & -425 \\
  & & & $\pm$ 40 & & $\pm$175 \\
DF-RPA & Carbone-10~\cite{Carbone2010} & & 60 & & \\
  & & & $\pm$ 30 & & \\
Correlation  & Danielewicz-14~\cite{Danielewicz2014} & & 53 & & \\
  & & & $\pm$ 20 & & \\
Correlation  & Newton-14~\cite{Newton2014} & & 70 & & \\
  & & & $\pm$ 40 & & \\
Correlation  & Lattimer-14~\cite{Lattimer2014a} & & 53 & & \\
  & & & $\pm$ 20 & & \\
GMR  & Sagawa-07~\cite{Sagawa2007b} & &  & & -500 \\
  & & & & & $\pm$ 50 \\
GMR  & Patel-14~\cite{Patel2014} & &  & & -550 \\
  & & & & & $\pm$ 100 \\
\hline
present & &  300 & 60 & -100 & -400 \\
estimation & & $\pm$400 & $\pm$15 & $\pm$100 & $\pm$ 100 \\
\end{tabular}
\end{ruledtabular}
\caption{Group B parameters: isoscalar skewness $Q_{sat}$, slope of the symmetry energy $L_{sym}$, 
isovector incompressibility $K_{sym}$.
The parameter $K_{\tau}$ is defined as $K_{\tau}=K_{sym}-6L_{sym}-Q_{sat}L_{sym}/K_{sat}$.
See text for more details.}
\label{table:empiric2}
\end{table}

We now discuss the parameters of group B given in Tab.~\ref{table:empiric2}: $Q_{sat}$, $L_{sym}$ and $K_{sym}$.
These parameters are not yet very well determined, but a better accuracy might be reached in the near future.
We first discuss the skewness parameter $Q_{sat}$.
This parameter is poorly known and there are very few experimental analysis which propose an estimation.
An analysis of charge and mass radii of Tin isotopes  concluded that either $Q_{sat}\approx 30$~MeV or $L_{sym}\approx 0$~MeV~\cite{Berdichevsky1988}.
Another analysis of the incompressibility modulus concluded that $Q_{sat}\approx -700\pm500$~MeV~\cite{Farine1997}.
This very large error bar reflects once again the model dependence of $Q_{sat}$,
induced by its correlation with the incompressibility modulus, as shown in Fig.~\ref{fig:plotKQ}.
It is therefore very difficult to estimate the value of this parameter and in the following, we shall explore a large domain.

The parameter $L_{sym}$ is much discussed nowadays and a large number of experiments aim at determining its value~\cite{BALi2014}.
Combining different constraints from neutron skin thickness, heavy-ion collisions, dipole polarizability, nuclear masses,
giant-dipole resonances and isobaric analog states, it was recently concluded that the value of $L_{sym}$ should
be between 33 and 72~MeV~\cite{BALi2013a,Lattimer2014a}.
Note that in Ref.~\cite{Lattimer2014a} the symmetry energy is comprised between 31 and 36~MeV, which is consistent with the
present estimation given in Tab.~\ref{table:empiric1}.
Other analysis predict slightly larger values for $L_{sym}$, and integrating all analysis, we come to the following
estimation: $L_{sym}=60\pm15$~MeV.

The isospin dependence of the ISGMR is a natural observable to determine the parameter $K_\tau$, defined as $K_{\tau}=K_{sym}-6L_{sym}-Q_{sat}L_{sym}/K_{sat}$~\cite{Piekarewicz2009}.
It represents the isoscalar curvature at the saturation density in asymmetric matter, $n_{sat}(\delta)\approx n_{sat}(1-3L_{sat}\delta^2/K_{sat})$.
The parameter $K_{sym}$ could, in principle, be deduced from $K_{\tau}$ if $L_{sym}$ and $Q_{sat}$ were well determined.
Considering the uncertainties in these parameters, we found a very naive estimation of the error-bar in $K_{sym}$, $\sigma\approx$600~MeV, which is certainly overestimated.
Waiting for better experimental analysis in the future, the value $K_{sym}=-100\pm100$~MeV given in Tab.~\ref{table:empiric2} is obtained from statistical analysis
of various model predictions, see Sec.~\ref{sec:empiricalparameters}.
It is mainly related to the expected values from chiral EFT approach and is comparable with the recent analysis from unitary gas constraint~\cite{Tews2017b}.
{Let us mention that this range for $K_{sym}$ is compatible with the one from Ref.~\cite{Constantinou2014} which is -100$\pm$200~MeV.
In our case we cover the same uncertainty range considering $2\sigma$ deviation from the central value.}

We now switch to the discussion of a quantity which is usually not considered as an empirical parameter, but 
enters nevertheless into the important quantities which characterize nuclear matter properties. 
The effective mass is a powerful concept used to characterize the propagation of quasiparticles inside a strongly 
interacting medium, such as nuclei or nuclear matter.
It reflects the non-locality in space and time of the quasiparticle self-energy.
The non-locality in space, also called the Landau effective mass or $k$-effective mass, is related to the momentum dependence of the nuclear interaction.
The Landau effective mass depends on the isoscalar and isovector densities and can be different for neutrons and protons, 
$m^*_q(n_0,n_1)$ where $q=n$, p.
In SM, it is generally assumed that $m^\ast_n=m^\ast_p$, while in AM the neutron and proton Landau effective mass can be different.
The isospin splitting of the Landau effective mass can then be expressed as,
\begin{equation}
\Delta m^*(n_0,n_1) = m^*_n(n_0,n_1)-m^*_p(n_0,n_1) .
\end{equation}
Two quantities are usually compared between various nuclear interactions: the Landau effective mass in SM at saturation
$m^*_{sat}$ and the isospin splitting taken for $n_0=n_1=n_{sat}$ in NM, $\Delta m^*_{sat}$.
A summary of the determination of $m^*_{sat}$ and $\Delta m^*_{sat}$ from nuclear experiments is shown in 
Table~\ref{table:empiric3}.

\begin{table}[tb]
\centering
\setlength{\tabcolsep}{2pt}
\renewcommand{\arraystretch}{1.2}
\begin{ruledtabular}
\begin{tabular}{ccccc}
Model    & Ref. &  $m_{sat}^*/m$ & $\kappa_v$ & $\Delta m_{sat}^*/m$ \\
\hline
DF-Skyrme~\cite{Lipparini1989a} & Lipparini-89 & & 0.2-0.54 \\
DF-Skyrme~\cite{Reinhard1999} & Reinhard-99 & 0.8$\pm$0.1 & 0.25$\pm$0.5 \\
DF-Skyrme~\cite{Lesinski2006} & Lesinski-06 & 0.75$\pm$0.05 & 0.6 & 0.17 \\
Opt. Pot.~\cite{Perey1962,Ma1983} & Perey-62 &  0.75$\pm$0.05 \\
Opt. Pot.~\cite{Dover1972} & Dover-72 &  0.75$\pm$0.03 \\
BHF~\cite{Hassaneen2004} & Hassaneen-04 & 0.65$\pm$0.05& & 0.1-0.2 \\
DBHF~\cite{Ma2004} & Ma-04 & 0.66 & & 0.1-0.2 \\
DBHF~\cite{VanDalen2005} & VanDalen-05 & 0.78 & & 0.1-0.2 \\
\hline
present                 &               & 0.75$\pm$0.1 & 0.4$\pm$0.2  & 0.1$\pm$0.1 \\
estimation \\
\end{tabular}
\end{ruledtabular}
\caption{Landau effective mass properties in nuclear matter at saturation density. 
From the estimated value of $m_{sat}^*/m$, we can deduce $\kappa_s$=0.43$\pm$0.1.
See text for more details.}
\label{table:empiric3}
\end{table}

The Landau effective mass can be extracted from the energy dependence of the
optical potential which is used in phenomenological analyses of nucleon scattering data.
By comparing the energy dependent term of the real optical potential in the energy range 10-30~MeV to the equivalent 
local potential from the Skyrme interaction, it was deduced that $m_{sat}^*$ is approximately
$(0.75\pm0.03) m$~\cite{Dover1972}.
A similar damping of the mass was also found by Perey based on phenomenological local and non-local potentials 
giving the same phase shifts~\cite{Perey1962,Ma1983}.
An apparently contradictory information comes from the measurement of level densities:
It was indeed observed in the 1960s that the experimental level density could be reproduced only if 
$m_{sat}^*\approx m$~\cite{Brown1963}.
The solution of this contradiction was found by recognizing that the mean field is not static, but it has also a
dynamic component~\cite{Bertsch1968}:
among the modes associated with the fluctuations of the field one finds vibrations of the nuclear 
surface~\cite{Bernard1979,Brown1979,Bortignon1987}, which is associated to an energy dependent effective mass (or $\omega-$mass, to be distinguished from the Landau effective mass).
This non-local in time property of the effective mass~\cite{VanGiai1983,Mahaux1985,Mahaux1992a}  however goes beyond the scope of the present model.
It is mentioned here only to illustrate the difficulty to accurately determine the effective mass from experiments.

The Landau effective mass in Skyrme models can be expressed as~\cite{Lesinski2006}
\begin{eqnarray}
\frac{m}{m^\ast_q} = 1+\kappa_s+\tau_3 (\kappa_s-\kappa_v) \delta ,
\end{eqnarray}
where $\kappa_s=m/m_{sat}^*-1$ in symmetric matter and $\kappa_v$ is the enhancement factor entering the 
Thomas-Reiche-Khun sum rule in the case of the iso-vector Giant Dipole Resonance E1 (IVGDR)~\cite{Lipparini1989a}.
There is a direct relation between $\kappa_s$ and the isoscalar Giant Quadrupolar Resonance (ISGQR)~\cite{Blaizot1980,Reinhard1999}, while
the value of $\kappa_v$ depends to a large extend on the energy region of the resonance energy~\cite{Lipparini1989a,Reinhard1999}.

So far, no experimental data from finite nuclei has allowed a determination of the effective mass splitting.
Microscopic approaches such as BHF and DBHF have been employed and predict in a non-ambiguous way that $m^*_n>m^*_p$ in neutron rich matter~\cite{Hassaneen2004,Ma2004,VanDalen2005}.
The sign of $\Delta m^*$ is solidly positive, but its amplitude is not yet clearly determined and believed to be around 0.1 to 0.2 $m$.
The neutron and proton Landau effective masses calculated within the Brueckner diagrammatic approach~\cite{Bombaci1991,Zuo1999a,Hofmann2001,Hassaneen2004,VanDalen2005,Satula2006}  
are reported in Table~\ref{table:empiric3}, see lines BHF (Brueckner-Hartree-Fock) and DBHF (Dirac-Brueckner-Hartree-Fock).

For small values of the isospin splitting $\Delta m_{sat}^*/m\ll 1$,   the following relation is approximately satisfied
\begin{equation}
\kappa_v\approx \kappa_s-\frac 1 2 \frac{\Delta m_{sat}^*}{m} \left( 1+\kappa_s\right)^2 .
\end{equation}

To summarize, the present estimation for these parameters can be expressed as $m^*_{sat}/m=0.75\pm0.1$, $\kappa_v=0.4\pm0.2$ and $\Delta m_{sat}^*/m=0.1\pm0.1$.

\begin{table*}[t]
\centering
\setlength{\tabcolsep}{2pt}
\renewcommand{\arraystretch}{1.2}
\begin{ruledtabular}
\begin{tabular}{cccccccccccccccccc}
    &   & $E_{sat}$ & $E_{sym}$ & $
n_{sat}$ &  $L_{sym}$ & $K_{sat}$ &  $K_{sym}$  & $Q_{sat}$  &  $Q_{sym}$ & $Z_{sat}$ & $Z_{sym}$  & 
$m^*_{sat}/m$ &$\Delta m^*_{sat}/m$ & $\kappa_{v}$ & $K_{\tau}$ \\
Model  & & MeV & MeV & fm$^{-3}$ & MeV & MeV & MeV & MeV & MeV & MeV & MeV & &  &  & MeV \\
($N_{\alpha}$) &  der. order   & 0 &0 & 1 & 1 & 2 & 2 & 3 & 3 & 4 & 4 & - & - & - & - \\
\hline
\multicolumn{16}{c}{Phenomenological approaches}\\
\hline
     Skyrme   & Average &   -15.88  &     30.25  &    0.1595  &      47.8  &     234  &    -130  &    -357  &     378  &    1500  &   -2219  &     0.73  &     0.08  &     0.46  &    -344\\
      (16)        & $\sigma$  &  0.15  &      1.70  &    0.0011  &      16.8  &      10  &      66  &      22  &     110  &     169  &     617  &     0.10  &     0.24  &     0.27   &      25 \\
 \hline
    Skyrme    & Average &  -15.87  &     30.82  &    0.1596  &      49.6  &     237  &    -132  &    -349  &     370  &    1448  &   -2175  &     0.77  &     0.127  &     0.44  &    -354\\
       (35)       & $\sigma$  &  0.18  &      1.54  &    0.0039  &      21.6  &      27  &      89  &      89  &     188  &     510  &    1069  &     0.14  &     0.310  &     0.37  &      45\\
 \hline
      RMF      & Average &    -16.24  &     35.11  &    0.1494  &      90.2  &     268  &      -5  &      -2  &     271  &    5058  &   -3672  &     0.67  &    -0.09  &     0.40 &    -549\\
        (11)      & $\sigma$  &    0.06  &      2.63  &    0.0025  &      29.6  &      34  &      88  &     393  &     357  &    2294  &    1582  &     0.02  &     0.03  &     0.06  &     153\\
 \hline
     RHF   &  Average  &   -15.97  &     33.97  &    0.1540  &      90.0  &     248  &     128  &     389  &     523  &    5269  &   -9956  &     0.74  &    -0.03  &     0.34 &    -572\\
       (4)     & $\sigma$ &      0.08  &      1.37  &    0.0035  &      11.1  &      12  &      51  &     350  &     237  &     838  &    4156  &     0.03  &     0.01  &     0.07  &     169\\
 \hline
 \hline
    Total & Average  &   -16.03  &     33.30  &    0.1543  &      76.6  &     251  &      -3  &      13  &     388  &    3925  &   -5268  &     0.72  &     0.01  &     0.39  &    -492\\
    (50)  & $\sigma_{tot}$  &    0.20  &      2.65  &    0.0054  &      29.2  &      29  &     132  &     431  &     289  &    2270  &    4282  &     0.09  &     0.20  &     0.22  &     166\\
             & Min  &     -16.35  &     26.83  &    0.1450  &       9.9  &     201  &    -394  &    -748  &     -86  &    -903  &  -16916  &     0.38  &    -0.47  &     0.00 &    -835 \\
             & Max  &    -15.31  &     38.71  &    0.1746  &     122.7  &     355  &     213  &     950  &     846  &    9997  &      -5  &     1.11  &     1.02  &     2.02 &    -254\\

\hline
\multicolumn{16}{c}{Ab-initio approaches}\\
\hline
APR & Average  & -16.0 & 33.12 & 0.16 & 50.0 & 270 & -199 & -665 & 923 & 337 & -2053 & 1.0 & 0.0 & 0.0 & -376 \\
(1)         & $\sigma$& -$^\dagger$ & 0.30 & -$^\dagger$ & 1.3   & 2   &   13   &    30 &  67  &  94  &   125  & -$^\dagger$ & -$^\dagger$ & -$^\dagger$ & 30 \\
\hline
chiral EFT & Average  &   -15.16  & 32.01 & 0.171 & 48.1  &  214 & -172  & -139  & -164  & 1306 &  -2317 & - & - & - & -428 \\
Drischler 2016 & $\sigma_{tot}$ &       1.24 &   2.09 & 0.016 &   3.6  &   22  &    40  &  104  &  234  &    214 &   379   & - & - & - &    63 \\
(7)    & Min                   &   -16.92 & 28.53  & 0.140 & 43.9  &  182 & -224 & -310  & -640  &    901 &  -2961 & - & - & - & -534 \\
        & Max                  &   -13.23 & 34.57  & 0.190 & 53.5  &  242 & -108 &    24  &     96 &  1537 &  -1750  & - & - & - & -334 \\
\end{tabular}
\end{ruledtabular}
$^\dagger$ This parameter is fixed.
\caption{binding energy $E_{sat}$, the symmetry energy $E_{sym}$, saturation density $n_{sat}$, 
slope of the symmetry energy $L_{sym}$, isoscalar incompressibility $K_{sat}$, isovector incompressibility  $K_{sym}$,
isoscalar skewness $Q_{sat}$, isovector skewness $Q_{sym}$, isoscalar kurtosis $Z_{sat}$,
isovector kurtosis $Z_{sym}$, the Landau effective mass at saturation $m^*_{sat}$, its isospin splitting $\Delta m^*_{sat}$. 
For the relativistic approaches, the effective mass is defined to be the Landau mass derived from the equivalent Schr\"odinger equation,
see Ref.~\cite{RHF:PKO} and references therein for more details.}
\label{tab:mean-table-models}
\end{table*}

\subsection{Generic determination of the empirical parameters}
\label{sec:empiricalparameters}

Besides the constraints determined from direct analysis of experimental data, we performed a complementary analysis of
the predictions for the empirical parameters determined from various relativistic and non-relativistic 
functionals.
We have investigated several types of relativistic and non-relativistic phenomenological models, namely 35 
Skyrme-type functionals, 11 models based on RMF effective Lagrangians,  4  RHF effective Lagrangians,
as well as two more ab-initio approaches, APR and chiral EFT.
{For simplicity, APR and chiral EFT EOS are grouped together since they are both based on the NN interaction in vacuum, at variance with the
so-called phenomenological approaches. The interactions on which they are based are however very different in nature, but this goes beyond the present analysis.}
Concerning the phenomenological models, we report in appendix~\ref{Sec:tables} the isoscalar and isovector empirical parameters up to the fourth
order, the Landau effective mass at saturation $m^*_{sat}$, and its isospin splitting $\Delta m^*_{sat}$, see Tabs.~\ref{table:empiricalsky}, \ref{table:empiricalrmf}, and \ref{table:empiricalrhf}.
Details for ab-initio chiral EFT approach are discussed in Section~\ref{Sec:ap:XEFT}.

In Tab.~\ref{tab:mean-table-models}, we present a summary of the detailed results shown in appendix~\ref{Sec:tables}
and Section \ref{Sec:ap:XEFT}: the average values for each type of model (Skyrme, RMF and RHF) are calculated as
well as the standard deviation $\sigma$ for each type of model, defined as 
$\sigma^2=\sum_i [ x_i^2-\langle x\rangle ]/N_{models}$, where $x$ stands for an empirical parameter
and $N_{models}$ is the number of parameter sets for each type of modeling ($N_{models}=N_\alpha$) 
given in Tab.~\ref{tab:mean-table-models}.
Note that for the Skyrme-type models, we present two different averages, over 16 and 35 models respectively.
Our sampling of Skyrme forces is more limited than in other analysis, see for instance Ref.~\cite{Dutra2012} and references therein.
The 35 Skyrme forces that we have considered here are among the mostly used forces. 
In addition, the reduced sampling of 16 Skyrme functionals contains the forces which are usually employed for finite nuclei.
Since some groups have produced many different forces, but with rather similar constraints, we have decided to 
consider only a few of these forces in our sample. 
In doing so, we limit as much as possible the bias which may come from the details of the fit and give almost equal weights to
different groups, thus increasing the meaning of the calculated average and standard deviations.
The test of the stability of our statistical analysis is performed by comparing the small sample to the wider one.
The central values are shown to be rather independent on the sampling, while the standard deviation $\sigma$ increases with the
number of models.

Considering the group of phenomenological approaches, the last lines of this group in Tab.~\ref{tab:mean-table-models} 
provide the average values, the standard deviations $\sigma_{tot}$, the minimal and maximal values found for 
all empirical parameters.
The average and the standard deviation could be influenced by the number of models belonging to
each type of model.
In order to reduce this influence, the average and standard deviation are weighted differently for the 
three different type of models.
The mean value is defined as
$\sum_\alpha 1/3 \sum_i x_i/N_{\alpha}$, where $\alpha$ runs over the three types of models 
(Skyrme, RMF and RHF) and $i$ over the models themselves.
This is strictly identical to take the arithmetic mean of the three first average values given in the first lines
of  Tab.~\ref{tab:mean-table-models}.
In a similar way, the standard deviation is defined as
$\sigma_{tot}^2=\sum_\alpha 1/3 \sum_i [ x_i^2-\langle x\rangle ]/N_{\alpha}$, where $N_\alpha$ is given in the first column of 
Tab.~\ref{tab:mean-table-models} and the mean value $\langle x\rangle$ is the one of the final average considering the 50 models.
By comparing different types of phenomenological approaches we expect that the final central values and central deviations
that we obtained are weakly impacted by the choice of the samples, provided only models used in finite nuclei are considered.

For the group of ab-initio approaches, the same statistical quantities are generated from the 7 chiral EFT
results. 
For APR, we have fitted Eq.~(\ref{eq:etotisiv}) to the APR symmetric and neutron matter EOS and we provide
in Tab.~\ref{tab:mean-table-models} the best fit and its associated error-bar.

It is clear from Tab.~\ref{tab:mean-table-models} that all the empirical parameters are model dependent, even
the lowest order ones: for instance the average saturation energies $E_{sat}$ and densities $n_{sat}$ are different between 
Skyrme, RMF and RHF type of models, and the difference between these average values are larger than
the standard deviations. 
The same remark applies also for the ab-initio approaches.
This indicates a model dependence for these quantities.
It is particularly interesting to remark the big deviations for the empirical parameters $K_{sym}$ and $Q_{sat}$ between the
Skyrme, RMF and RHF models.
These two quantities are predicted negative for Skyrme interactions, compatible with zero for RMF, and
positive (almost equal in absolute value to the Skyrme models) for the RHF approaches.
This is an indication that these values are weakly constrained by their fitting protocol,
which is mostly based on nuclear masses and charge radii.
The higher order empirical parameters ($Q_{sym}$, $Z_{sat}$, $Z_{sym}$) are quite unknown, as shown by the fact that
their standard deviations are comparable to their average values.
Finally, there is also a quite large model dependence for the effective mass $m^*_{sat}$ and the isospin
splitting $\Delta m^*_{sat}$.

It is interesting to note from Tab.~\ref{tab:mean-table-models} that the values obtained for the empirical parameters
$E_{sat}$, $E_{sym}$, $n_{sat}$, $K_{sat}$,  $L_{sym}$ and $K_\tau$ are rather close to the ones extracted from an analysis of experimental data, as discussed in Sec.~\ref{Sec:empirical:exp}.
This is also the case for the so-called ab-initio approaches, except for $n_{sat}$ which is slightly too high for the chiral EFT case.
This is however a general issue shared by ab-initio approaches~\cite{Baldo2012}.
Except for the value of $Q_{sym}$, the average Skyrme and chiral EFT predictions match in a satisfactory way.
The value for $Q_{sat}$ is systematically lower for Skyrme and chiral EFT than for the relativistic phenomenological
approaches. This makes the non-relativistic EOS generally softer than the relativistic ones.
This good matching between the low order empirical parameters deduced from the statistical average and shown
in  Tab.~\ref{tab:mean-table-models} with the experimental data discussed in Sec.~\ref{Sec:empirical:exp}  indicates that the estimated values provided by  Tab.~\ref{tab:mean-table-models} are reasonably
well constrained. 
It appears therefore reasonable to take the values of Tab.~\ref{tab:mean-table-models} also for the empirical parameters for which there are no experimental data.


\begin{table}[t]
\centering
\setlength{\tabcolsep}{2pt}
\renewcommand{\arraystretch}{1.2}
\begin{ruledtabular}
\begin{tabular}{ccccccc}
Model & & $E_{NM}$ & $L_{sym}$ & $K_{NM}$  & $Q_{NM}$  & $Z_{NM}$\\
($N_\alpha$) &  & MeV & MeV & MeV & MeV & MeV\\
\hline
\multicolumn{7}{c}{Phenomenological approaches}\\
\hline
Skyrme & Average   &     14.95  &      49.6  &      106  &       21  &     -727\\
 (35)            & $\sigma$  &      1.72  &      21.6  &      116  &      276  &     1580\\
RMF & Average  &     18.86  &      90.2  &      263  &      269  &     1386\\
(11)         & $\sigma$ &      2.69  &      29.6  &      121  &      750  &     3876\\
RHF & Average   &     17.99  &      90.0  &      376  &      912  &    -4686\\
(4)        & $\sigma$  &      1.46  &      11.1  &       63  &      587  &     4994\\
\hline
\multicolumn{7}{c}{Ab-initio approaches}\\
\hline
APR           & Average  & 17.27 & 50.0 & 71 & 258 & -1716 \\
(1)                  & $\sigma$ &   0.30 &  1.3  & 15 &  97  &   219 \\
GCR 2012 &  Average & 16.76 & 45.8 & 77 &  80 & -131 \\
(7)                     & $\sigma$&    1.39 & 9.7 & 43 & 29 & 15 \\
chiral EFT & Average &  16.39 & 56.4 & 119  & - & - \\
Tews 2013 & $\sigma$&    2.97 &  11.0 & 101 & - & - \\
chiral EFT &  Average & 16.93 & 48.3 & 41 & -314 & -991 \\
Drischler 2016 (7) & $\sigma$&    0.92 & 3.5 & 33 & 226 & 349\\
\end{tabular}
\end{ruledtabular}
\caption{Neutron matter energy per nucleon $E_{NM}$, slope of the symmetry energy $L_{sym}$, neutron matter 
incompressibility $K_{NM}$, neutron matter skewness $Q_{NM}$ and neutron matter kurtosis $Z_{NM}$ for
phenomenological and ab-initio approaches.
See text for more details as well as Appendix~\ref{Sec:ap:XEFT}.
deduced from for GCR~2012~\cite{Gandolfi2012} and chiral EFT up to N$^3$LO~\cite{Tews2013}.}
\label{tab:mean-table-models-EFT}
\end{table}

Some ab-initio calculations provide only the neutron matter (NM) EOS, since it does not present the extra complication of  the spinodal 
instability at low density.
The NM EOS is obtained in the metamodel by taking the value $\delta=1$ in Eq.(\ref{eq:etotisiv}).
At each order, the two isoscalar and isovector coefficients become a single coefficient that we indicate in the
following with index $NM$.
Note that since the pressure of symmetric matter at saturation density is zero, $L_{NM}=L_{sym}$.
In Tab.~\ref{tab:mean-table-models-EFT}, we show the predictions for the NM empirical parameters of the 
same approaches as in Tab.~\ref{tab:mean-table-models} plus a couple of other ab-initio predictions:
GCR~2012~\cite{Gandolfi2012} and chiral EFT Tews~2013~\cite{Tews2013}.
Details on how these numbers have been obtained for GCR~2012 and chiral EFT Tews~2013 are given
in Section~\ref{Sec:ap:XEFT}.
It is worth notifying that the different ab-initio approaches give consistent estimation for $K_{NM}$ 
between 120 and 40~MeV.
Since $K_{NM}=K_{sat}+K_{sym}$, and $K_{sat}=230\pm 20$~MeV, we have approximately
$K_{sym}\approx -100\pm100$~MeV (including also the preferred values from RMF and RHF approaches).
The Skyrme and ab-initio approaches prefer values $K_{sym}\approx -200, -150$ while the relativistic
approaches prefer $K_{sym}\approx 0,100$~MeV.

{Let us mention another phenomenological approach, the so-called two-loop quantum hadrodynamics, which is based on RMF with an adjunction of the two-loop exchange diagrams~\cite{Zhang2016}.
For fixed values of $E_{sat}$, $n_{sat}$, $K_{sat}$ and $E_{sym}$ comparable with the ones in Tab.~\ref{tab:mean-table-models},
this approach predicts $L_{sym}\approx$83-85~MeV and $K_{sym}\approx -20$~MeV, which is in the range of values that we explore.}

To conclude this analysis, we now discuss the total average and total standard deviation $\sigma_{tot}$ 
shown in Tab.~\ref{tab:mean-table-models}.
They provide a global estimation for the empirical parameters including the systematic error-bar induced by
the model dependence, as previously discussed.
From these global results, it is possible to separate the empirical quantities into four groups:
\begin{enumerate}
\item The parameters which are known within a few percent: $E_{sat}$ and $n_{sat}$.
\item The parameters which are known within about 10 percent: $E_{sym}$, $K_{sat}$ and $m^*_{sat}$.
\item The parameters which are known within about 50 percent: $L_{sym}$.
\item The parameters which are almost unknown: $Q_{sat}$, $Z_{sat}$, $K_{sym}$, $Q_{sym}$, $Z_{sym}$ and $\Delta m^*_{sat}$.
\end{enumerate}

We may hope that the empirical parameters in the three first groups will be better constrained from
nuclear physics experiments in the future, considering in addition to the masses and charge radii constraints the
ones provided by the neutron skin radii, the collective modes in neutron rich nuclei, and possibly large deformations
in the ground state.
It is however hard to imagine that the parameters from the last group will ever be constrained from the properties
of finite nuclei around saturation density.
To be better determined, they require the knowledge of the properties of systems at densities and
asymmetries different from those of finite nuclei.
It could be expected that Heavy Ion Collision (HIC) will provide some constraints, as well as the observed 
properties of compact stars.
This will be further discussed in the following sections.

\section{A metamodeling for the nuclear equation of state}
\label{Sec:eeos}

In this section, we investigate to which extend a series expansion of the same kind
as the one given by Eqs.~(\ref{eq:eis})-(\ref{eq:bindingenergy}) can generate a realistic
equation of state (EOS).
There are two questions to answer, which are i) is the density and isospin dependence
rich enough, and ii) what is the convergence in density and
isospin parameter of such series expansions.

A purely polynomial density expansion as in Eqs.~(\ref{eq:eis})-(\ref{eq:bindingenergy}) is 
too simple to provide realistic results because it does not catch the 
natural density and isospin dependence of the kinetic term~\cite{Ducoin2011}.  For this reason, we will 
separate the kinetic term from the potential one~(\ref{eq:eis})-(\ref{eq:bindingenergy}).

To fully cover the parameter space of both relativistic and 
non-relativistic models, the best treatment of the kinetic term 
would be an expansion in powers of the Fermi momentum $k_F$~\cite{Serot1997,Margueron2007a}.
This would however introduce a high number of extra poorly constrained parameters. We have therefore chosen to limit ourselves to a non-relativistic treatment for this paper, such that the kinetic term can be exactly handled and the expansion only concerns the Landau effective mass. It is important to remark that, even if the kinetic energy density is treated non-relativistically, the functional is still flexible enough to satisfactorily reproduce also the density dependence of relativistic models. This point will be demonstrated in Section \ref{sec:analytical}

{Let us also mention that an expansion of the energy per nucleon in terms of the Fermi momentum $k_F$ is also possible~\cite{Bulgac2015}. In our present study, we aim at keeping a simple relation between the empirical parameters and the parameters of the model. This determines our choice for an expansion in powers of the density.}

The metamodel on which the EOS is based on has therefore four requirements:
\begin{enumerate}
\item The nuclear potential is quadratic in the isospin asymmetry parameter $\delta$.
\item The EOS is analytic in the parameter $x$,  and possible phase transitions are not accounted for.
\item The energy per nucleon satisfies the following limit: $\lim_{n_0\rightarrow 0} e(n_0,n_1)=0$.
\end{enumerate}

From the functional form of the energy per nucleon $e(n_0,n_1)$, it is possible to calculate analytically its first and second
order derivatives, which are related to the nucleon pressure and to the nucleon sound velocity as,
\begin{eqnarray}
P_n(n_0,n_1)&=&n_0^{2}\frac{\partial e}{\partial n_0}\, ,
\label{Pressure} \\
\bigg(\frac{v_{s,n}}{c}\bigg)^{2}&=&\frac{K_{is}(n_0,n_1)}{9\bigg[mc^{2}+e+\frac{P(n_0,n_1)}{n_0} \bigg] } \, ,
\label{eq:Sound-velocity}
\end{eqnarray}
where the isoscalar compressibility $K_{is}(n_0,n_1)$ is defined as, 
 \begin{eqnarray}\label{Incompressibility-modulus}
K_{is}(n_0,n_1)=9n_0^{2}\frac{\partial^{2} e }{\partial n_0^{2}}
+18\frac{P(n_0,n_1)}{n_0} .
 \end{eqnarray}
Note that $K_{is}(n_{sat},0)=K_{sat}$.

The symmetry energy is defined as
\begin{equation}
S(n_0) = \frac 1 2 \frac{\partial^2 e(n_0,n_1)}{\partial \delta^2}\vert_{n_1=0}  \, .
\label{eq:symmetryenergy}
\end{equation}

In the following, we first express the kinetic energy contribution to the total energy, then
we explore  various approximations for the potential energy.

\subsection{The kinetic energy term}

For a non-relativistic free Fermi Gas (FG) the kinetic energy per particle is simply given by
\begin{equation}
t^{FG}(n_0,n_1)=\frac{t_{sat}^{FG}}{2}\left(\frac{n_0}{n_{sat}}\right)^{2/3} f_1(\delta)
\end{equation}
where $t_{sat}^{FG}= 3\hbar^{2}/(10m)\left(3\pi^{2}/2\right)^{2/3}n_{sat}^{2/3}$ is
the kinetic energy per nucleons in SM and at saturation,
$m$ is the nucleonic mass taken identical for neutrons and protons 
($m=(m_n+m_p)/2=938.919$~MeV/c$^2$),
giving $t_{sat}^{FG}\approx 22.1$~MeV,
and the function $f_1$ is defined as
\begin{equation}
f_1(\delta) = (1+\delta)^{5/3}+(1-\delta)^{5/3} .
\end{equation}

The momentum dependence of the nuclear interaction gives rise to the concept of effective
mass: an average effect of the in-medium nuclear interaction is to modify the
inertial mass of the nucleons.
The Landau effective mass can be parameterized in the following way ($\tau$=n or p),
\begin{equation}
\frac{m}{m^*_\tau(n_0,n_1)} = 1 + \left( \kappa_{sat} + \tau_3 \kappa_{sym} \delta \right) \frac{n_0}{n_{sat}} ,
\label{eq:effmass}
\end{equation}
where $\tau_3=1$ for neutrons and -1 for protons, and where the parameters $\kappa_{sat}$ and $\kappa_{sym}$ are functions of $m^*_{sat}$ and $\Delta m^*_{sat}$ previously discussed, see Sec.~\ref{Sec:empirical:exp}.
We have for both $\kappa_{sat/sym}$ the following expressions taken at $n_0=n_{sat}$,
\begin{eqnarray}
\kappa_{sat}&=&\frac{m}{m_{sat}^*} - 1=\kappa_s \hbox{ in SM ($\delta=0$)}  , \nonumber \\
\kappa_{sym}&=&\frac 1 2 \left[ \frac{m}{m^*_n} - \frac{m}{m^*_p}  \right] \hbox{ in NM ($\delta=1$)} .
\end{eqnarray}
Introducing the parameters $\kappa_s$ and $\kappa_v$, we have $\kappa_{sat}=\kappa_s$
and $\kappa_{sym}=\kappa_s-\kappa_v$~\cite{Lesinski2006}.
The functional form~(\ref{eq:effmass}) for the in-medium effective mass is the simplest form of a density series expansion.
Truncating the expansion at first order as in ~(\ref{eq:effmass}) 
allows to recover the expression used in standard
Skyrme functionals.  
For simplicity, we do not generalize Eq.~(\ref{eq:effmass}) with a more complete polynomial in this work.
Anticipating results presented in Sec.~\ref{sec:eos}, it will be shown that the impact of the effective mass on the
equation of state is very weak (at zero temperature), justifying our present approximation.

Considering the functional form~(\ref{eq:effmass}) for the nucleonic effective mass,
the new expression for the kinetic energy in nuclear matter reads,
\begin{eqnarray}
t^{FG^*}(n_0,n_1)&=&\frac{t_{sat}^{FG}}{2}\left(\frac{n_0}{n_{sat}}\right)^{2/3} 
\bigg[ \left( 1+\kappa_{sat}\frac{n_0}{n_{sat}} \right) f_1(\delta) \nonumber \\
&& \hspace{2.5cm} + \kappa_{sym}\frac{n_0}{n_{sat}}f_2(\delta)\bigg] ,
\label{eq:effmassms2}
\end{eqnarray}
where the new function $f_2$ is defined as
\begin{equation}
f_2(\delta) = \delta \left( (1+\delta)^{5/3}-(1-\delta)^{5/3} \right) .
\end{equation}

In the following, the kinetic energy contribution to the density functional will be given by Eq.~(\ref{eq:effmassms2}),
which is the simplest way to consider the contribution of the momentum dependence of the nuclear interaction.

This expression gives the exact kinetic energy density only if we want to reproduce models with non-relativistic kinematics. 
In the case of relativistic EOS models,  it would be more natural to employ a relativistic formulation for the kinetic energy density 
and use Dirac masses instead than Landau masses as non-local empirical parameters. 
This will certainly be necessary if we want to adress specific observables which are especially sensitive to the kinetic energy term.
We also expect that isolating a relativistic kinetic
energy density term from the polynomial expansion will improve the convergence of the series when reproducing relativistic models,
and such an extension towards a relativistic metamodeling is planned for the next future. 

Concerning the energy per particle and the pressure of homogeneous matter, which are our main concern here,
we will see in Section  \ref{sec:system} that RMF and RHF models are also satisfactorily reproduced by our metamodeling, even if the 
degree of reproduction is less accurate than for non-relativistic ones.
 
We now discuss the functional form for the potential energy.

\subsection{Metamodeling ELFa: the simplest approach}

Once the kinetic energy density is sorted out via Eq.(\ref{eq:effmassms2}), the energy per nucleon can be written as:
\begin{eqnarray}
e^N_{ELFa}(n_0,n_1)=t^{FG*}(n_0,n_1)+v^N_{ELFa}(n_0,n_1),
\label{eq:ELFa}
\end{eqnarray}
where the potential energy is expressed as a series expansion in the parameter 
$x$,
\begin{eqnarray}
v^N_{ELFa}(n_0,n_1)=\sum_{\alpha\geq0}^N \frac{1}{\alpha!}v_{\alpha}(\delta) x^\alpha .
\label{eq:vELFa}
\end{eqnarray}
Since Eq.~(\ref{eq:ELFa}) provides an Empirical Local density Functional (ELF), this
metamodeling is called ELFa.

Supposing a quadratic approximation for the potential energy, as suggested by 
microscopic Bruckner calculations~\cite{Vidana2009}, we have
\begin{equation}
v_{\alpha}(\delta)=v_{\alpha}^{is}+v_{\alpha}^{iv}\delta^2 .
\end{equation}

Simple relations can be obtained between the model parameters $v_{\alpha}^{is}$ and
$v_{\alpha}^{iv}$ and the empirical parameters.
We have for the isoscalar parameters,
\begin{eqnarray}
v_{\alpha=0}^{is} &=& E_{sat}-t_{sat}^{FG}(1+\kappa_{sat}), \label{eq:vis}\\ 
v_{\alpha=1}^{is} &=& -t_{sat}^{FG}(2+5\kappa_{sat}) ,\\ 
v_{\alpha=2}^{is} &=& K_{sat}-2t_{sat}^{FG}(-1+5\kappa_{sat}) , \\ 
v_{\alpha=3}^{is} &=& Q_{sat}-2t_{sat}^{FG}(4-5\kappa_{sat}) ,\\ 
v_{\alpha=4}^{is} &=& Z_{sat}-8t_{sat}^{FG}(-7+5\kappa_{sat}) ,
\\ \nonumber
\end{eqnarray}
and the isovector parameters,
\begin{eqnarray}
v_{\alpha=0}^{iv}&=&E_{sym}-\frac{5}{9}t_{sat}^{FG}[1+(\kappa_{sat}+3\kappa_{sym})] , \label{eq:viv}\\ 
v_{\alpha=1}^{iv}&=&L_{sym}-\frac{5}{9}t_{sat}^{FG}[2+5(\kappa_{sat}+3\kappa_{sym})] , \\ 
v_{\alpha=2}^{iv}&=&K_{sym}-\frac{10}{9}t_{sat}^{FG}[-1+5(\kappa_{sat}+3\kappa_{sym})] , \\ 
v_{\alpha=3}^{iv}&=&Q_{sym}-\frac{10}{9}t_{sat}^{FG}[4-5(\kappa_{sat}+3\kappa_{sym})] , \\ 
v_{\alpha=4}^{iv}&=&Z_{sym}-\frac{40}{9}t_{sat}^{FG}[-7+5(\kappa_{sat}+3\kappa_{sym})] . \label{eq:viv2}
\end{eqnarray}

The simple one-to-one correspondence between the model parameters and the empirical parameter
is coming from the expansion of the potential energy in the parameter $x$.
It is therefore related to the fact that only integer powers of the density are considered.
Another interesting aspect of the series expansion~(\ref{eq:vELFa}) is that we have a clear
control on the derivative up to which we think we can determine to potential contribution.
For derivative of higher order than the limit $N$ there is no contribution from the potential energy and 
only the kinetic energy contributes.

\begin{figure}[tb]
\begin{center}
\includegraphics[angle=0,width=1.0\linewidth]{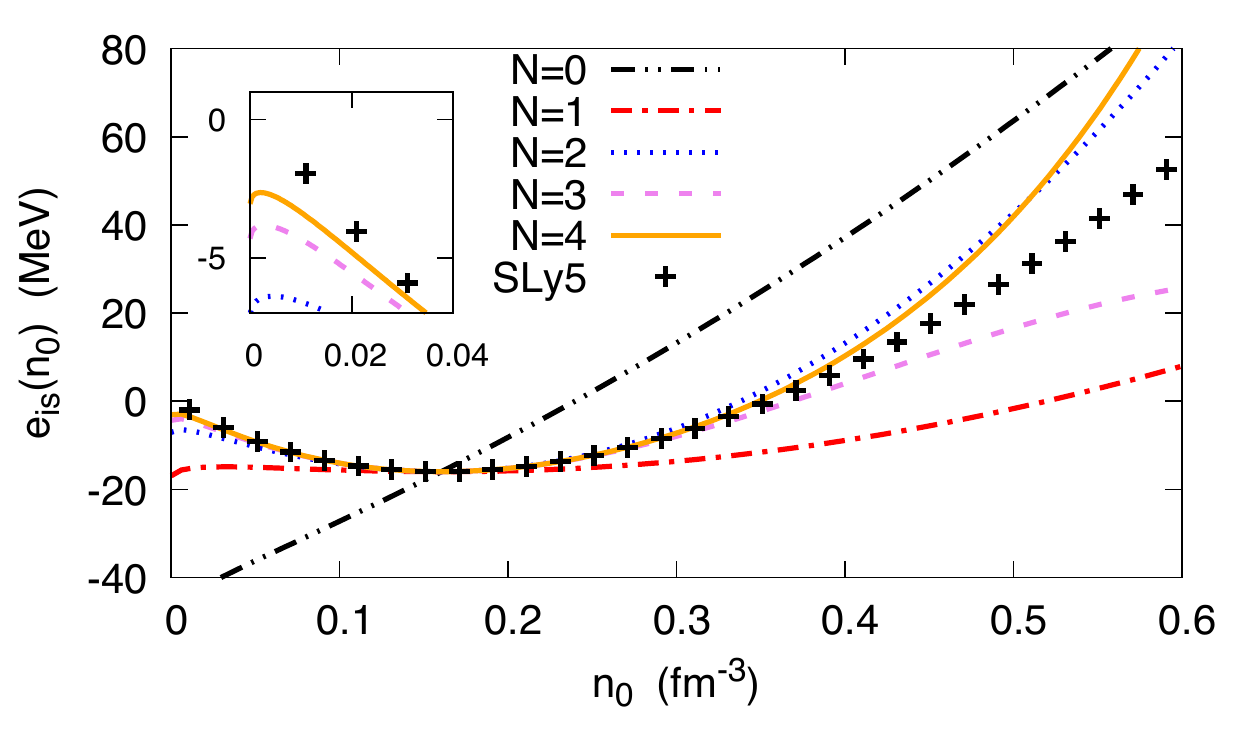}
\end{center}
\caption{(Color online) Comparison of the energy per nucleon in symmetric matter between Skyrme SLy5 and ELFa 
metamodeling, where the empirical parameters of Skyrme SLy5 has been used, as a function of the 
order $N$ (lines with different colors).
The crosses show the reference value given by SLy5.}
\label{fig:elfa1} 
\end{figure}

In the metamodel ELFa, the parameter $N$ (integer) defines the highest power in $x$. 
It can vary from 0 to $\infty$.
The larger $N$, the better the series expansion.
In order to illustrate the contribution of the different orders in $N$, we show in Fig.~\ref{fig:elfa1} the 
energy per nucleon in symmetric matter as function of $N$ going from 0 to 4 at maximum, and for a
range of densities going from $n_0$=0 to 0.6~fm$^{-3}$.
The values for the empirical parameters are taken from the Skyrme interaction SLy5, and for comparison 
the energy per nucleon given by SLy5 is also shown in Fig.~\ref{fig:elfa1}.
We first remark that the different metamodels ELFa pass through the saturation point, but only the models
with $N\ge 2$ reproduce the saturation properties with a minimum value for the energy per nucleon at the
saturation point.
As the density departs from the saturation density, the models with largest value in $N$ get closer to
the reference model given by SLy5 (crosses in Fig.~\ref{fig:elfa1}).

The inset in Fig.~\ref{fig:elfa1} shows in more details the low density behavior of the energy per nucleon
given by the model ELFa for various $N$.
The main default of the model is that the potential energy is not zero at $n_0=0$.
Since the model is a series expansion around the saturation density $n_{sat}$, it is indeed not given
that the energy per nucleon goes to zero when the density goes to zero.
In the following, we propose two modifications of ELFa, namely ELFb and ELFc, for curing
this issue at zero density.

\subsection{Metamodeling ELFb: a correction at zero density}

In this section, we still express the energy per nucleon in the following way,
\begin{eqnarray}
e^N_{ELFb}(n_0,n_1)=t^{FG*}(n_0,n_1)+v^N_{ELFb}(n_0,n_1).
\label{eq:ELFb}
\end{eqnarray}

A way to ensure that the zero density limit is verified is to change the Taylor expansion around $n_{sat}$
from metamodel ELFa to a polynomial expansion in terms of the density $n_0$, as
\begin{eqnarray}
v^N_{ELFb}(n_0,n_1)=\sum_{\alpha\geq 1}^N ( p_{\alpha}^{is}+ p_{\alpha}^{iv} \delta^2 ) n_0^\alpha .
\label{eq:vELFb_poly}
\end{eqnarray}
The expression (\ref{eq:vELFb_poly}) has been used in various functionals, see for instance 
Refs.~\cite{Baldo2004d,Baldo2008}.
It can however be shown that this expansion is strictly equivalent to an expansion around the saturation density,
\begin{eqnarray}
v^N_{ELFb}(n_0,n_1)=\sum_{\alpha\geq0}^N \frac{1}{\alpha!} ( v_{\alpha}^{is}+ v_{\alpha}^{iv} \delta^2 ) x^\alpha ,
\label{eq:vELFb}
\end{eqnarray}
where the parameters $v_N^{is}$ and $v_N^{iv}$ are fixed by the zero density limit, to be
\begin{eqnarray}
v_N^{is/iv} &=& -  \sum_{\alpha\geq0}^{N-1} \frac{N!}{\alpha!}  v_{\alpha}^{is/iv} (-3)^{N-\alpha} ,
\label{eq:vNELFb}
\end{eqnarray}
while the model parameters $v_\alpha^{is/iv}$ for $\alpha<N$ are still related to the empirical parameters according to
Eqs.~(\ref{eq:vis})-(\ref{eq:viv2}).

Combining Eqs.~(\ref{eq:vELFb}) and (\ref{eq:vNELFb}) together, the potential energy can be rewritten as,
\begin{eqnarray}
v^N_{ELFb}(n_0,n_1)=\sum_{\alpha\geq0}^{N-1} \frac{1}{\alpha!}( v_{\alpha}^{is}+ v_{\alpha}^{iv} \delta^2) x^\alpha u^N_{ELFb,\alpha}(x) ,
\label{eq:vELFb2}
\end{eqnarray}
where $u^N_{ELFb,\alpha}(x)=1-(-3x)^{N-\alpha}$.

\begin{figure}[tb]
\begin{center}
\includegraphics[angle=0,width=1.0\linewidth]{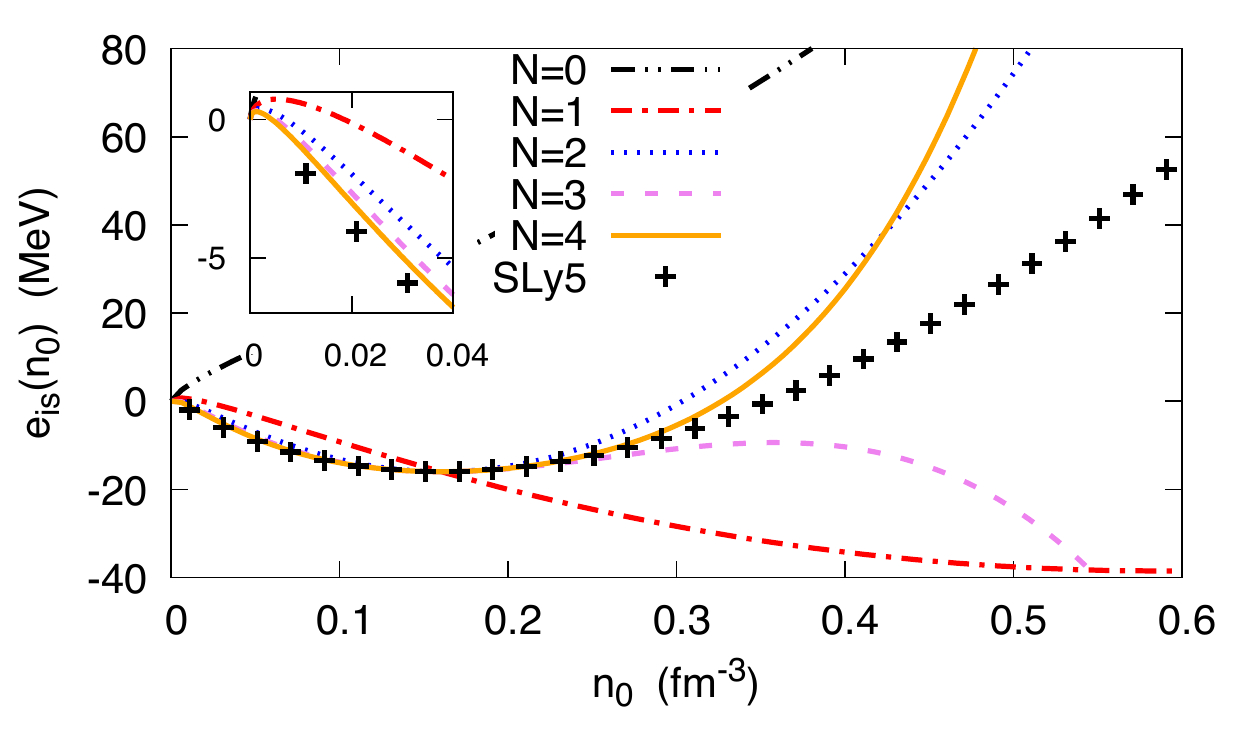}
\end{center}
\caption{(Color online) Same as Fig.~\ref{fig:elfa1} for ELFb metamodeling.}
\label{fig:elfb1} 
\end{figure}

The energy per nucleon deduced from the metamodeling ELFb is shown in Fig.~\ref{fig:elfb1} and can be compared to the previous
Fig.~\ref{fig:elfa1} for the metamodel ELFa.
The zero density limit is now well satisfied, as shown in the inset figure, however, the convergence ordering with $N$ beyond
saturation density is missing with metamodeling ELFb: there is no improvement of the convergence by increasing $N$.
This breaking of the convergence ordering observed with metamodeling ELFb is not very surprising since the parameters
$v_N^{is}$ and $v_N^{iv}$, which govern the high density behavior of the energy per nucleon, are now uniquely determined
by the zero density limit.
Fig.~\ref{fig:elfb1} illustrates that the density dependence of the energy per nucleon below and above saturation density is not symmetric,
and a condition improving the low density behavior of the energy per nucleon can strongly deteriorate the properties of the
EOS above saturation density.

\subsection{Metamodeling ELFc: an improved correction at zero density}

\begin{figure}[tb]
\begin{center}
\includegraphics[angle=0,width=1.0\linewidth]{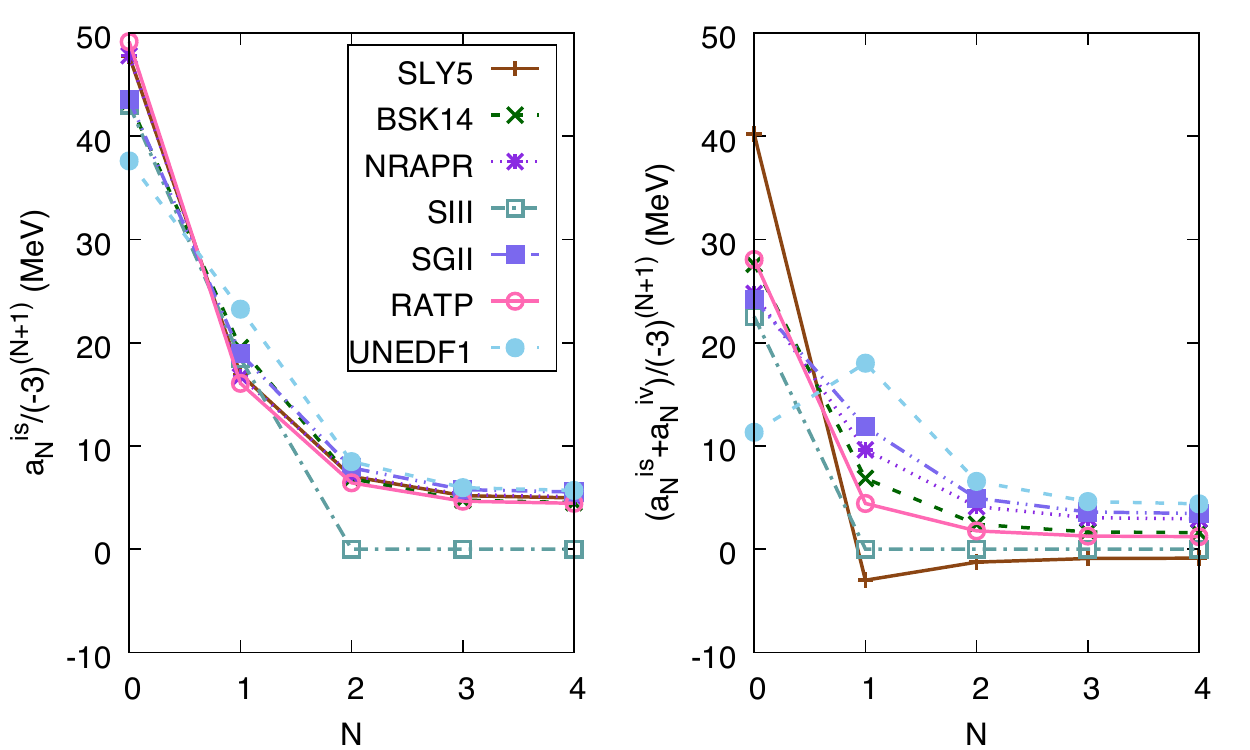}
\includegraphics[angle=0,width=1.0\linewidth]{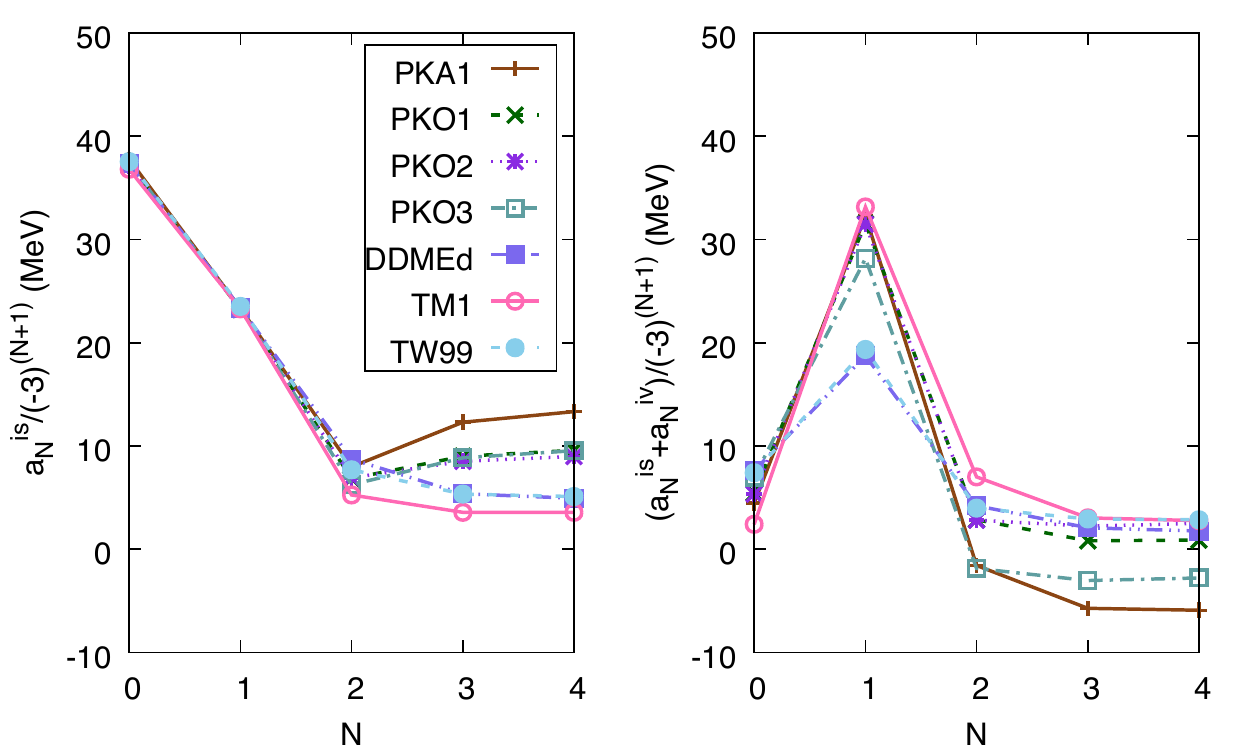}
\end{center}
\caption{(Color online) Coefficients $a_N^{is}/(-3)^{N+1}$ (left) and $a_N^{iv}/(-3)^{N+1}$ (right)
as a function of the order $N$ for various relativistic (bottom panels) and non-relativistic (top panels) nuclear interactions.}
\label{Nuclear-New3-SNM-fit1}
\end{figure}

Since metamodeling ELFb has shown that the behavior below and above saturation density of the energy per nucleon are essentially disconnected,
we investigate with metamodeling ELFc an alternative approach which breaks the symmetry around saturation density.
In metamodeling ELFc, a density dependent term is added at low density in order to satisfy the zero density limit, and this term drops to zero as the density increases. 
Since the small correction acts only at very low density, the properties of ELFa around and above saturation density 
are entirely conserved.
The small term is determined such that i) it decreases quickly with the density, ii) it
does not modify the relation between the model parameters and the empirical quantities
(\ref{eq:vis})-(\ref{eq:viv2}), and iii) it is fixed to annihilate the finite value given by an expansion (\ref{eq:vELFa}). 
Considering these requirements, we consider the following expression for the potential energy,
\begin{eqnarray}
v^N_{ELFc}(n_0,n_1)&=&\sum_{\alpha\geq0}^N \frac{1}{\alpha!}( v_{\alpha}^{is}+ v_{\alpha}^{iv} \delta^2) 
x^\alpha \nonumber \\
&& \hspace{0.1cm}-(a_N^{is}+a_N^{iv}\delta^2) x^{N+1} \exp \left(-b\frac{n_0}{n_{sat}}\right).
\label{eq:vELFc}
\end{eqnarray}
The numerical values for the coefficients $a_N^{is}$ and $a_N^{iv}$ are fixed such that the potential
energy in Eq.~(\ref{eq:vELFc}) is zero at zero density.
These parameters are functions of the order $N$ of the expansion and are defined as,
\begin{eqnarray}
a_N^{is}&=&-\sum_{\alpha\ge 0}^{N}\frac{1}{\alpha!}v_{\alpha}^{is}(-3)^{N+1-\alpha} , \label{eq:anis} \\
a_N^{iv}&=&-\sum_{\alpha\ge 0}^{N}\frac{1}{\alpha!}v_{\alpha}^{iv}(-3)^{N+1-\alpha} . \label{eq:bniv} 
\end{eqnarray}

In FIG.~\ref{Nuclear-New3-SNM-fit1} the $N$-dependence of the coefficients $a_N^{is}$ and $a_N^{iv}$ 
is shown for a set of relativistic (bottom panels) and non relativistic (top panels) interactions.
It is interesting to remark that as $N$ increases, the absolute value of $a_N^{is}$ and $a_N^{iv}$ decreases 
to a small number.
Therefore, the larger $N$, the smaller the correction at low density.
 
The condition at zero density determining the value of the coefficients $a_N^{is}$ and $a_N^{iv}$ 
does not fix the parameter $b$, which remains  free to determine.
In the following, it is fixed such that the correction quickly vanish at small but finite density.
Imposing that the exponential function in Eq.~\ref{eq:vELFc} is 1/2 at $n_0=0.1n_{sat}$, 
we obtain $b=10\ln 2 \approx 6.93$.
Note that the results shown in this paper are not impacted by the choice for the constant $b$ since the correction term
plays a role only at low density and we focus our study above saturation density.

Finally, the following compact form for the potential energy can be obtained,
\begin{eqnarray}
v^N_{ELFc}(n_0,n_1)=\sum_{\alpha\geq0}^N \frac{1}{\alpha!}( v_{\alpha}^{is}+ v_{\alpha}^{iv} \delta^2) x^\alpha u^N_{ELFc,\alpha}(x) ,
\label{eq:vELFc2}
\end{eqnarray}
where $u^N_{ELFc,\alpha}(x)=1-(-3x)^{N+1-\alpha}\exp(-bn_0/n_{sat})$, and the energy per particle is defined as
\begin{eqnarray}
e^N_{ELFc}(n_0,n_1)=t^{FG*}(n_0,n_1)+v^N_{ELFc}(n_0,n_1).
\label{eq:ELFc}
\end{eqnarray}

\begin{figure}[tb]
\begin{center}
\includegraphics[angle=0,width=1.0\linewidth]{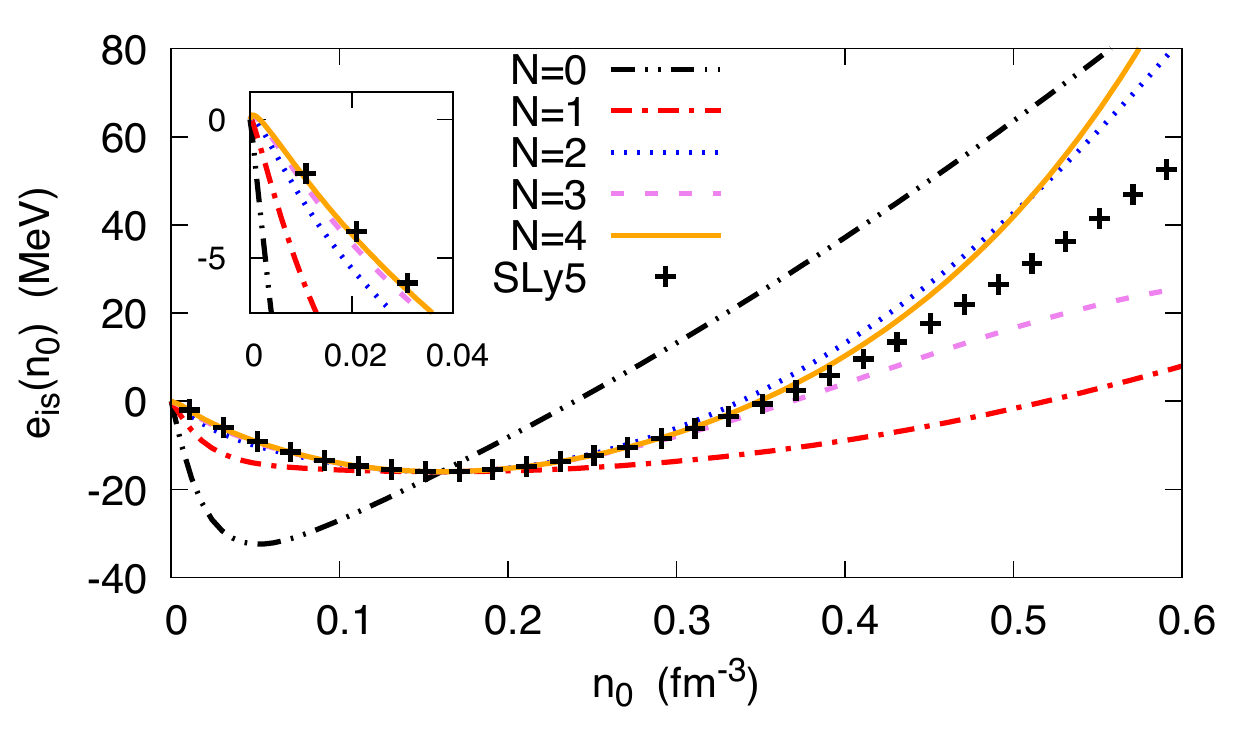}
\end{center}
\caption{(Color online) Same as Fig.~\ref{fig:elfa1} for ELFc metamodeling.}
\label{fig:elfc1} 
\end{figure}

The energy per nucleon provided by metamodeling ELFc is shown in Fig.~\ref{fig:elfc1} for various orders $N$.
We can see that the convergence at high density is the same as in
the metamodeling ELFa (see  Fig.~\ref{fig:elfa1}), while the low density behavior is now correct.
In the following, we now study in more detail the convergence of the energy per nucleon above saturation density
and we propose a way to fix the high order parameters 
such as to reproduce existing nuclear interaction predictions.

\subsection{Metamodeling ELFd: a faster convergence at high density}
\label{sec:elfd}

The high density convergence of the metamodeling ELFc with $N$ is not entirely satisfying, as illustrated in Fig.~\ref{fig:elfc1}.
First, it shall be remarked that the empirical parameters increase (in absolute value) as the order $\alpha$ 
increases, see Tab.~\ref{tab:mean-table-models}.
Second, the sign of the empirical parameters alternates at each order for $\alpha\ge 3$.
These two properties of the empirical parameters make the convergence lengthy, meaning that an important number of high order derivatives are needed to determine the high density behavior.
Moreover, the impact of the high order empirical parameters around saturation is extremely small. 
This means that, as we have already discussed, nuclear structure data and ab-initio calculations will hardly be able to provide reliable values for these parameters. 
In this respect, the situation might look hopeless. However, we can observe that
the density behavior of a purely nucleonic equation of state above saturation  is rather smooth and
does not show any complicated structure in all existing models, see for instance model SLy5 represented in Fig.~\ref{fig:elfc1}.
This observation implies that the globality of the high order parameters would be pretty much under control if we 
would fit these high order parameters directly to the EOS. Equivalently, we could also impose
the value of the EOS at a single high-density point, in addition to the empirical information around saturation.
We have checked that there is no major differences between the fit of the high density behavior of the EOS and the choice of a single high-density 
point, except that the latter allows for analytical expressions.

\begin{figure}[tb]
\begin{center}
\includegraphics[angle=0,width=1.0\linewidth]{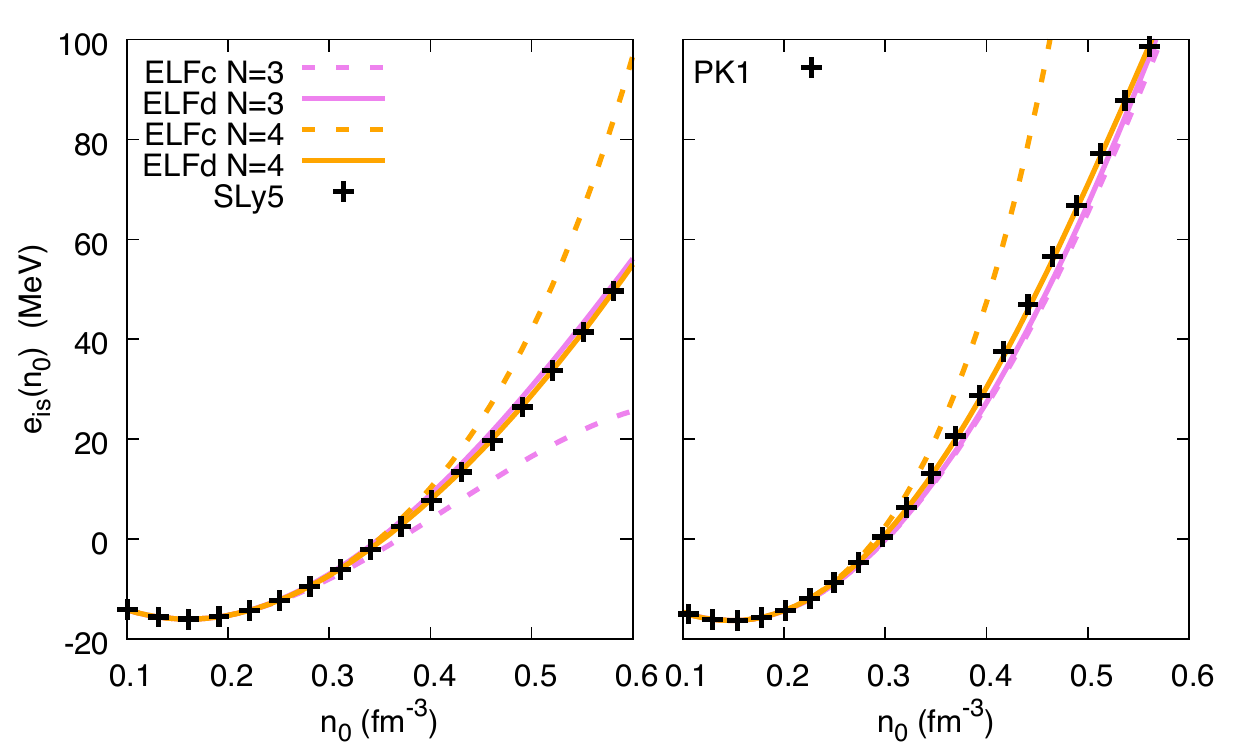}
\end{center}
\caption{(Color online) Comparison of the ELFc (dashed lines) and ELFd (solid lines) metamodels against the reference models SLy5 (left panel) and PK1 (right panel).
Only the orders $N$=3 and 4 are plotted.}
\label{fig:elfd1} 
\end{figure}

\begin{table*}[t]
\centering
\setlength{\tabcolsep}{2pt}
\renewcommand{\arraystretch}{1.2}
\begin{ruledtabular}
\begin{tabular}{cccccccccccccccccc}
                            &                    &  & $Q_{sat}$     &  $Q_{sym}$ & $Q_{sym}$  & $Z_{sat}$  & $Z_{sym}$  & $Z_{sym}$    & $\sigma_{e} (SM) $  & $\sigma_{e}(NM)$    & $\sigma_{e}(NM)$  \\
                            &                    &  &                      &                     & ($v_{4}$=0) &                   &                    & ($v_{4}$=0)  &                                  &                                  &  ($v_{4}$=0)\\       
 Model                 &                    &  & MeV              &  MeV           & MeV            & MeV           & MeV            & MeV             &             MeV              &         MeV                    &       MeV          \\
($N_{\alpha}$)    & der. order    &  & 3                    &  3                & 3                 & 4                 & 4                  & 4                  &                                  &                                  &                      \\
\hline
Skyrme   & 10 & Average & -197 & 283 & 199 & -477 & -802 & -472 &  0.23 &  0.10 &  0.67\\
(10)   &     & $\sigma$ & 81 & 95 & 81 & 212 & 289 & 341 &  0.07 &  0.06 &  0.46 \\
\hline                                                  
RMF   & 11 & Average & 452 & -90 & 16 & -2266 & -601 & -1023 &  0.79 &  1.22 &  1.82\\
 (11)  &     & $\sigma$ & 580 & 392 & 232 & 1266 & 656 & 12 &  0.44 &  1.14 &  1.09\\
\hline                                                  
RHF   & 4 & Average & 606 & -565 & -281 & -2553 & 92 & -1043 &  0.91 &  0.63 &  2.59\\
 (4)  &     & $\sigma$ & 167 & 373 & 185 & 432 & 737 & 15.91 &  0.49 &  0.26 &  1.90\\
\hline                                                  
TOTAL   & 25 & Average & 287 & -124 & -22 & -1765 & -437 & -846 &  0.64 &  0.65 &  1.69\\
   &     & $\sigma$ & 352 & 317 & 178 & 782 & 594 & 197 &  0.38 &  0.67 &  1.29\\
   &     & Min & -369 & -1178 & -585 & -4478 & -1369 & -1070 \\
   &     & Max & 1488 & 400 & 334 & 185 & 1298 & 167 \\

\end{tabular}
\end{ruledtabular}
\caption{Modification of the empirical parameters, $Q_{sat/sym}$ and $Z_{sat/sym}$ in the metamodeling ELFd adjusted
to reproduce reference models. The residual difference between the energy per particle given by the EOS and the one given by the associated metamodeling is shown as $\sigma_e$ in SM and NM. This distribution of the residual difference is encoded in terms of an average value and a standard deviation.
It is evaluated between $n_{sat}$ and $4n_{sat}$ considering 45 densities in total.}
\label{tab:system}
\end{table*}

To give an illustration of this statement, let us suppose that  the energy per nucleon and the pressure are known at $n_0=4n_{sat}=n_{hd}$, where $hd$ means high density.
The reference density $n_{hd}$ is quite arbitrary in this illustration, it could also have been chosen at a lower density.
The advantage of the present choice is in the simplification of the equations.

Considering $N=3$, the parameters $Q_{sat/sym}$ can then be fixed such that
$e^N_{ELFd}(n_0=n_{hd},n_1=0)=e_{hd}^{SM}$ and $e^N_{ELFd}(n_0=n_{hd},n_1=n_0)=e_{hd}^{NM}$, giving
\begin{eqnarray}
v_3^{is} &=& 6 ( e_{hd}^{SM}-t_{hd}^{SM}) -6 v_0^{is}-6 v_1^{is} - 3 v_2^{is}  ,\nonumber \\
v_3^{iv} &=& -6 ( e_{hd}^{SM}-e_{hd}^{NM} - t_{hd}^{SM}+t_{hd}^{NM}) -6 v_0^{iv}-6 v_1^{iv} - 3 v_2^{iv}  , \nonumber 
\label{eq:elfdN3}
\end{eqnarray}
while considering $N=4$, the parameters $Q_{sat/sym}$ and $Z_{sat/sym}$ are given by
$e^N_{ELFd}(n_0=n_{hd},n_1=0)=e_{hd}^{SM}$, 
$e^N_{ELFd}(n_0=n_{hd},n_1=n_0)=e_{hd}^{NM}$, 
and
$p^N_{ELFd}(n_0=n_{hd},n_1=0)=p_{hd}^{SM}$, 
$p^N_{ELFd}(n_0=n_{hd},n_1=n_0)=p_{hd}^{NM}$, 
where $e_{hd}^{SM/NM}$ and $p_{hd}^{SM/NM}$ are the energy per nucleon and pressure at the known reference point. 
 
These conditions lead to the following expressions,
\begin{eqnarray}
v_3^{is} &=& 24 \Big[ e_{hd}^{SM}-t_{hd}^{SM} \Big] - \frac{9}{8n_{sat}}\Big[p_{hd}^{SM} - p^{kin, SM}_{hd}\Big] \nonumber \\
&& - 6 \Big[ 4v_0^{is} + 3v_1^{is} + v_2^{is} \Big]  , \label{eq:elfd1}\\
v_3^{iv} &=& -24 \Big[ e_{hd}^{SM}-e_{hd}^{NM}-t_{hd}^{SM}+t_{hd}^{NM} \Big] \nonumber \\
&& + \frac{9}{8n_{sat}}\Big[p_{hd}^{SM}-p_{hd}^{NM} - p^{kin, SM}_{hd}+p^{kin, NM}_{hd}\Big] \nonumber \\
&& - 6 \Big[ 4v_0^{iv} + 3v_1^{iv} + v_2^{iv} \Big]  , \label{eq:elfd1iv}\\
v_4^{is} &=& -72 \Big[ e_{hd}^{SM}-t_{hd}^{SM} \Big] + \frac{9}{2n_{sat}}\Big[p_{hd}^{SM} - p^{kin, SM}_{hd}\Big] \nonumber \\
&& +12 \Big[ 6v_0^{is} + 4v_1^{is} + v_2^{is} \Big]  , \label{eq:elfd2}\\
v_4^{iv} &=& 72 \Big[ e_{hd}^{SM}-e_{hd}^{NM}-t_{hd}^{SM}+t_{hd}^{NM} \Big] \nonumber \\
&& - \frac{9}{2n_{sat}}\Big[p_{hd}^{SM}-p_{hd}^{NM} - p^{kin, SM}_{hd}+p^{kin, NM}_{hd}\Big] \nonumber \\
&& +12 \Big[ 6v_0^{iv} + 4v_1^{iv} + v_2^{iv} \Big]  , \label{eq:elfd2iv}
\end{eqnarray}
where $p^{kin}$ is the kinetic contribution to the pressure. 

This high density reference point {\it hd} should ideally be taken from empirical information, such as might be given in the future by high energy heavy ion collisions with exotic beams. For the time being, such an empirical reference does not exist, and we will take for $e_{hd}^{SM/NM}$ and $p_{hd}^{SM/NM}$ the values given by a reference model. This introduces again some model dependence in the EOS, which is exactly what we want to avoid with the empirical treatment. To circumvent this problem, in the calculation of nuclear and astrophysical observables we will consider huge uncertainties for the high order parameters from Tab. \ref{tab:system}, such as to cover the whole domain of density dependence at high density. In this sense, the reference values 
for $Q_{sat/sym}$, $Z_{sat/sym}$ given by Eqs.(\ref{eq:elfd1})-(\ref{eq:elfd2iv}) should only be considered as the central values of a large prior distribution which has to be filtered through constraints from astrophysical or laboratory observables.

\begin{figure}[tb]
\begin{center}
\includegraphics[angle=0,width=1.0\linewidth]{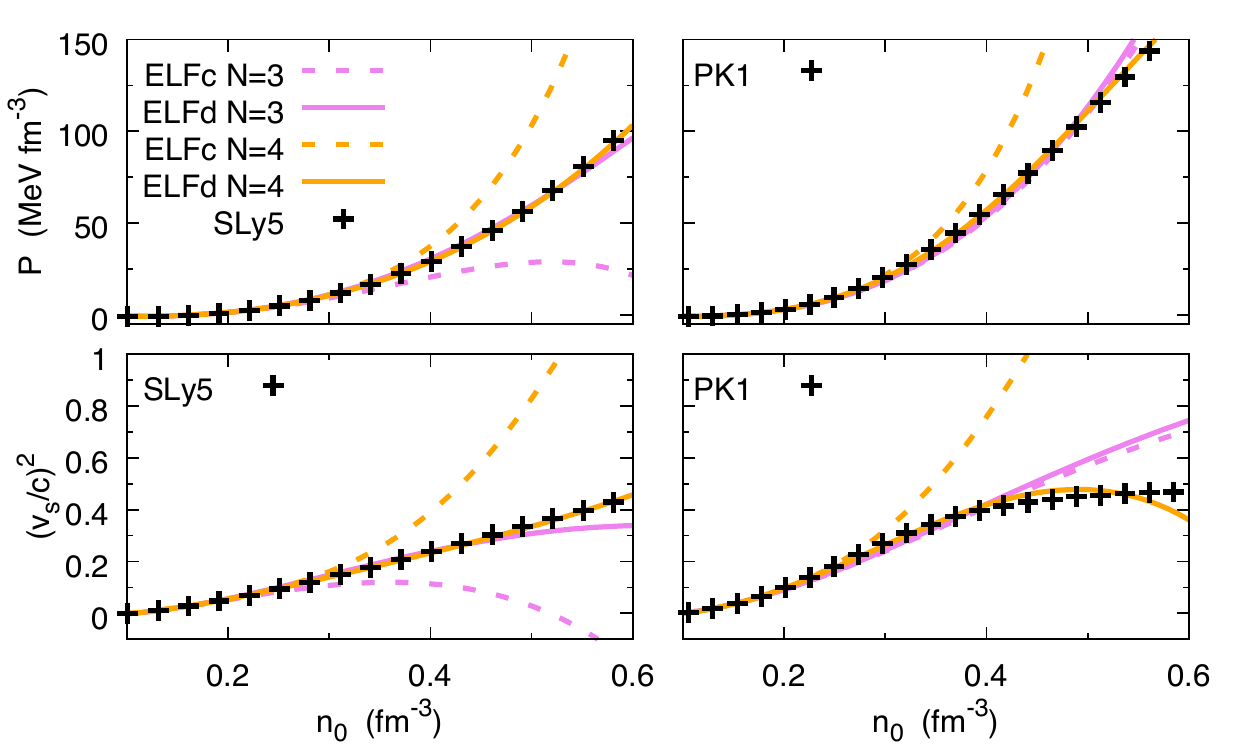}
\end{center}
\caption{(Color online) Comparison of the pressure (top panels) and sound velocity (bottom panels)
for the metamodels ELFc (dashed lines) and ELFd (solid lines) against the reference models SLy5 (left panels) and PK1 (right panels).
Only the orders $N$=3 and 4 are plotted.}
\label{fig:elfd2} 
\end{figure}

The new metamodeling ELFd is shown in Fig.~\ref{fig:elfd1} and compared to the metamodeling ELFc for $N$=3 and 4. 
We are using the non-relativistic Skyrme SLy5~\cite{SKY:SLY} and the RMF PK1~\cite{RMF:PK} as reference models for this illustration.
We can see that the reference model is very accurately reproduced by ELFd already for $N=4$.
A quantitative comparison of the metamodeling ELFd and other reference models is given in Sec.~\ref{sec:system}.
The pressure and the sound velocity which account for first and second derivative of the EOS, are shown
in Fig.~\ref{fig:elfd2}, where metamodelings ELFc and ELFd are compared to the reference models for $N$=3 and 4.
An excellent agreement between metamodeling ELFd with $N=4$ with the reference model can be remarked.
This is also reflected in the standard deviations between the binding energies predicted by the models and the ones of the metamodeling, which are shown in  Tab. \ref{tab:system} under the columns $\sigma_e$ in SM and NM.

To summarize, we have shown that, in order to better reproduce the density dependence of the binding 
energy and its derivatives up to about 4$n_{sat}$, it is necessary to re-adjust the empirical parameters
$Q_{sat/sym}$ and $Z_{sat/sym}$ by explicitly introducing some EOS information at another reference density. The introduction of this other reference density $n_{hd}$ stands for the necessity to complement the information determined
at saturation density, and could potentially be obtained by different ways, such as for instance HIC or from the properties 
of NS.
We have decided to take the reference density $n_{hd}=4n_{sat}$.
It is indeed an arbitrary choice, made only to simplify equations~(\ref{eq:elfd1})-(\ref{eq:elfd2}), but we have checked that the final result is largely unaffected by the choice of the reference density, provided it is 
larger than $2-3n_{sat}$.

More systematic comparisons and more quantitative criterion for the comparison between metamodeling ELFd and
reference models shall now be presented.

\section{Systematical comparison of the metamodeling ELFd with existing EOS}
\label{sec:system}

Our final proposition for the meta-EOS is ELFd presented in Sec.~\ref{sec:elfd}. 
We now turn to show the main advantages of this metamodeling, namely (i) the model is sufficiently flexible to be able to reproduce most existing phenomenological and ab-initio functionals, (ii) it can also accommodate density dependences which might correspond to a physical behavior, but are forbidden in the existing phenomenological EOS because of the assumed functional form. 

\subsection{Extracting high order parameters from the density behavior of existing analytical models} \label{sec:analytical}

\begin{table*}[t]
\centering
\setlength{\tabcolsep}{2pt}
\renewcommand{\arraystretch}{1.2}
\begin{ruledtabular}
\begin{tabular}{ccccccccccccc}
$P_{\alpha}$ & $E_{sat}$ & $E_{sym}$ & $n_{sat}$ & $L_{sym}$ & $K_{sat}$ & $K_{sym}$ & $Q_{sat}$ & $Q_{sym}$ & $Z_{sat}$ & $Z_{sym}$ & $m^*_{sat}/m$ & $\Delta m^*_{sat}/m$\\
                      & MeV         & MeV           & fm$^{-3}$  &  MeV           & MeV           & MeV           & MeV           & MeV           & MeV           & MeV \\
\hline
$\langle P_{\alpha} \rangle$ & -15.8           & 32           & 0.155            & 60           & 230          & -100           & 300                 & 0                 & -500           & -500          & 0.75           & 0.1\\ 
$\sigma_{P_{\alpha}}$         &     $\pm$0.3 &  $\pm$2  &  $\pm$0.005 & $\pm$15 & $\pm$20  & $\pm$100  & $\pm$400 & $\pm$400 & $\pm$1000 & $\pm$1000  & $\pm$0.1 & $\pm$0.1\\
\end{tabular}
\end{ruledtabular}
\caption{Synthesis of the expected values for the empirical parameters and their associated uncertainties. They are extracted from experimental analysis, except for the empirical parameters $Q_{sat/sym}$ and $Z_{sat/sym}$ which are estimated from Tab.~\ref{tab:system}.}
\label{tab:epvar}
\end{table*}

In this section, the values of the empirical parameters $Q_{sat/sym}$ and $Z_{sat/sym}$ are calculated from
Eqs.~(\ref{eq:elfd1})-(\ref{eq:elfd2iv}) for a large number of nuclear relativistic and non-relativistic interactions.
We ran over the same models as the one already considered in the study presented in Tab.~\ref{tab:mean-table-models} for instance.
Detailed results are given in Tab.~\ref{table:Nuclear-New3-func-Table2} and summarized in Tab.~\ref{tab:system}.
In order to evaluate the effect induced by the highest order term in the series expansion, $v_4$, on the empirical parameters 
$Q_{sym}$ and $Z_{sym}$ we have added a column in Tab.~\ref{tab:system} where these empirical parameters are evaluated
imposing $v_4=0$.
The last columns in Tab.~\ref{tab:system} stand for the average dispersion between the fitted EOS and the original data for
symmetric and neutron matter, considering densities from $n_{sat}$ up to $4n_{sat}$. It is particularly interesting to remark that the quality of EOS reproduction is very similar for relativistic and non-relativistic models, even if we have employed a completely classical treatment of the kinetic energy density and no Dirac mass has been introduced.

In the isoscalar channel, it is interesting to remark that the parameter $Q_{sat}$ is systematically shifted up by about 200~MeV,
and the average value of the parameter $Z_{sat}$ is shifted down by a large amount.
There is therefore a compensation between the parameters $Q_{sat}$ and $Z_{sat}$.
In absolute value the parameters $Z_{sat}$ in Tab.~\ref{tab:system} are lower than in Tab.~\ref{tab:mean-table-models},
and the issue related to alternative series is now pushed to densities well above $4n_{sat}$, where our approach is
not any more well suited.

In the isovector channel, both $Q_{sym}$ and $Z_{sym}$ are shifted down and in absolute value the parameters $Z_{sym}$ in 
Tab.~\ref{tab:system} are lower than in Tab.~\ref{tab:mean-table-models}, in the same way as for $Z_{sat}$.
The issues induced by alternative series expansion are also reduced in NM.

The average dispersion in symmetric and neutron matter given in Tab.~\ref{tab:system} show the excellent reproduction of
the original EOS up to $4n_{sat}$. It is also shown the important role of $v_4$ since removing it shifts up the average standard 
deviation.
In conclusion, Tab.~\ref{tab:system} shows that considering the shifted value given in Tab.~\ref{tab:system} and 
Eqs.~(\ref{eq:elfd1})-(\ref{eq:elfd2}) it is possible to reproduce very accurately all the tested EOS with the ELFd metamodeling.

In addition to having proven that all the considered EOS can be well reproduced by the ELFd metamodel, we can use
the average values given in Tab.~\ref{tab:system} to estimate the average values and uncertainties of the highest order
empirical parameters $Q_{sat/sym}$ and $Z_{sat/sym}$.
In Tab.~\ref{tab:epvar}, we present a global summary of all empirical parameters and estimated uncertainties based on
our analysis presented in previous sections.
The low order empirical parameters $E_{sat/sym}$, $L_{sym}$, $K_{sat/sym}$ as well as the effective masses are 
compatible with the experimental analysis presented in sec.~\ref{Sec:empirical:exp} and model average predictions
in sec.~\ref{sec:empiricalparameters}, and synthesized in Tab.~\ref{tab:mean-table-models} and \ref{tab:mean-table-models-EFT}.
For the highest order empirical parameters $Q_{sat/sym}$ and $Z_{sat/sym}$, we have considered the shifted values
given in Tab.~\ref{tab:system} with large estimated uncertainties $\sigma_{P_{\alpha}}$.
The large estimated uncertainties are chosen such as to compensate the fact that the average values taken for these
empirical parameters are only determined from the existing models, without any empirical information.
The average value of $Z_{sat}$ is taken identical to $Z_{sym}$ for simplicity.
In the following, we consider average values and uncertainties for the empirical parameters which are summarized in 
Tab.~\ref{tab:epvar}.

\subsection{Extracting empirical parameters from chiral EFT results}
\label{Sec:ap:XEFT}

Neutron matter calculations based on modern potentials have recently been performed up to saturation 
density, see for instance Refs.~\cite{Gandolfi2012,Tews2013,Drischler2016}.
In these approaches, either advanced many-body technics have been employed, such as auxiliary field diffusion
Monte-Carlo (AFDMC) with hard-core potentials~\cite{Gandolfi2012} or with soft potentials from chiral EFT~\cite{Roggero2014,Wlazlowski2014,Lynn2017}, or many-body perturbation theory
based on chiral EFT~\cite{Tews2013,Drischler2016}.
The perturbative convergence of the chiral potential can be studied in detail in this approach and
the equation of state is usually provided within a band representing the uncertainties in the nuclear interaction as
well as in the convergence in the expansion.
In this section, we perform an analysis of these recent predictions, based either on the fits provided by the authors 
or on their numerical results.

\begin{table}[t]
\centering
\setlength{\tabcolsep}{2pt}
\renewcommand{\arraystretch}{1.2}
\begin{ruledtabular}
\begin{tabular}{cccccccc}
Model & 3BF &  $E_{NM}$ & $L_{sym}$ & $K_{NM}$ &  $Q_{NM}$ &  $Z_{NM}$  \\
 &  & MeV & MeV & MeV & MeV & MeV  \\
\hline
GCR-1 & none & 14.48 & 30.7 & 17   & 55   & -117 \\
GCR-2 & $V_{2\pi}^{PW}+V_{\mu=150}^R$ &16.15 & 40.2 & 45   & 52   & -117 \\
GCR-3 & $V_{2\pi}^{PW}+V_{\mu=300}^R$ &15.99 & 39.8 & 47   & 59   & -121 \\
GCR-4 & $V_{3\pi}+V_R$ & 16.21 & 42.9 & 76   & 93   & -137 \\
GCR-5 & $V_{2\pi}^{PW}+V_{\mu=150}^R$ & 17.76 & 50.8 & 82   & 56   & -121 \\
GCR-6 & $V_{3\pi}+V_R$ & 17.71 & 54.7 & 128 & 121 & -149 \\
GCR-7 & UIX & 19.02 & 61.7 & 147 & 122 & -155 \\
\hline
Average  & &  16.76 & 45.8 & 77   & 80 & -131 \\ 
$\sigma$ & &  1.39   & 9.7 & 43   & 29 &  15  \\
Min           & & 14.48  & 30.7 & 17   & 52 & -155  \\
Max          & & 19.02  & 61.7 & 147 & 122 & -117  \\
\end{tabular}
\end{ruledtabular}
\caption{Neutron matter energy per nucleon $E_{NM}$, slope of the symmetry energy $L_{sym}$, 
incompressibility $K_{NM}$, $Q_{NM}$ and kurtosis $Z_{NM}$
for the seven Hamiltonians studied in Ref.~\cite{Gandolfi2012}.
These Hamiltonian are treated within the AFDMC many-body method and they are
based on AV8' NN potential supplemented by various three-body forces (3BF).
They are named GCR in reference to the names of the authors.}
\label{tab:mean-table-models-AI}
\end{table}

Considering the accurate fit for the energy per nucleon in neutron matter provided in Ref.~\cite{Gandolfi2012},
\begin{equation}
e(n_0)=a\left(\frac{n_0}{n_{sat}}\right)^\alpha+b\left(\frac{n_0}{n_{sat}}\right)^\beta,
\end{equation}
where the parameters $a$, $\alpha$, $b$, and $\beta$ are provided in Tab.~I of Ref.~\cite{Gandolfi2012}
for various prescriptions for the 3BF, we have calculated the empirical parameters in neutron matter
at the fixed density $n_{sat}=0.16$~fm$^{-3}$:
$E_{NM}=E_{sat}+E_{sym}$ (energy per nucleon),
$L_{sym}$ (slope of the symmetry energy),
$K_{NM}=K_{sat}+K_{sym}$ (curvature),
$Q_{NM}=Q_{sat}+Q_{sym}$ (skewness),
$Z_{NM}=Z_{sat}+Z_{sym}$ (kurtosis).
These quantities are given in Tab.~\ref{tab:mean-table-models-AI} for the 7 different Hamiltonians
explored in Ref.~\cite{Gandolfi2012}.
As in previous analyses, we have also extracted an average value and standard deviation for these 7 Hamiltonians.

\begin{figure}[tb]
\begin{center}
\includegraphics[angle=0,width=1.0\linewidth]{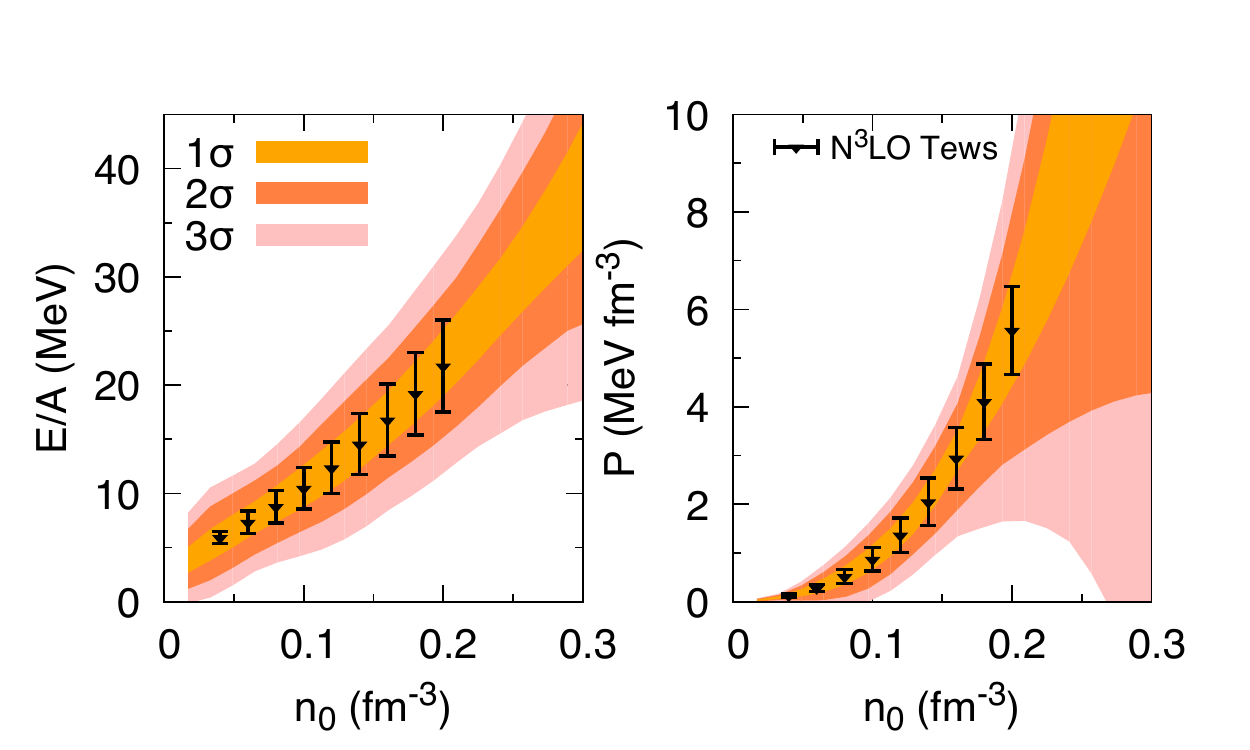}
\end{center}
\caption{(Color online) Equations of state at $1\sigma$, $2\sigma$ and $3\sigma$ obtained from
the first calculation of neutron matter with chiral EFT at N$^3$LO (points with error-bars) and given in 
Ref.~\cite{Tews2013}. See text for more details.}
\label{fig:Ingo1} 
\end{figure}

\begin{figure*}[tb]
\begin{center}
\includegraphics[angle=0,width=0.9\linewidth]{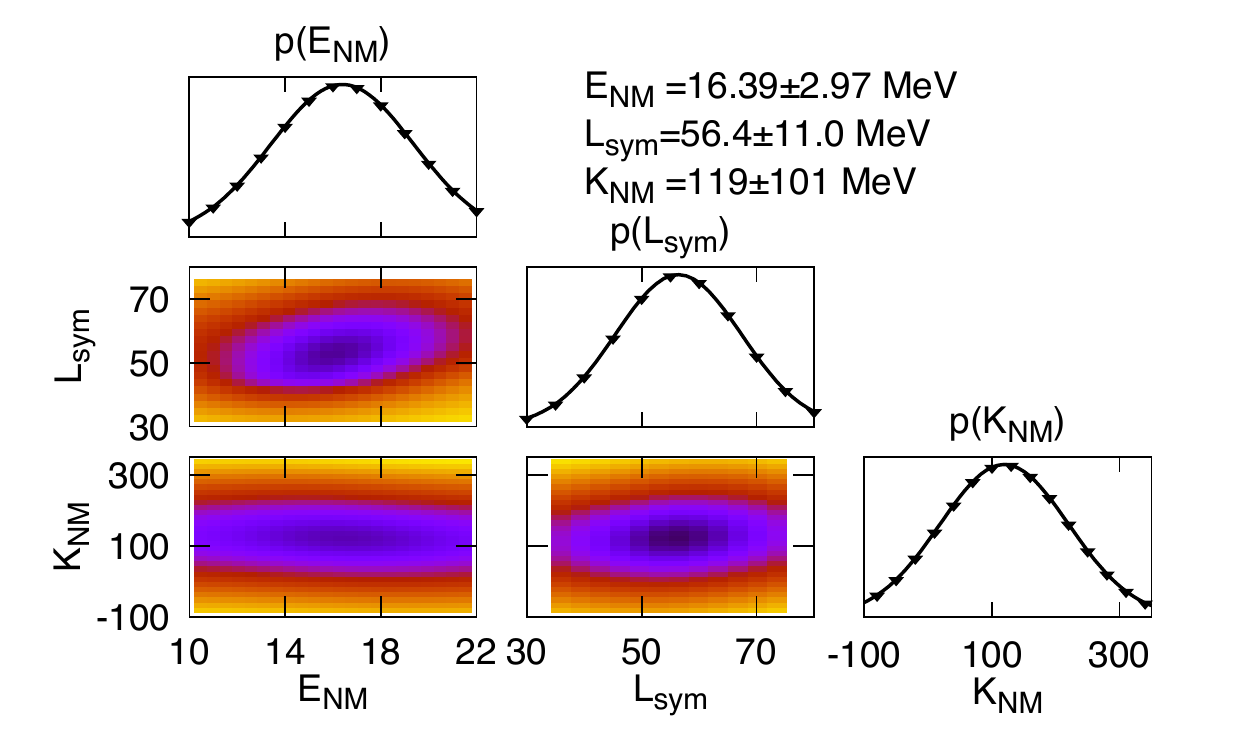}
\end{center}
\caption{(Color online) Analysis of the probability distribution $p(E_{NM}, L_{sym}, K_{NM})$ deduced from the best fits to neutron matter calculations shown in Fig.~\ref{fig:Ingo1}.
The centroids and standard deviations for the parameters $E_{NM}$, $L_{sym}$ and $K_{NM}$ are given in the figure.
See text for more details.}
\label{fig:Ingo2} 
\end{figure*}

We have also analyzed the results for the first  complete N$^3$LO calculation of the neutron matter energy~\cite{Tews2013}.
In this case, the energy and the pressure in neutron matter are provided as tabulated numbers with error-bars, up to the
density $n_0=0.2$~fm$^{-3}$.
We have here considered the metamodeling ELFc with $N=2$ since only a small domain of density is provided (the empirical
parameters associated to $N=3$ and 4 are undetermined by the provided data).
To extract the empirical parameters and their error-bars from these results, we have associated to each set of
empirical parameters ($E_{NM}$, $L_{sym}$, $K_{NM}$) a $\chi^2$ defined as,
\begin{eqnarray}
\chi^2 &=&\frac{1}{2M-3} \sum_{i=1}^M \left( \frac{e_i-e_{ELFc}(n_0^i)}{\epsilon_i^e}\right)^2 \nonumber \\
&&+ \left( \frac{p_i-p_{ELFc}(n_0^i)}{\epsilon_i^p}\right)^2 ,
\end{eqnarray}
where $(e_i,\epsilon_i^e)$ and $(p_i,\epsilon_i^p)$ are the average values and error-bars for the energies and pressures predicted in Ref.~\cite{Tews2013}.
A likelihood probability is then associated to each set of empirical parameters, as
$p(E_{NM}, L_{sym}, K_{NM})=\exp(-\chi^2/2)/p_{norm}$, and the distance to the maximum likelihood probability ($p_{max}$) is measured
in terms of $\sigma$, where the models at $k\sigma$ are those for which $p/p_{max}>\exp(-k^2/2)$.

The $1\sigma$, $2\sigma$, and $3\sigma$ domains for ELFc meta-EOS passing through the binding energies and 
pressures calculated in Ref.~\cite{Tews2013} are represented in Fig.~\ref{fig:Ingo1}, while the representation of the probability
distribution showing the correlations among the empirical parameters is shown in the corner-Fig.~\ref{fig:Ingo2}.
The two-variables probabilities are defined as $p(A,B)=\sum_C p(A,B,C)$, while the single-variable probabilities are
$p(A)=\sum_{B,C} p(A,B,C)$, where $(A,B,C)$ can be any permutation of $(E_{NM}, L_{sym}, K_{NM})$.

It is interesting to note the nice correlation between $E_{NM}$ and $L_{sym}$, as well as between
$K_{NM}$ and $L_{sym}$, while $K_{NM}$ is almost independent of $E_{NM}$.
The single-variable probabilities are shown on the diagonal of the corner-Fig.~\ref{fig:Ingo2}.
The value of the centroids, defined as $\langle A \rangle = \sum_A A p(A)/\sum_A p(A)$
and of the standard deviation, defined as $\sigma_A =  \sqrt{\langle A^2 \rangle - \langle A \rangle^2 }$,
are given in Fig.~\ref{fig:Ingo2} as well as in Tab.~\ref{tab:mean-table-models-EFT}.

Let us remind that the following constraints have been obtained in Ref.~\cite{Tews2013}: $E_{sym}=31.9\pm3$~MeV and
$L_{sym}=54.5\pm 11.5$~MeV. Assuming $E_{sat}=-16.0\pm0.5$~MeV, we deduce $E_{NM}=15.9\pm3.5$~MeV
from Ref.~\cite{Tews2013}.
These results are compatible with the ones we obtain from a different analysis.
In addition, we could obtain a constraint for the neutron matter incompressibility $K_{NM}$ which was not given in Ref.~\cite{Tews2013}.
Let us finally remark that the determination of the parameter $K_{NM}$ is possible since the pressure was 
introduced into the definition of the $\chi^2$.
With the energies only, we would have obtained a much larger spread in the probability distribution of $K_{MN}$.
The fit of the energy per nucleon and pressure at the same time offers therefore very interesting constraints for the nuclear 
equation of state and its derivatives.

\begin{table*}[t]
\centering
\setlength{\tabcolsep}{2pt}
\renewcommand{\arraystretch}{1.2}
\begin{ruledtabular}
\begin{tabular}{ccccccccccccccccc}
    &   $E_{sat}$ & $E_{sym}$ & $
n_{sat}$ &  $L_{sym}$ & $K_{sat}$ &  $K_{sym}$  & $Q_{sat}$  &  $Q_{sym}$ & $Z_{sat}$ & $Z_{sym}$  & $K_{\tau}$ \\
Model  & MeV & MeV & fm$^{-3}$ & MeV & MeV & MeV & MeV & MeV & MeV & MeV & MeV \\
\hline
\#1 & -16.92 & 34.57 & 0.189 & 48.5  &  241 & -224 & -125  & -311  & 1281 &  -1974 & -490 \\ 
\#2 & -15.73 & 32.81 & 0.178 & 46.9  &  216 & -192 & -176  & -182  & 1454 &  -2283 & -435 \\
\#3 & -15.18 & 32.18 & 0.174 & 53.0  &  224 & -108 & 24     &  96    & 1299 &  -2961  & 432 \\
\#4 & -14.84 & 31.63 & 0.170 & 46.1  &  198 & -170 & -223  & -82    & 1534 &  -2464  & -395 \\
\#5 & -13.80 & 29.83 & 0.158 & 44.5  &  182 & -143 & -310  &  69    & 1537 &  -2638  & -334 \\
\#6 & -16.44 & 34.54 & 0.190 & 53.5  &  242 & -220 & -31    & -640  & 1139 &  -1750  & -534 \\
\#7 & -13.23 & 28.53 & 0.140 & 43.9  &  192 & -144 & -132  &  -95   & 901   &  -2149  & -378 \\
\hline
Average  &    -15.16 & 32.01 & 0.171 & 48.1  &  214 & -172  & -139  & -164  & 1306 &  -2317 & -428 \\
$\sigma$ &       1.24 &   2.09 & 0.016 &   3.6  &   22  &    40  &  104  &  234  &    214 &   379  & 63 \\
Min          &   -16.92 & 28.53  & 0.140 & 43.9  &  182 & -224 & -310  & -640  &    901 &  -2961 & -534 \\
Max         &   -13.23 & 34.57  & 0.190 & 53.5  &  242 & -108 &    24  &     96 &  1537 &  -1750 & -334 \\
 \end{tabular}
\end{ruledtabular}
\caption{Binding energy $E_{sat}$, the symmetry energy $E_{sym}$, saturation density $n_{sat}$, 
slope of the symmetry energy $L_{sym}$, isoscalar incompressibility $K_{sat}$, isovector incompressibility  $K_{sym}$,
isoscalar skewness $Q_{sat}$, isovector skewness $Q_{sym}$, isoscalar kurtosis $Z_{sat}$,
isovector kurtosis $Z_{sym}$, and $K_\tau$ for chiral EFT given in Ref.~Ref.~\cite{Drischler2016}.}
\label{tab:mean-table-models-XEFT}
\end{table*}

Finally, we analyze the very recent calculations of asymmetric nuclear matter based on chiral two- and three-body
interactions at $N^3LO$~\cite{Drischler2016}.
The empirical parameters shown in Tab.~\ref{tab:mean-table-models-XEFT} are deduced from the fit of the binding
energy given in Ref.~\cite{Drischler2016},
\begin{equation}
e(n_0,n_1)=\sum_{\nu=2,3,4,5,6} (C_{0\nu}+C_{2\nu} \delta^2) \left( \frac{n_0}{0.16} \right)^{\nu/3} ,
\end{equation}
where the values of the parameters $C_{0\nu}$ and $C_{2\nu}$ are given in Tab.~II 
for the seven Hamiltonians detailed in Tab.~I of Ref.~\cite{Drischler2016}.

The reported ranges for $K_{sat}$ and $E_{sym}$ in Ref.~\cite{Drischler2016} are very close to ours,
$K_{sat}=218\pm 36$~MeV and $E_{sym}=32.05\pm 3.65$~MeV, if we consider the values we obtained
from the minimal (Min) and maximal (Max) values for these parameters. 
The slight difference we obtained may come from the fact that the free Hartree-Fock spectrum was also considered
in Ref.~\cite{Drischler2016} for the estimation of the error-bars.

The empirical parameters of neutron matter given in Tab.~\ref{tab:mean-table-models-EFT} are obtained
from the combining of isoscalar and isovector parameters given in Tab.~\ref{tab:mean-table-models-XEFT}.

\subsection{Behavior of the symmetry energy around saturation density}

In the previous sections we have shown that our meta-EOS can very accurately reproduce existing models of very different types (relativistic and non-relativistic, phenomenological and ab-initio).
It can also be used to make complete statistical analysis of such models, for example to evaluate the confidence interval of a nuclear or astrophysical observable compatible with chiral EFT at any chosen confidence level. We now turn to show that the parameter space is large enough to accommodate density behaviors which cannot be explored by existing functionals. To this aim, we will take the example of the density dependence of the symmetry energy.

Many theoretical and experimental studies of the density dependence of the symmetry energy consider models with very limited
number of parameters.
This leads to some strong correlations in the density dependence of the symmetry energy, e.g. iso-soft behavior below $n_{sat}$
and iso-stiff behavior above $n_{sat}$.
We have discussed for instance the opposite impact of $L_{sym}$ and $K_{sym}$ in the analysis of Fig.~\ref{fig:vary2} and
\ref{fig:vary3}.
Combining together the effect of $L_{sym}$ and $K_{sym}$ a rich behavior for the symmetry energy can be explored.

\begin{figure*}[tb]
\begin{center}
\includegraphics[angle=0,width=0.7\linewidth]{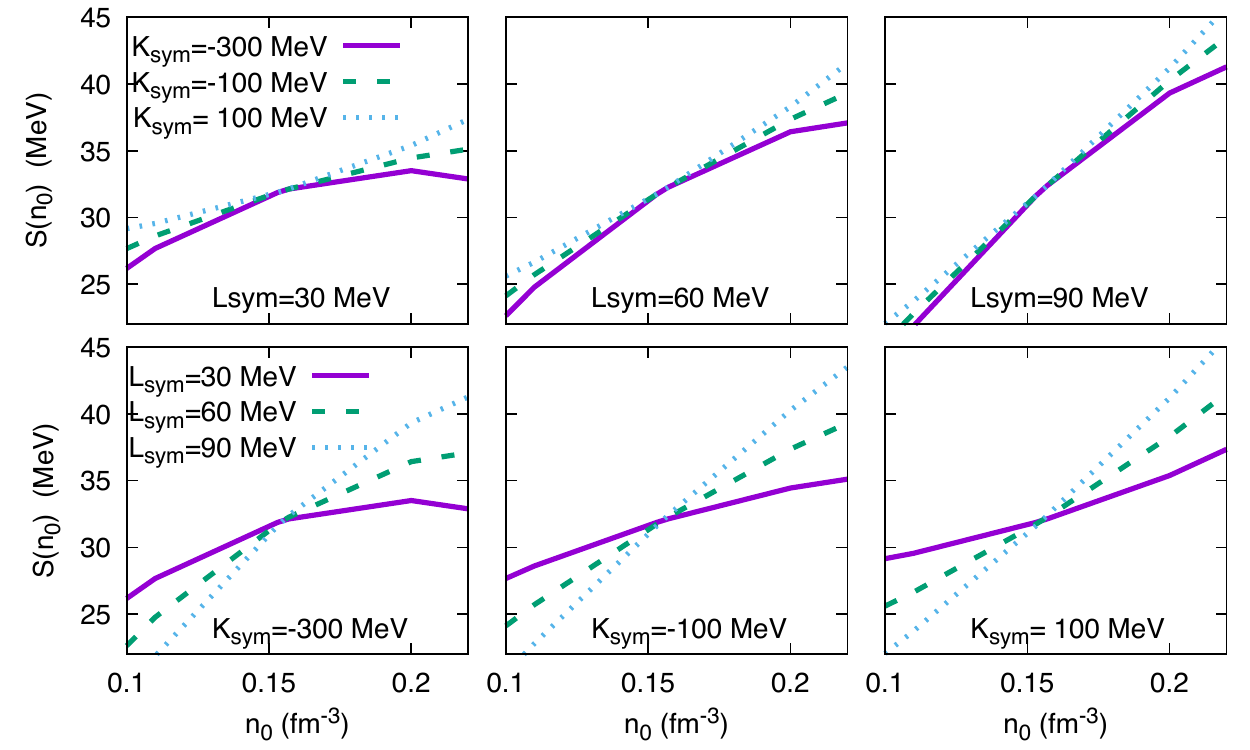}
\end{center}
\caption{(Color online) Effect of varying the value of $L_{sym}$ and $K_{sym}$ on the symmetry energy around saturation density.
Note that we vary the parameters by $\pm2\sigma$ to enhance the effects.
Top panels: varying $K_{sym}$ at fixed $L_{sym}$.
Bottom panels: varying $L_{sym}$ at fixed $K_{sym}$.}
\label{fig:vary_iso} 
\end{figure*}

We illustrate our purpose in Fig.~\ref{fig:vary_iso} where the symmetry energy $S(n_0)$ is represented in various cases:
$K_{sym}$ is varied for fixed values of $L_{sym}$ (top panels), and $L_{sym}$ is varied for fixed values of $K_{sym}$ 
(bottom panels).
We observe on the top panels of Fig.~\ref{fig:vary_iso} that for a fixed value of $L_{sym}$, changing the value of
$K_{sym}$ changes the curvature of the symmetry energy around $n_{sat}$: negative values of $K_{sym}$ produce a
concave density dependence of the symmetry energy, e.g. iso-soft below $n_{sat}$ and iso-soft above $n_{sat}$,
while positive values of $K_{sym}$ produce a convex density dependence of the symmetry energy, e.g. iso-stiff below 
$n_{sat}$ and iso-stiff above $n_{sat}$.
We now turn to the bottom panels of Fig.~\ref{fig:vary_iso} where we have varied $L_{sym}$ at fixed value of $K_{sym}$.
varying $L_{sym}$ generates different slopes of the symmetry energy which are more or less iso-stiff below $n_{sat}$ and 
iso-soft above $n_{sat}$.

In conclusion, we have shown that it is possible to play with the two most important empirical parameters $L_{sym}$ and
$K_{sym}$ to modify the behavior of the symmetry energy around $n_{sat}$ and explore a wide range of possible density dependence.

\begin{figure*}[tb]
\begin{center}
\includegraphics[angle=0,width=0.7\linewidth]{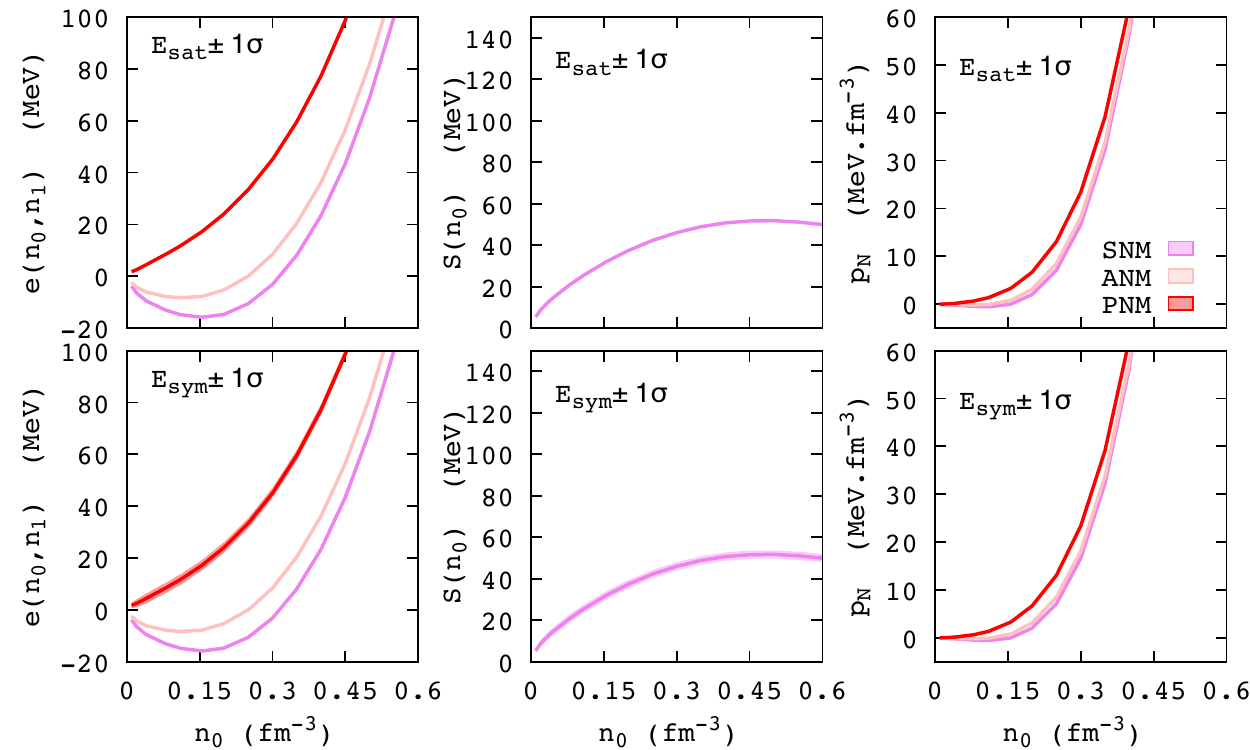}
\end{center}
\caption{(Color online) Effect of varying the value of $E_{sat}$ (top panels) and $E_{sym}$ (bottom panels) around the mean value
given in Tab.~\ref{tab:epvar} considering $1\sigma$ deviation. Here $1\sigma$ is also taken from Tab.~\ref{tab:epvar}.
From left to right: energy per nucleon, symmetry energy and pressure in SNM ($\delta=0$),  ANM ($\delta=0.5$), and  PNM ($\delta=1$).}
\label{fig:vary1} 
\end{figure*}

\begin{figure*}[tb]
\begin{center}
\includegraphics[angle=0,width=0.7\linewidth]{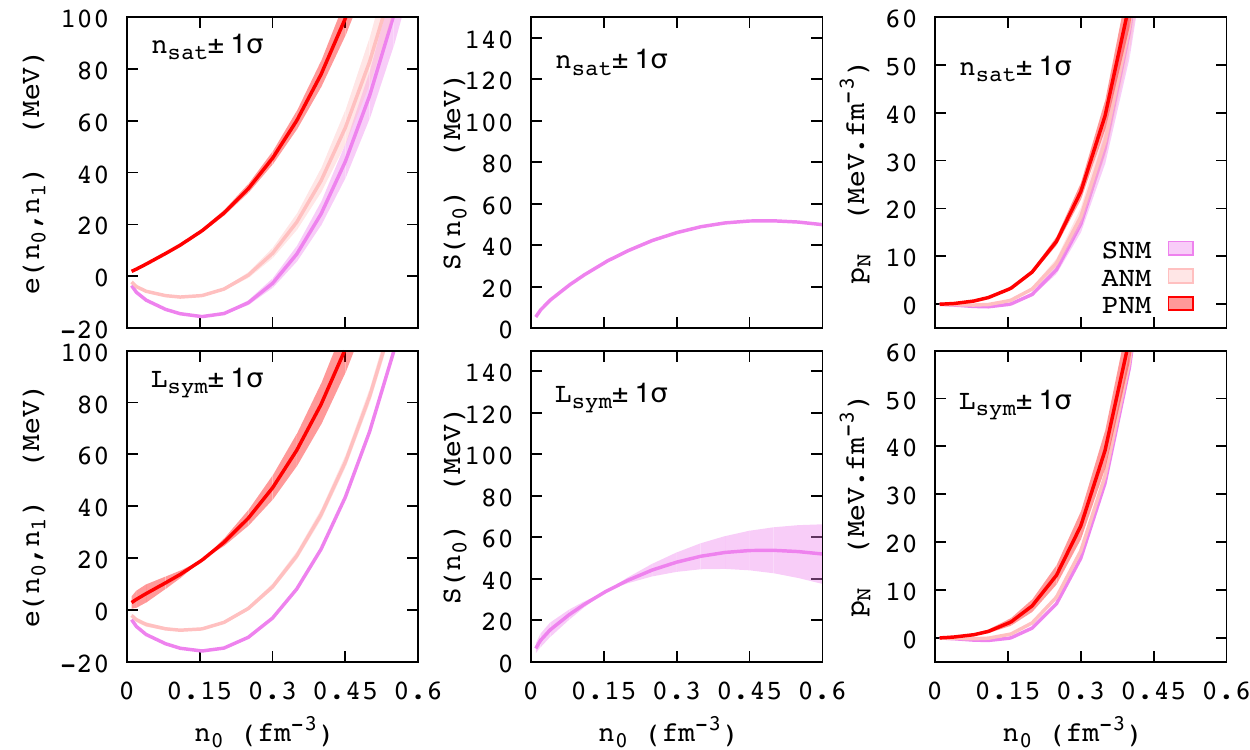}
\end{center}
\caption{(Color online) Same as Fig.~\ref{fig:vary1} for $n_{sat}$ (top panel) and $L_{sym}$ (bottom panel).}
\label{fig:vary2} 
\end{figure*}

\begin{figure*}[tb]
\begin{center}
\includegraphics[angle=0,width=0.7\linewidth]{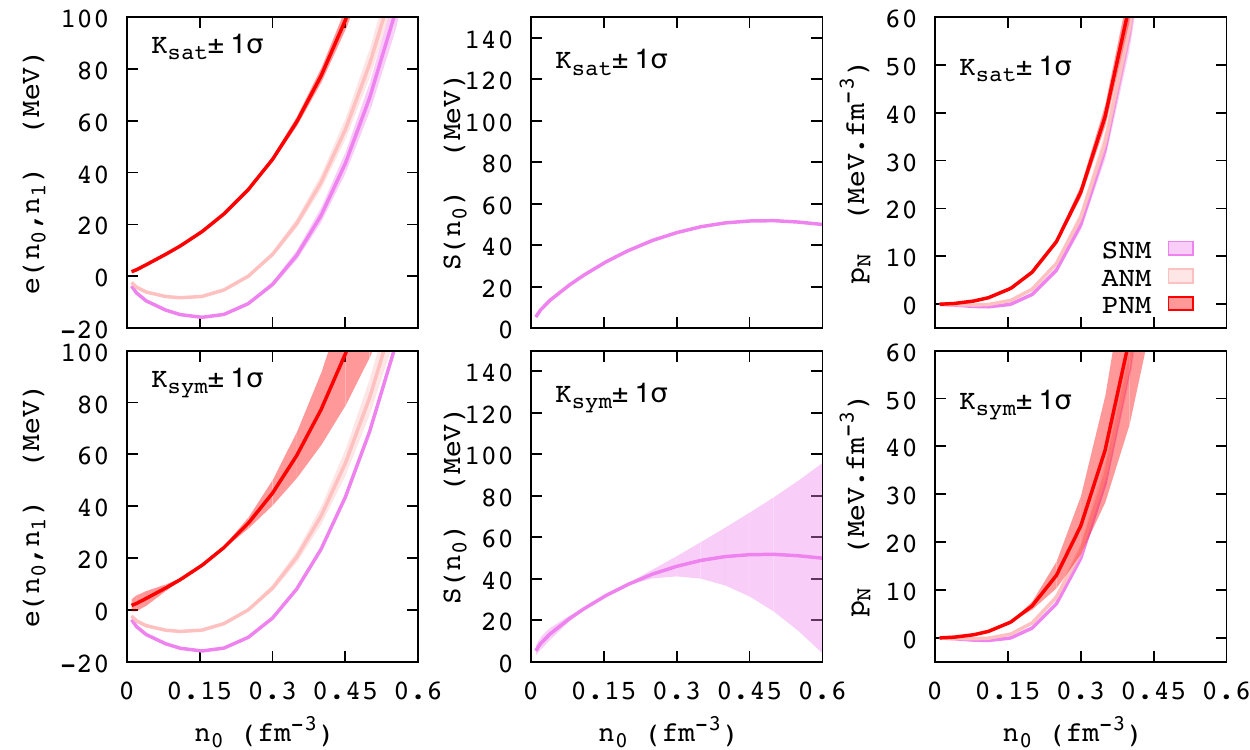}
\end{center}
\caption{(Color online) Same as Fig.~\ref{fig:vary1} for $K_{sat}$ (top panel) and $K_{sym}$ (bottom panel).}
\label{fig:vary3} 
\end{figure*}

\begin{figure*}[tb]
\begin{center}
\includegraphics[angle=0,width=0.7\linewidth]{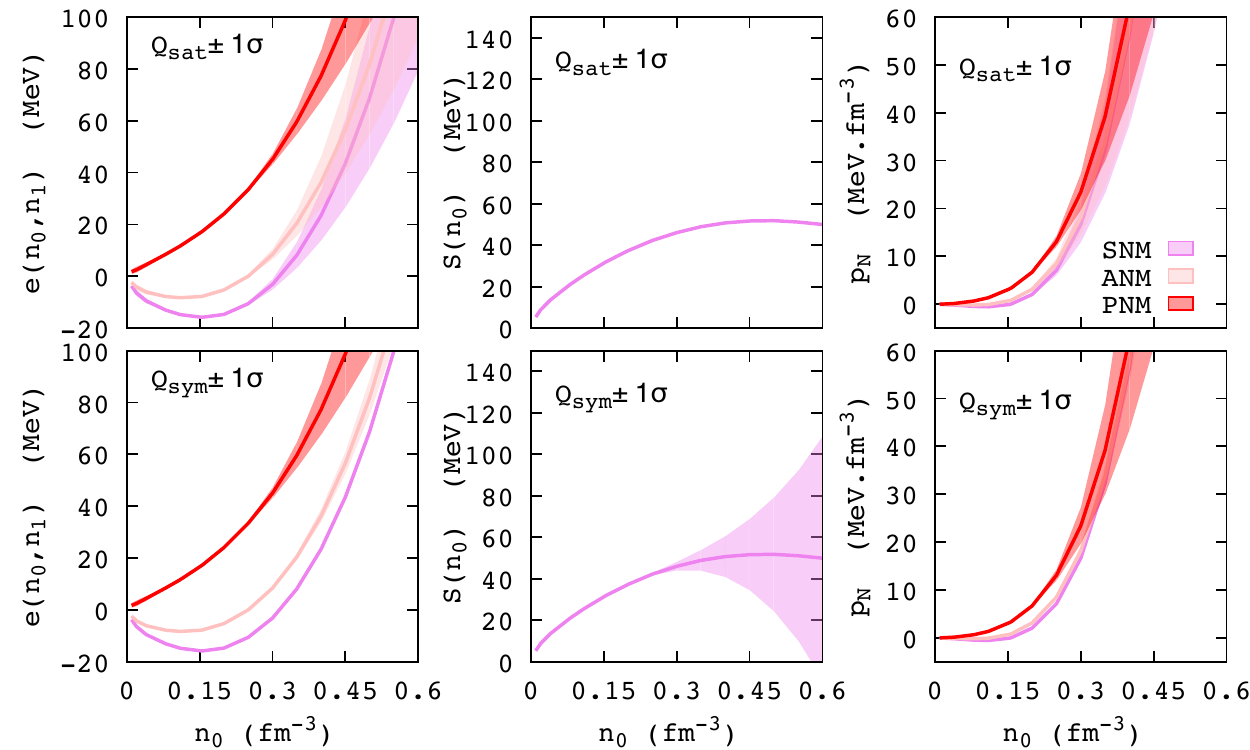}
\end{center}
\caption{(Color online) Same as Fig.~\ref{fig:vary1} for $Q_{sat}$ (top panel) and $Q_{sym}$ (bottom panel).}
\label{fig:vary4} 
\end{figure*}

\begin{figure*}[tb]
\begin{center}
\includegraphics[angle=0,width=0.7\linewidth]{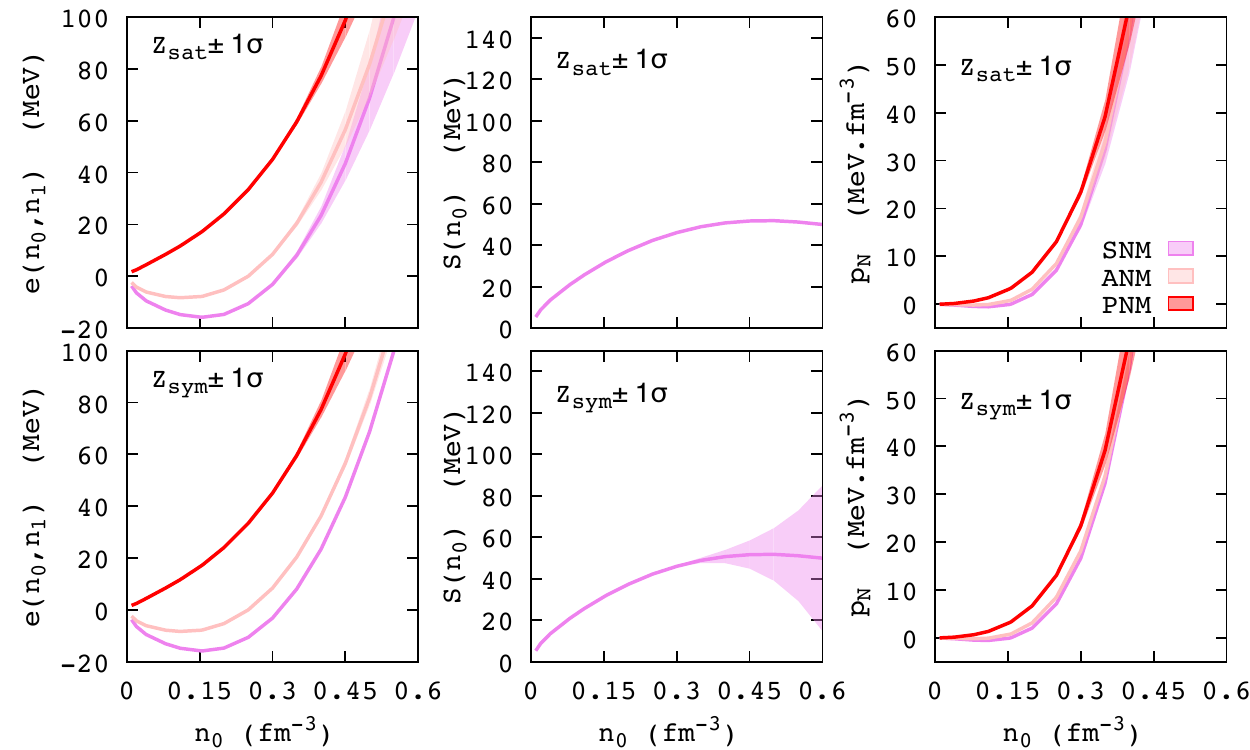}
\end{center}
\caption{(Color online) Same as Fig.~\ref{fig:vary1} for $Z_{sat}$ (top panel) and $Z_{sym}$ (bottom panel).}
\label{fig:vary5} 
\end{figure*}

\begin{figure*}[tb]
\begin{center}
\includegraphics[angle=0,width=0.7\linewidth]{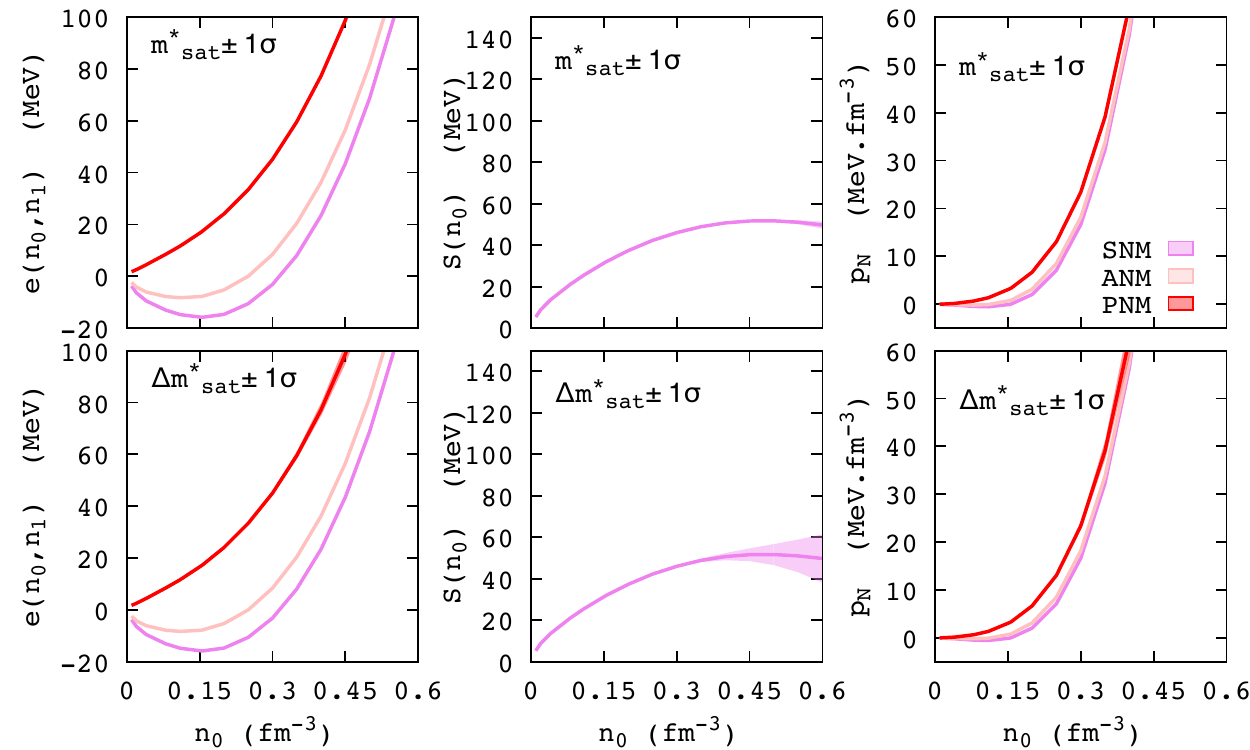}
\end{center}
\caption{(Color online) Same as Fig.~\ref{fig:vary1} for $m^*_{sat}$ (top panel) and $\Delta m^*_{sat}$ (bottom panel).}
\label{fig:vary0} 
\end{figure*}

\section{Effect of the present uncertainties on the nuclear EOS}
\label{sec:eos}

In this section, we will explore another advantage of the proposed formalism, namely the possibility of studying the effect of the different empirical parameters independently from each other.
This allows performing a sensitivity study on the meta-EOS and recognizing the most influential empirical parameters.
 
We consider Tab.~\ref{tab:epvar} as a reference for fixing an average nuclear meta-EOS and probing the impact of
varying the empirical parameters within their estimated uncertainties.

\subsection{Effect of varying a single parameter}

In this analysis, we define an average meta-EOS determined by the average empirical parameters given in 
Tab.~\ref{tab:epvar}, then we vary the empirical parameters one-after-the-other to estimate the impact
of the estimated uncertainties on the prediction of the EOS.
This impact is obtained as a combined effect between the real influence of the considered parameter and its estimated
uncertainty. 
For instance, an influential empirical parameter known very accurately will have a weak influence, while a less influential but very poorly known empirical parameter might have a large influence.
In the following, we will also observe that the different order empirical parameters play a role a different densities.
In general, the higher the order, the farther from saturation density it has an impact.

In Figs.~\ref{fig:vary1}-\ref{fig:vary0}, we have grouped together the isoscalar and isovector empirical parameters 
order-by-order. On the top panels are shown the effects induced by the change of the isoscalar empirical parameters
while on the bottom panels, it is the impact isovector empirical parameters which are tested.

Let us start with Fig.~\ref{fig:vary1} where we show the impact of varying $E_{sat/sym}$ considering a $1\sigma$ deviation
around the average meta-EOS given in Tab.~\ref{tab:epvar}.
The energy per particle and the pressure for SNM, ANM (defined as $\delta=0.5$) and PNM are shown as a function of the density, and the symmetry
energy $S(n_0)$ is also represented in Fig.~\ref{fig:vary1}.
The impact of our uncertainty on $E_{sat/sym}$ is rather weak for the energy per particle, the symmetry energy, and the pressure.
Let us note that since the pressure is the derivative of the energy per particle, it is not impacted at all by $E_{sat/sym}$.

More interestingly we show in Fig.~\ref{fig:vary2} the impact of varying $n_{sat}$ and $L_{sym}$ in the same way as
in Fig.~\ref{fig:vary1}.
The impact of our uncertainty on $n_{sat}$ is quite small, while the effect of varying $L_{sym}$ has a noticeable impact above
and below saturation density for the energy per particle, the symmetry energy, and the pressure. 
In asymmetric and neutron matter, the effect of $L_{sym}$ below saturation density is opposite to above saturation density.
So a large value of $L_{sym}$ will strongly reduce (resp. increase) the energy per particle below (resp. above) saturation density.
The crossing as saturation density of these quantities is expected since the meta-EOS is a series expansion taking
$n_{sat}$ as the reference density.
The slope of the symmetry energy is sensitive to $L_{sym}$. The largest value of $L_{sym}$ that we have considered induces a
positive slope until $n_{sat}$ while the lowest value produce a negative slope.
For the pressure, since it is obtained as the derivative of the energy per particle, the uncertainty in $L_{sym}$ propagates almost
constantly through the densities.

The impact of our uncertainty in $K_{sat/sym}$ is shown in Fig.~\ref{fig:vary3}.
While the uncertainty in the value of $K_{sat}$ has a weak impact on the nuclear EOS (almost un-noticeable), the large 
uncertainty of $K_{sym}$ has a large impact on the nuclear EOS.
At variance with $L_{sym}$, the effect of $K_{sym}$ below and above saturation density is similar: an increase of
$K_{sym}$ increases the energy per particle both below and above $n_{sat}$.
Large negative values of $K_{sym}$ force the symmetry energy $S$ to bend down at high density.
For very large values of $K_{sym}$, $S$ could almost become negative below $4n_{sat}$.
$K_{sym}$ is therefore the most influential empirical parameter governing the density dependence of the symmetry energy
from $2n_{sat}$ to $4n_{sat}$.

\begin{figure*}[tb]
\begin{center}
\includegraphics[angle=0,width=0.7\linewidth]{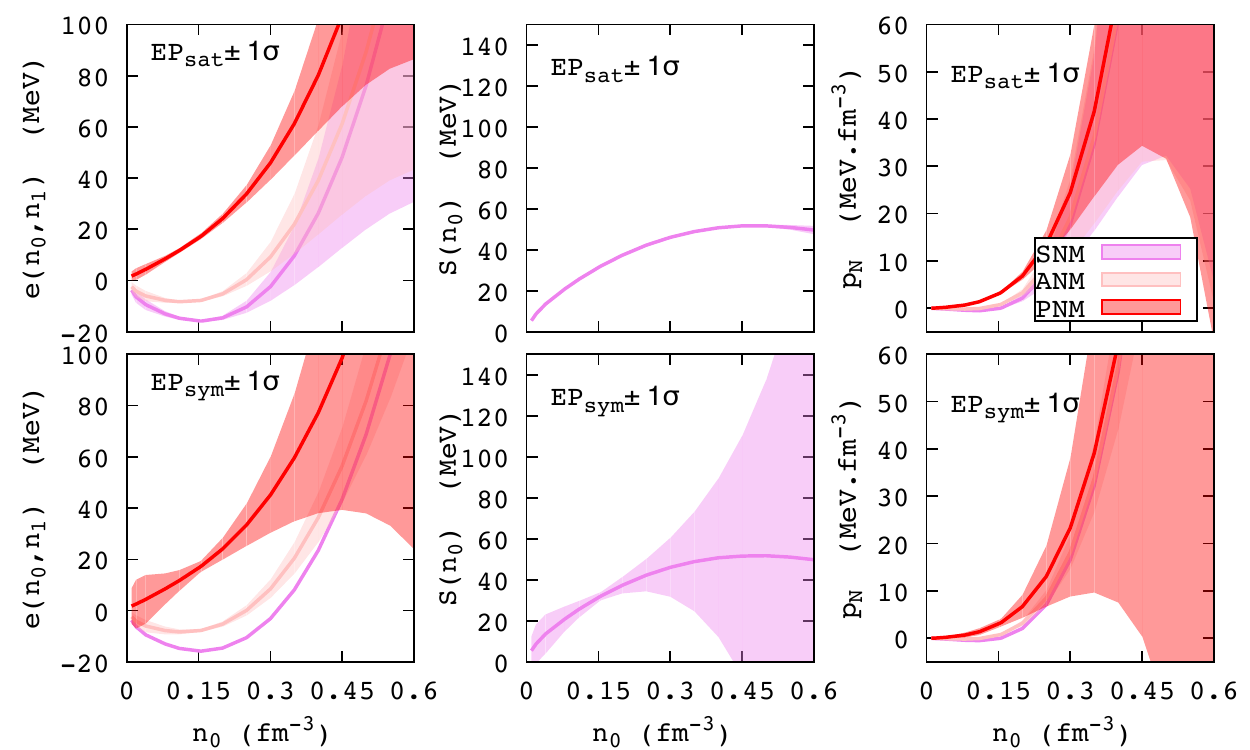}
\end{center}
\caption{(Color online) Effect of varying the value of all empirical parameters in the isoscalar (top panels) and isovector (bottom panels) channels.
From left to right: energy per nucleon, symmetry energy and pressure in SNM ($\delta=0$),  ANM ($\delta=0.5$), and  PNM ($\delta=1$).}
\label{fig:vary_all1} 
\end{figure*}

The impact of the variation of $Q_{sat/sym}$ is shown in  Fig.~\ref{fig:vary4}.
The uncertainty in $Q_{sat}$ has a quite large impact on the EOS and this is clearly the most important empirical parameter in
the isoscalar channel which is yet weakly known.
The uncertainty in $Q_{sym}$ has also an important impact on the EOS.
Its impact starts above $2n_{sat}$.

We show in Fig.~\ref{fig:vary5} the impact of our uncertainty in the parameters $Z_{sat/sym}$.
Both $Z_{sat}$ and $Z_{sym}$ have an impact on the EOS.
By comparing Figs.~~\ref{fig:vary2} to Fig.~\ref{fig:vary5}, it is interesting to observe that the higher the empirical parameter
the farther from $n_{sat}$ it has an impact.
This effect is limited by the increasing uncertainty in the empirical parameters as their order increases, but nevertheless
it is rather visible.
It is a consequence of the series expansion since the different order terms in the series expansion have a weight which goes
decreasing around saturation density as their order increases.

The last figure of this series, Fig.~\ref{fig:vary0}, shows the impact of the isoscalar and isovector splitting of the Landau effective mass,
e.g. $m^*_{sat}$ and $\Delta m^*_{sat}$.
It is very interesting to observe that the Landau effective mass $m^*_{sat}$ and $\Delta m^*_{sat}$ have a very weak impact on the EOS.
We remind that in our EOS, we are able to probe the impact of each empirical parameter separately, leaving the
other parameters unchanged.
It is clear from Fig.~\ref{fig:vary0} that the uncertainties in the Landau effective mass are very limited and almost un-observable for
the EOS. It is however expected to impact more crucially the single particle properties and the density of state, as well as 
the temperature dependence of the EOS.
This effect will be studied in a future work.


\begin{figure*}[tb]
\begin{center}
\includegraphics[angle=0,width=0.7\linewidth]{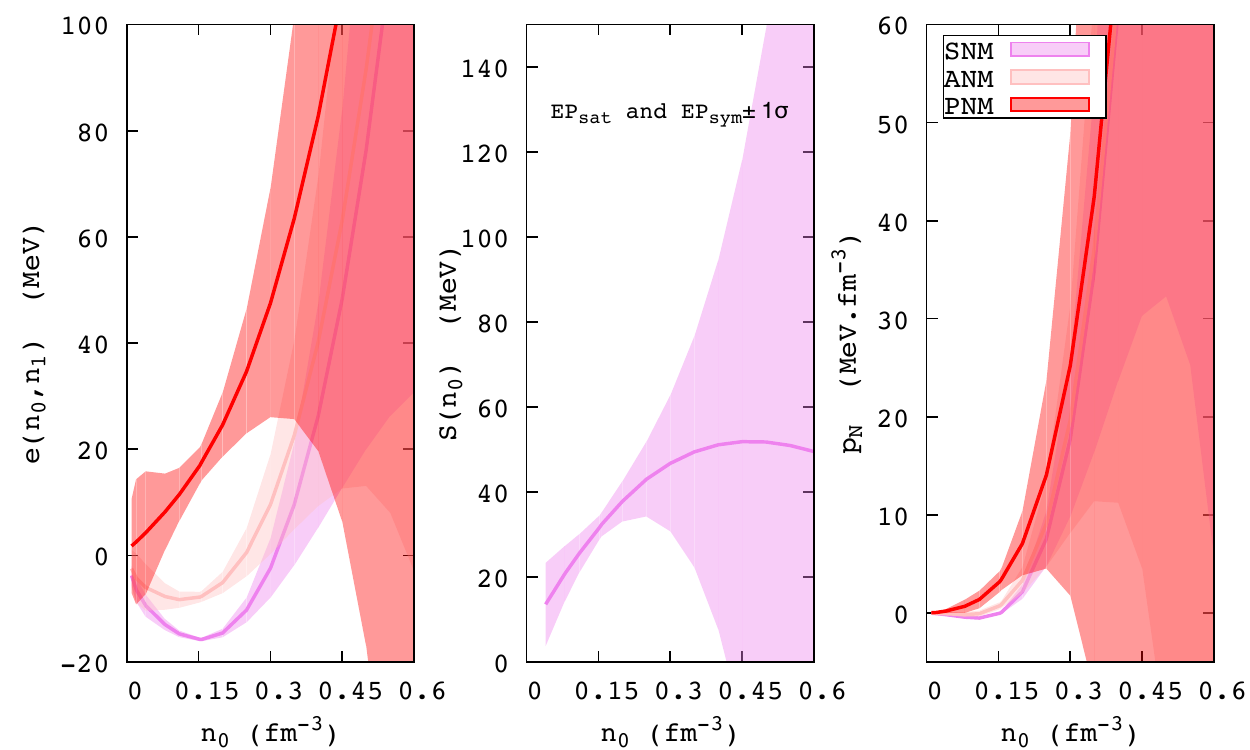}
\end{center}
\caption{(Color online) Same as Fig.~\ref{fig:vary_all1} varying the value of the empirical parameters all together.}
\label{fig:vary_all2} 
\end{figure*}


\subsection{Global effect varying isoscalar and isovector empirical parameters}

We now discuss globally the impact of the isoscalar and isovector empirical parameters.
They are varied independently all together within their own uncertainties, and the impact is shown in Fig.~\ref{fig:vary_all1} (for the isoscalar empirical parameters)
and in Fig.~\ref{fig:vary_all2} (for the isovector empirical parameters).

Looking at Figs.~\ref{fig:vary_all1} and \ref{fig:vary_all2} one has the impression that, in spite of the tremendous effort of the community since that last years,  the EOS above {2-3$n_{sat}$} is completely unknown. 
This impression is however not fully correct for at least two reasons.
We have seen in the preceding sections that the meta-EOS is sufficiently flexible to explore all possible density dependencies, thus being able to reproduce all existing models. The drawback of this flexibility is that it can also allow unphysical parameterizations. 
This can be clearly seen in  Figs.~\ref{fig:vary_all1} and \ref{fig:vary_all2}: 
{models with negative pressures or negative symmetry energies can be explored by the meta-EOS but are indeed forbidden by the stability requirement.} 
The speed of sound is not represented in these figures, but superluminal models also belong to our possible parameter sets. It is clear that these unphysical models have to be excluded before reasonable error bars on the equation of state can be calculated.

The second reason why the bands of Figs.~\ref{fig:vary_all1} and \ref{fig:vary_all2} cannot be interpreted as error bars on the EOS is that by construction the different parameters of the meta-EOS
are independent. This is a quality of the model, because it allows to do sensitivity studies and it avoids spurious correlations due to a too limited number of parameters in existing functionals. However, physical correlations among the empirical parameters can exist, like the ones shown by the ab-initio calculation in Fig.\ref{fig:Ingo2}. These correlations limit the parameter space and therefore the uncertainty intervals on the EOS. Such correlations can only be found if the ensemble of meta-EOS is filtered through the requirement of reproduction  of experimental or observational data. 
This statistical analysis of the meta-EOS can be performed with standard Bayesian techniques and is the object of a forthcoming paper~\cite{Margueron2017b}.

\section{Conclusions}
\label{Sec:Conclusions}

In this paper, we have presented a meta-EOS for uniform matter which is very simply related to the empirical parameters
characterizing the density and isospin density dependence of the EOS.
This meta-EOS is based on nucleonic degrees of freedom and assumes that they are non-relativistic.
It allows a natural implementation of our best knowledge of the nuclear empirical parameters
which is based on nuclear physics experiments, that we have discussed in some detail.

We have analyzed the confidence intervals of the empirical parameters obtained with different methods, namely from the direct analysis of experimental data, and from a statistical analysis of various theoretical modelings.
Phenomenological and ab-initio approaches, as well as relativistic and non-relativistic interactions, have been
analyzed in detail, and from this study, we have proposed a set of average values and estimated uncertainties
for all empirical parameters from $E_{sat/sym}$ up to the kurtosis ones $Z_{sat/sym}$, see Tab.~\ref{tab:epvar}.

Finally, we have analyzed the impact of the uncertainties on the empirical parameters on the meta-EOS of nucleonic matter.
We have deduced that the lowest order empirical parameters which require better determination in the future are
the skewness parameter $Q_{sat}$ in the isoscalar channel, and the slope of the symmetry energy $L_{sym}$ and
its curvature $K_{sym}$ in the isovector channel.
The determination of these parameters needs to depart substantially from saturation density, either below or above.
They could be determined either by relativistic heavy ion collision, see for instance Ref.~\cite{Danielewicz2001}, or as 
we already discussed by the mass and radii constraints of 
NS~\cite{Ozel2009,Ozel2010,Steiner2010a,Guillot2011,Guillot2013,Steiner2013,Lattimer2014}.
The observation of the gravitational waves coming from NS merging are also expected to give tight constraints on the
properties of dense matter~\cite{Abbott2017}.
Most probably, it is a combining of these different constraints which will be able to provide a better knowledge on these
empirical parameters.

We have also discussed the density dependence of the symmetry energy which is provided by
our meta-EOS, and we have shown that a rich behavior of the symmetry energy below and above 
$n_{sat}$ is possible by combining both $L_{sym}$ and $K_{sym}$.
If $K_{sym}$ is negligible, $L_{sym}$ alone provides the well known iso-soft (below) / iso-stiff (above)
or vice-versa, while if $K_{sym}$ is large enough, different behaviors can be generated: iso-soft (below) / iso-soft (above) 
or iso-stiff (below) / iso-stiff (above).

The meta-EOS which is presented in this paper {is an example of possible metamodeling.
Further extensions are possible, such as for instance implementing quartic dependence in the asymmetry parameter, considering relativistic kinetic energies instead of non-relativistic, or replacing the polynomial expansion in density by an expansion in Fermi momentum or other quantity. 
Considering the present meta-EOS, it}
will be applied to  $\beta$-equilibrium matter in neutron stars in a forthcoming
paper~\cite{Margueron2017b}.
In the next future, we want also to use this meta-EOS as a density functional and apply it to the determination of the global structure 
of finite nuclei~\cite{Chatterjee2017}.
In our future application of this meta-EOS, its simple relation to the empirical parameters is expected to easily highlight the
role of the empirical parameters in the nuclear properties.
Empirical parameters are indeed a simple way to encode the basic properties of nuclear matter and provides a link between
nuclear physics experiments and astrophysical applications. 
They therefore constitute an interesting and promising tool for a better synergy between nuclear physics and astrophysics.
Finally, the interesting aspect of this meta-EOS is its simplicity and richness. 
Since it is simple, it shall be easy to implement it in many different modelings, going from the global structure of finite nuclei to more complex dynamical simulations. 
For neutron star physics, it offers the possibility to provide a simple approach which can encode many existing EOS for uniform matter, as well as new EOS, which properties could be tested against  observations.
It is therefore an interesting theoretical tool which could bring more consistency in the exploration of the nuclear EOS from different branches of nuclear physics and astrophysics.

\section{Acknowledgments}

This work was partially supported by CNPq (Brazil), processes (209440/2013-9) and (400877/2015-5), by the SN2NS project ANR-10-BLAN-0503 and by New-CompStar COST action MP1304.
J.M. thanks N. Baillot, G. Bertsch, K. Bennaceur, A. Bulgac, G. Col\`o, S\'ebastien Guillot and J. Meyer, stimulating discussions and interesting suggestions
during the completion of this work.
J.M. thanks Ingo Tews as well for providing him the data for the energy per nucleon and unpublished pressure from his
work~\cite{Tews2013}, and J.-J. Li for providing tables of empirical parameters for relativistic approaches.

\appendix

\section{Detailed tables of empirical parameters}
\label{Sec:tables}

In this appendix, we provide more details on the general analysis of the empirical parameters deduced from
nuclear interactions or effective Lagrangians.

\begin{table*}[tb]
\begin{center}
\setlength{\tabcolsep}{2pt}
\renewcommand{\arraystretch}{1.2}
\begin{ruledtabular}
 \begin{tabular}{cccccccccccccccccc}
    & &  $E_{sat}$ & $E_{sym}$ & $
n_{sat}$ &  $L_{sym}$ & $K_{sat}$ &  $K_{sym}$  & $Q_{sat}$  &  $Q_{sym}$ & $Z_{sat}$ & $Z_{sym}$  & 
$m^*_{sat}/m$ &$\Delta m^*_{sat}/m$ & $\kappa_{v}$ & $K_\tau$ \\
Model  & $N_{models}$ & MeV & MeV & fm$^{-3}$ & MeV & MeV & MeV & MeV & MeV & MeV & MeV & &  & & MeV\\
\hline
SGII~\cite{SKY:SGII}         &   1  &    -15.59  &     26.83  &    0.1583  &      37.6  &      215  &     -146  &     -381  &      330  &     1742  &    -1891  &      0.79  &      0.28  &      0.49  &     -305 \\
RATP~\cite{SKY:RATP}        &   2  &    -16.05  &     29.26  &    0.1598  &      32.4  &      240  &     -191  &     -350  &      440  &     1452  &    -2477  &      0.67  &      0.26  &      0.78  &     -338 \\
SKM*~\cite{SKY:SKMs}        &   3  &    -15.75  &     30.04  &    0.1602  &      45.8  &      216  &     -156  &     -386  &      330  &     1766  &    -1868  &      0.79  &      0.34  &      0.53  &     -349 \\
SKI2~\cite{SKY:SKI}        &   4  &    -15.76  &     33.37  &    0.1575  &     104.3  &      241  &       71  &     -339  &       52  &     1349  &     -610  &      0.69  &     -0.21  &      0.24  &     -408 \\
SKI4~\cite{SKY:SKI}        &   5  &    -15.93  &     29.50  &    0.1601  &      60.4  &      248  &      -41  &     -331  &      351  &     1278  &    -2235  &      0.65  &     -0.25  &      0.25  &     -322 \\
BSK14~\cite{SKY:BSK14}       &   6  &    -15.85  &     30.00  &    0.1586  &      43.9  &      239  &     -152  &     -359  &      389  &     1435  &    -2191  &      0.80  &      0.03  &      0.28  &     -350 \\
BSK16~\cite{SKY:BSK16}       &   7  &    -16.05  &     30.00  &    0.1586  &      34.9  &      242  &     -187  &     -364  &      462  &     1460  &    -2566  &      0.80  &      0.04  &      0.28  &     -344 \\
BSK17~\cite{SKY:BSK17}       &   8  &    -16.05  &     30.00  &    0.1586  &      36.3  &      242  &     -182  &     -364  &      451  &     1460  &    -2508  &      0.80  &      0.04  &      0.28  &     -345 \\
SLY4~\cite{SKY:SLY}        &   9  &    -15.97  &     32.01  &    0.1595  &      46.0  &      230  &     -120  &     -363  &      521  &     1587  &    -3197  &      0.69  &     -0.19  &      0.25  &     -323 \\
SLY5~\cite{SKY:SLY}        &  10  &    -15.98  &     32.03  &    0.1604  &      48.3  &      230  &     -112  &     -364  &      501  &     1592  &    -3087  &      0.70  &     -0.18  &      0.25  &     -326 \\
T44~\cite{SKY:T44}         &  11  &    -16.02  &     32.00  &    0.1612  &      50.0  &      230  &     -107  &     -366  &      481  &     1603  &    -2972  &      0.70  &     -0.18  &      0.25  &     -327 \\
LNS1~\cite{SKY:LNS}        &  12  &    -15.90  &     29.91  &    0.1616  &      30.9  &      244  &     -211  &     -325  &      444  &     1299  &    -2471  &      0.60  &      0.34  &      1.09  &     -356 \\
LNS5~\cite{SKY:LNS}        &  13  &    -15.56  &     29.15  &    0.1599  &      50.9  &      240  &     -119  &     -316  &      286  &     1255  &    -1671  &      0.60  &      0.23  &      0.97  &     -358 \\
SAMI~\cite{SKY:SAMI}        &  14  &    -15.93  &     28.16  &    0.1587  &      43.7  &      245  &     -120  &     -339  &      372  &     1331  &    -2179  &      0.68  &      0.02  &      0.51  &     -322 \\
UNEDF1~\cite{SKY:UNEDF1}      &  15  &    -15.80  &     28.99  &    0.1587  &      40.0  &      220  &     -179  &     -404  &      324  &     1781  &    -1744  &      1.01  &      0.56  &      0.25  &     -346 \\
NRAPR~\cite{SKY:NRAPR}       &  16  &    -15.85  &     32.78  &    0.1606  &      59.7  &      226  &     -123  &     -363  &      312  &     1611  &    -1838  &      0.69  &      0.21  &      0.66  &     -385 \\
 \hline
     Average  &  &    -15.88  &     30.25  &    0.1595  &      47.8  &      234  &     -130  &     -357  &      378  &     1500  &    -2219  &      0.73  &      0.08  &      0.46  &     -344 \\
    $\sigma$  &  &      0.15  &      1.70  &    0.0011  &      16.8  &       10  &       66  &       22  &      110  &      169  &      618  &      0.10  &      0.24  &      0.27  &       25 \\
         Min  &  &    -16.05  &     26.83  &    0.1575  &      30.9  &      215  &     -211  &     -404  &       52  &     1255  &    -3197  &      0.60  &     -0.25  &      0.24  &     -408 \\
         Max  &  &    -15.56  &     33.37  &    0.1616  &     104.3  &      248  &       71  &     -316  &      521  &     1781  &     -610  &      1.01  &      0.56  &      1.09  &     -305 \\
 
\hline
FPL~\cite{SKY:FPLYON}         &  17  &    -15.92  &     30.93  &    0.1619  &      42.8  &      217  &     -136  &     -399  &      486  &     1833  &    -2913  &      0.84  &     -0.23  &      0.03  &     -314 \\
SKGSIGMA~\cite{SKY:E_ES_GS_RS}    &  18  &    -15.59  &     31.37  &    0.1576  &      94.0  &      237  &       14  &     -349  &      -27  &     1379  &       -5  &      0.78  &      0.25  &      0.48  &     -412 \\
SKRSIGMA~\cite{SKY:E_ES_GS_RS}    &  19  &    -15.59  &     30.58  &    0.1577  &      85.7  &      237  &       -9  &     -348  &       22  &     1377  &     -255  &      0.78  &      0.25  &      0.48  &     -397 \\
SKX~\cite{SKY:SKX}         &  20  &    -16.05  &     31.10  &    0.1554  &      33.2  &      271  &     -252  &     -297  &      379  &      904  &    -1889  &      0.99  &      0.72  &      0.33  &     -415 \\
SIII~\cite{SKY:SIII}        &  21  &    -15.85  &     28.16  &    0.1453  &       9.9  &      355  &     -394  &      101  &      131  &     -903  &     -799  &      0.76  &      0.26  &      0.53  &     -456 \\
SV~\cite{SKY:SIII}          &  22  &    -16.05  &     32.82  &    0.1551  &      96.1  &      306  &       24  &     -176  &       48  &      183  &     -481  &      0.38  &      0.12  &      2.02  &     -497 \\
SLY230A~\cite{SKY:SLY230}     &  23  &    -15.99  &     31.99  &    0.1600  &      44.3  &      230  &      -98  &     -364  &      603  &     1594  &    -3786  &      0.70  &     -0.47  &      0.00  &     -294 \\
SLY230B~\cite{SKY:SLY230}     &  24  &    -15.97  &     32.01  &    0.1595  &      46.0  &      230  &     -120  &     -363  &      521  &     1587  &    -3198  &      0.69  &     -0.19  &      0.25  &     -323 \\
F0~\cite{SKY:F0}          &  25  &    -16.03  &     32.00  &    0.1617  &      42.4  &      230  &     -113  &     -405  &      658  &     1705  &    -3870  &      0.70  &      0.00  &      0.43  &     -293 \\
F+~\cite{SKY:F0}          &  26  &    -16.04  &     32.00  &    0.1618  &      41.5  &      230  &     -118  &     -406  &      661  &     1710  &    -3875  &      0.70  &      0.17  &      0.60  &     -294 \\
F-~\cite{SKY:F0}          &  27  &    -16.02  &     32.00  &    0.1616  &      43.8  &      230  &     -105  &     -405  &      655  &     1702  &    -3869  &      0.70  &     -0.28  &      0.15  &     -291 \\
LNS~\cite{SKY:LNS}         &  28  &    -15.31  &     33.43  &    0.1746  &      61.5  &      211  &     -127  &     -383  &      303  &     1749  &    -1766  &      0.83  &      0.23  &      0.38  &     -385 \\
UNEDF0~\cite{SKY:UNEDF0}      &  29  &    -16.06  &     30.54  &    0.1605  &      45.1  &      230  &     -190  &     -404  &      288  &     1707  &    -1495  &      1.11  &      1.02  &      0.25  &     -381 \\
SKMP~\cite{SKY:SKMP}        &  30  &    -15.56  &     29.89  &    0.1570  &      70.3  &      231  &      -50  &     -338  &      159  &     1424  &    -1020  &      0.65  &      0.16  &      0.71  &     -369 \\
SKO~\cite{SKY:SKO}         &  31  &    -15.83  &     31.97  &    0.1605  &      79.1  &      223  &      -43  &     -393  &      131  &     1720  &     -851  &      0.90  &      0.09  &      0.17  &     -379 \\
SKOP~\cite{SKY:SKO}        &  32  &    -15.75  &     31.95  &    0.1602  &      68.9  &      222  &      -79  &     -391  &      223  &     1710  &    -1349  &      0.90  &      0.05  &      0.15  &     -371 \\
SKP~\cite{SKY:SKP}         &  33  &    -15.95  &     30.00  &    0.1625  &      19.7  &      201  &     -267  &     -436  &      508  &     2128  &    -2748  &      1.00  &      0.80  &      0.35  &     -342 \\
Skz2~\cite{SKY:SKZ2}        &  34  &    -16.00  &     32.01  &    0.1600  &      16.8  &      230  &     -260  &     -365  &      682  &     1598  &    -3894  &      0.70  &      0.14  &      0.57  &     -334 \\
T6~\cite{SKY:T6}          &  35  &    -15.96  &     29.97  &    0.1609  &      30.9  &      236  &     -212  &     -383  &      473  &     1561  &    -2562  &      1.00  &      0.00  &      0.00  &     -347 \\
 \hline
     Average  &  &    -15.87  &     30.82  &    0.1596  &      49.6  &      237  &     -132  &     -349  &      370  &     1448  &    -2175  &      0.77  &      0.13  &      0.43  &     -354 \\
    $\sigma$  &  &      0.18  &      1.54  &    0.0039  &      21.6  &       27  &       89  &       88  &      188  &      510  &     1069  &      0.14  &      0.31  &      0.37  &       45 \\
         Min  &  &    -16.06  &     26.83  &    0.1453  &       9.9  &      201  &     -394  &     -436  &      -27  &     -903  &    -3894  &      0.38  &     -0.47  &      0.00  &     -497 \\
         Max  &  &    -15.31  &     33.43  &    0.1746  &     104.3  &      355  &       71  &      101  &      682  &     2128  &       -5  &      1.11  &      1.02  &      2.02  &     -291 \\
 
 \end{tabular}
\end{ruledtabular}
 \caption{Empirical properties of nuclear matter for Skyrme-type interactions.}
\label{table:empiricalsky}
\end{center}
\end{table*}

\begin{table*}[tb]
\begin{center}
\setlength{\tabcolsep}{2pt}
\renewcommand{\arraystretch}{1.2}
\begin{ruledtabular}
 \begin{tabular}{cccccccccccccccccc}
    & &  $E_{sat}$ & $E_{sym}$ & $
n_{sat}$ &  $L_{sym}$ & $K_{sat}$ &  $K_{sym}$  & $Q_{sat}$  &  $Q_{sym}$ & $Z_{sat}$ & $Z_{sym}$  & 
$m^*_{sat}/m$ &$\Delta m^*_{sat}/m$ & $\kappa_{v}$ & $K_\tau$ \\
Model  & $N_{models}$ & MeV & MeV & fm$^{-3}$ & MeV & MeV & MeV & MeV & MeV & MeV & MeV & &  &  & MeV \\
\hline
DDME1~\cite{RMFDD:DDME1}            &  36  &    -16.20  &     33.07  &    0.1520  &      55.5  &      245  &     -101  &      317  &      705  &     4867  &    -5717  &      0.66  &     -0.06  &      0.45  &     -506 \\
DDME2~\cite{RMFDD:DDME2}            &  37  &    -16.14  &     32.31  &    0.1520  &      51.3  &      251  &      -87  &      479  &      777  &     4448  &    -7048  &      0.65  &     -0.06  &      0.47  &     -493 \\
DDME$\delta$~\cite{RMFDD:DDMEd}     &  38  &    -16.12  &     32.35  &    0.1520  &      52.8  &      219  &     -118  &     -748  &      846  &     3950  &    -3545  &      0.69  &     -0.17  &      0.27  &     -255 \\
NL3~\cite{RMFNL:NL3}              &  39  &    -16.24  &     37.35  &    0.1480  &     118.3  &      271  &      101  &      198  &      182  &     9302  &    -3961  &      0.67  &     -0.08  &      0.40  &     -696 \\
NL3s~\cite{RMFNL:NL3s}             &  40  &    -16.31  &     38.71  &    0.1500  &     122.7  &      259  &      106  &      124  &      224  &     9997  &    -3920  &      0.67  &     -0.09  &      0.39  &     -690 \\
NL-SH~\cite{RMFNL:NLSH}            &  41  &    -16.35  &     36.12  &    0.1460  &     113.7  &      355  &       80  &      602  &      -23  &     5061  &    -4264  &      0.67  &     -0.08  &      0.40  &     -795 \\
PK1~\cite{RMF:PK}              &  42  &    -16.27  &     37.59  &    0.1480  &     115.7  &      282  &       55  &      -29  &      -86  &     4008  &    -2866  &      0.68  &     -0.08  &      0.38  &     -627 \\
PK1R~\cite{RMF:PK}             &  43  &    -16.27  &     37.78  &    0.1480  &     116.3  &      283  &       56  &      -21  &      -86  &     4032  &    -2902  &      0.68  &     -0.08  &      0.38  &     -634 \\
PKDD~\cite{RMF:PK}             &  44  &    -16.27  &     31.19  &    0.1495  &      79.5  &      261  &      -50  &     -119  &      -28  &     4213  &    -1315  &      0.65  &     -0.08  &      0.44  &     -491 \\
TM1~\cite{RMFNL:TM1}              &  45  &    -16.26  &     36.94  &    0.1450  &     111.0  &      281  &       34  &     -285  &      -67  &     2014  &    -1546  &      0.71  &     -0.09  &      0.32  &     -520 \\
TW99~\cite{RMFDD:TW99}             &  46  &    -16.25  &     32.77  &    0.1530  &      55.3  &      240  &     -125  &     -540  &      539  &     3749  &    -3307  &      0.64  &     -0.06  &      0.49  &     -332 \\
 \hline
     Average  &  &    -16.24  &     35.11  &    0.1494  &      90.2  &      268  &       -5  &       -2  &      271  &     5058  &    -3672  &      0.67  &     -0.08  &      0.40  &     -549 \\
    $\sigma$  &  &      0.06  &      2.63  &    0.0025  &      29.6  &       34  &       88  &      393  &      357  &     2294  &     1582  &      0.02  &      0.03  &      0.06  &      153 \\
         Min  &  &    -16.35  &     31.19  &    0.1450  &      51.3  &      219  &     -125  &     -748  &      -86  &     2014  &    -7048  &      0.64  &     -0.17  &      0.27  &     -795 \\
         Max  &  &    -16.12  &     38.71  &    0.1530  &     122.7  &      355  &      106  &      602  &      846  &     9997  &    -1315  &      0.71  &     -0.06  &      0.49  &     -255 \\
 
 \end{tabular}
\end{ruledtabular}
\caption{Same as Table~\ref{table:empiricalsky} for RMF models.}
 \label{table:empiricalrmf}
\end{center}
\end{table*}

\begin{table*}[tb]
\begin{center}
\setlength{\tabcolsep}{2pt}
\renewcommand{\arraystretch}{1.2}
\begin{ruledtabular}
 \begin{tabular}{cccccccccccccccccc}
    & &  $E_{sat}$ & $E_{sym}$ & $
n_{sat}$ &  $L_{sym}$ & $K_{sat}$ &  $K_{sym}$  & $Q_{sat}$  &  $Q_{sym}$ & $Z_{sat}$ & $Z_{sym}$  & 
$m^*_{sat}/m$ &$\Delta m^*_{sat}/m$ & $\kappa_{v}$ & $K_\tau$ \\
Model  & $N_{models}$ & MeV & MeV & fm$^{-3}$ & MeV & MeV & MeV & MeV & MeV & MeV & MeV & &  &  & MeV \\
\hline
PKA1~\cite{RHF:PKA1}        &  47  &    -15.83  &     36.02  &    0.1600  &     103.5  &      230  &      213  &      950  &      292  &     4935  &   -16916  &      0.68  &     -0.02  &      0.45  &     -835 \\
PKO1~\cite{RHF:PKO}        &  48  &    -16.00  &     34.37  &    0.1520  &      97.7  &      250  &      106  &      262  &      290  &     4857  &    -5993  &      0.75  &     -0.03  &      0.31  &     -583 \\
PKO2~\cite{RHF:PKO}        &  49  &    -16.03  &     32.49  &    0.1510  &      75.9  &      250  &       77  &      -10  &      821  &     6703  &    -7993  &      0.76  &     -0.02  &      0.30  &     -375 \\
PKO3~\cite{RHF:PKO}        &  50  &    -16.04  &     32.99  &    0.1530  &      83.0  &      262  &      116  &      355  &      690  &     4581  &    -8921  &      0.76  &     -0.03  &      0.29  &     -494 \\
 \hline
     Average  &  &    -15.97  &     33.97  &    0.1540  &      90.0  &      248  &      128  &      389  &      523  &     5269  &    -9955  &      0.74  &     -0.02  &      0.34  &     -572 \\
    $\sigma$  &  &      0.08  &      1.37  &    0.0035  &      11.1  &       12  &       51  &      350  &      237  &      838  &     4156  &      0.03  &      0.00  &      0.07  &      169 \\
         Min  &  &    -16.04  &     32.49  &    0.1510  &      75.9  &      230  &       77  &      -10  &      290  &     4581  &   -16916  &      0.68  &     -0.03  &      0.29  &     -835 \\
         Max  &  &    -15.83  &     36.02  &    0.1600  &     103.5  &      262  &      213  &      950  &      821  &     6703  &    -5993  &      0.76  &     -0.02  &      0.45  &     -375 \\
 
 \end{tabular}
\end{ruledtabular}
\caption{Same as Table~\ref{table:empiricalsky} for RHF models.}
 \label{table:empiricalrhf}
\end{center}
\end{table*}

The empirical parameters provided by Skyrme interactions are given in Tab.~\ref{table:empiricalsky}.
This table is separated into two parts, the first 16th Skyrme models and the others.
The reason for this separation is explain in sec.~\ref{sec:empiricalparameters}.
We briefly summarize it: The first 16th Skyrme models have been selected since there are widely used interactions.
In addition, we have limited the number of models per groups generating these interactions, in order to mix
as much as possible the various assumptions in the fitting protocols.
The other Skyrme interactions, from 17th to 35th, are also Skyrme interactions widely used either in finite nuclei (for most of them),
either in nuclear matter.
They are there to test the sensitivity of the statistical analysis based on the first 16th.

In Tab.~\ref{table:empiricalrmf} we have listed the RMF effective Lagrangians that we have studied in this
paper.
Let us note that the effective mass reported in this table is the Landau (non-relativistic) one deduced from the
momentum dependence of the non-relativistic energy-density.

Finally, we list in Tab.~\ref{table:empiricalrhf} a few number of RHF effective Lagrangians.
The small number of models in this table is due to the very recent development of such approaches.
It shall be remarked that these effective Lagrangians have been determined by a single group, and it
would be interesting in the future to see more of these modelings.

\begin{table*}[t]
\centering
\setlength{\tabcolsep}{2pt}
\renewcommand{\arraystretch}{1.2}
\begin{ruledtabular}
\begin{tabular}{cccccccccccccccccc}
    &  $Q_{sat}$     &  $Q_{sym}$ & $Q_{sym}$  & $Z_{sat}$  & $Z_{sym}$  & $Z_{sym}$    & $\sigma_{e} (SM) $  & $\sigma_{e}(NM)$    & $\sigma_{e}(NM)$  \\
    &                      &                     & ($v_{4}$=0) &                   &                    & ($v_{4}$=0)  &                                  &                                  &  ($v_{4}$=0)\\       
 Model    &   MeV              &  MeV           & MeV            & MeV           & MeV            & MeV             &             -                    &             -                    &        -            \\
SGII~\cite{SKY:SGII}   &  -225.98 & 272.22 & 229.89 & -535.75 & -1040.95 & -871.63 &  0.28 &  0.17 &  0.27\\
RATP~\cite{SKY:RATP}   &  -222.18 & 347.83 & 279.10 & -414.59 & -1126.54 & -851.61 &  0.24 &  0.07 &  0.58\\
SKM*~\cite{SKY:SKMs}   &  -228.35 & 277.66 & 239.32 & -542.05 & -1097.83 & -944.48 &  0.29 &  0.19 &  0.24\\
SKI2~\cite{SKY:SKI}   &  -214.11 & 25.49 &  5.56 & -447.81 & -223.15 & -143.41 &  0.23 &  0.18 &  0.10\\
SKI4~\cite{SKY:SKI}   &  -212.90 & 224.29 & 127.37 & -410.38 & -383.25 &  4.42 &  0.22 &  0.04 &  0.90\\
BSK14~\cite{SKY:BSK14} & -217.69 & 297.78 & 227.08 & -578.08 & -884.03 & -601.20 &  0.26 &  0.09 &  0.58\\
BSK16~\cite{SKY:BSK16} & -221.14 & 350.01 & 262.83 & -573.66 & -952.84 & -604.13 &  0.26 &  0.05 &  0.76\\
BSK17~\cite{SKY:BSK17} & -221.14 & 341.95 & 257.21 & -573.68 & -941.63 & -602.66 &  0.26 &  0.06 &  0.74\\
SLY4~\cite{SKY:SLY}    & -225.01 & 350.68 & 226.01 & -443.11 & -690.35 & -191.68 &  0.26 &  0.05 &  1.19\\
SLY5~\cite{SKY:SLY}    & -223.62 & 337.37 & 217.90 & -451.22 & -679.62 & -201.75 &  0.25 &  0.05 &  1.13\\
\hline                                                  
DDME1~\cite{RMFDD:DDME1}    & 564.80 & 328.92 & 263.14 & -2377.81 & -1297.60 & -1034.45 &  0.62 &  0.81 &  1.38\\
DDME2~\cite{RMFDD:DDME2}    & 657.79 & 318.84 & 248.12 & -2611.93 & -1317.33 & -1034.45 &  0.80 &  1.12 &  1.74\\
DDMEd~\cite{RMFDD:DDMEd}    & -368.60 & 384.67 & 333.79 & -353.91 & -1237.96 & -1034.45 &  0.76 &  1.23 &  1.01\\
NL3~\cite{RMFNL:NL3}    & 1307.33 & -571.12 & -266.85 & -4049.23 & 200.86 & -1016.22 &  1.21 &  1.00 &  2.07\\
NL3$^*$~\cite{RMFNL:NL3s}   & 1488.08 & -675.42 & -305.43 & -4478.35 & 454.57 & -1025.36 &  1.71 &  1.14 &  2.74\\
NL-SH~\cite{RMFNL:NLSH} & 965.63 & -472.62 & -207.75 & -3542.60 & 52.43 & -1007.05 &  1.36 &  1.21 &  1.46\\
PK1~\cite{RMF:PK}   & 240.19 & -316.97 & -125.65 & -2007.56 & -250.93 & -1016.22 &  0.57 &  0.70 &  1.40\\
PK1R~\cite{RMF:PK}  & 249.68 & -320.53 & -127.58 & -2031.73 & -244.39 & -1016.22 &  0.58 &  0.72 &  1.40\\
PKDD~\cite{RMF:PK}  & 175.33 & 125.02 & 95.48 & -1556.55 & -1141.26 & -1023.08 &  0.34 &  4.69 &  4.82\\
TM1~\cite{RMFNL:TM1}   & -105.20 & -187.08 & -51.79 & -1188.79 & -461.28 & -1002.45 &  0.19 &  0.31 &  1.43\\
TW99~\cite{RMFDD:TW99}   & -206.47 & 400.18 & 317.66 & -729.57 & -1369.06 & -1038.98 &  0.56 &  0.46 &  0.52\\
\hline                                                  
PKA1~\cite{RHF:PKO}   & 877.58 & -1176.76 & -584.77 & -3249.25 & 1297.52 & -1070.43 &  1.74 &  1.05 &  5.72\\
PKO1~\cite{RHF:PKO}   & 518.91 & -489.42 & -238.03 & -2342.01 & -28.93 & -1034.45 &  0.62 &  0.41 &  2.35\\
PKO2~\cite{RHF:PKO}   & 432.64 & -169.80 & -85.68 & -2088.47 & -693.41 & -1029.91 &  0.55 &  0.42 &  0.64\\
PKO3~\cite{RHF:PKO}   & 595.24 & -424.75 & -216.59 & -2533.23 & -206.36 & -1038.98 &  0.72 &  0.63 &  1.68\\

\end{tabular}
\end{ruledtabular}
\caption{Modification of the empirical parameters, $Q_{sat/sym}$ and $Z_{sat/sym}$ in the model ELFd adjusted
to reproduce reference models.}
\label{table:Nuclear-New3-func-Table2}
\end{table*}

In Tab.~\ref{table:Nuclear-New3-func-Table2} are shown the empirical parameters $Q_{sat/sym}$ and 
$Z_{sat/sym}$ which are obtained from the meta-EOS ELFd.
The standard deviation in the energy per nucleon in symmetric matter $\sigma_e(SM)$ and in neutron matter $\sigma_e(NM)$
between ELFd and the original prediction from each of the listed models are also given, showing the
very good agreement between the meta-EOS ELFd and the original interaction.

The number of parameters in the meta-EOS, including the effective mass parameters, is 12. 
It could be interesting to perform further analysis to reduce a bit this number.
For this purpose, we have explored the impact of removing the highest order isovector empirical parameter, by
setting $v_{4}^{iv}=0$, on the comparison between the EOS ELFd and the original EOS.
The results are given in Tab.~\ref{table:Nuclear-New3-func-Table2}.
It is clear that the impact is extremely small, showing that this parameter has a weak influence on the EOS below $4n_{sat}$.

\bibliographystyle{apsrev}

\end{document}